\documentclass[12pt]{article}
\usepackage{amsmath,amsthm,amssymb,bm}
\usepackage{graphicx,psfrag,epsf,algorithm,algpseudocode}
\usepackage{enumerate,multirow,array,color,natbib,url,booktabs,setspace}
\usepackage[font=small,labelfont=bf]{caption}
\usepackage[total={6.45in,8.75in}]{geometry}
\newcommand{\blind}{1}

\newtheorem{thm}{Theorem}
\newtheorem{coro}{Corollary}
\newtheorem{prop}{Proposition}
\newtheorem{lemma}{Lemma}

\newtheoremstyle{exampstyle}
{\topsep} 
{\topsep} 
{} 
{} 
{\bfseries} 
{.} 
{.5em} 
{} 

\let\proglang=\textsf

\newcommand{\Mean}{{\mbox{E}}}
\newcommand{\Var}{{\mbox{Var}}}
\newcommand{\Cov}{{\mbox{cov}}}
\newcommand{\Corr}{{\mbox{corr}}}

\newcommand{\prob}{{\mbox{Pr}}}
\def\eop{\hfill $\Box$}

\DeclareMathOperator*{\argmin}{arg\,min}

\def\spacingset#1{\renewcommand{\baselinestretch}%
	{#1}\small\normalsize} \spacingset{1}

\newcommand{\change}[1]{{\leavevmode\color{black}{#1}}}

\UseRawInputEncoding

\begin{document}


\if1\blind
{
\title{\Large{\textbf{Testing Mediation Effects Using Logic of\\}}
\Large{\textbf{Boolean Matrices}}}
\author{
\large{Chengchun Shi and  Lexin Li} \\
\\
\normalsize{\textit{London School of Economics and Political Science}}\\
\normalsize{\textit{and University of California at Berkeley}} 
}
\date{}
\maketitle
} \fi

\if0\blind
{
\title{\Large{\textbf{Testing Mediation Effects Using Logic of\\}}
\Large{\textbf{Boolean Matrices}}}
\author{
}
\date{}
\maketitle
} \fi

\bigskip
\begin{abstract}
A central question in high-dimensional mediation analysis is to infer the significance of individual mediators. The main challenge is that the total number of potential paths that go through any mediator is super-exponential in the number of mediators. Most existing mediation inference solutions either explicitly impose that the mediators are conditionally independent given the exposure, or ignore any potential directed paths among the mediators. In this article, we propose a novel hypothesis testing procedure to evaluate individual mediation effects, while taking into account potential interactions among the mediators. Our proposal thus fills a crucial gap, and greatly extends the scope of existing mediation tests. Our key idea is to construct the test statistic using the logic of Boolean matrices, which enables us to establish the proper limiting distribution under the null hypothesis. We further employ screening, data splitting, and decorrelated estimation to reduce the bias and increase the power of the test. We show that our test can control both the size and false discovery rate asymptotically, and the power of the test approaches one, while allowing the number of mediators to diverge to infinity with the sample size. We demonstrate the efficacy of the method through simulations and a neuroimaging study of Alzheimer's disease. A \proglang{Python} implementation of the proposed procedure is available at \url{https://github.com/callmespring/LOGAN}. 
\end{abstract}

\vfill

\newpage
 
\baselineskip=20pt

\section{Introduction}

Mediation analysis is an important tool in scientific studies. It seeks to identify and explain the mechanism, or pathway, that underlies an observed relationship between an exposure and an outcome variable, through the inclusion of an intermediary variable, known as a mediator. It decomposes the effect of exposure on the outcome into a direct effect and an indirect effect, the latter of which is often of primary interest and has important intervention consequences \citep{Pearl2001}. Mediation analysis was first proposed with a single mediator in social science \citep{Baron1986}. In recent years, it is receiving increasing attention, and has been extended to the settings of multivariate and high-dimensional mediators. It is now widely used in a large variety of scientific applications, including psychology \citep{MacKinnon2009}, genomics \citep{Huang2016}, genetic epidemiology \citep{Huang2018}, and neuroscience \citep{ZhaoLuo2016}. 

In mediation analysis with high-dimensional mediators, a fundamental but challenging question is how to infer the significance of individual mediators. The main difficulty is the sheer number of possible paths that go through all combinations of the mediators. Consequently, the total number of potential paths that go through any mediator is super-exponential in the number of mediators, rendering almost any existing testing procedure ineffective. To circumvent this issue, most existing mediation inference solutions either explicitly impose that the mediators are conditionally independent given the exposure, or simply ignore any potential directed paths among the mediators. Such simplifications substantially reduce the complexity of the hypotheses to test. Adopting this conditional independence assumption, \cite{boca2014testing} proposed a permutation test with family-wise error control, while \cite{zhang2016estimating} proposed a screening-and-testing assisted approach. \cite{Huang2016} proposed a transformation model and assumed conditional independence for the transformed mediators. \citet{sampson2018fwer} and \citet{Djor2019} directly tested whether each mediator is independent of the exposure or conditionally independent of the outcome given the exposure, ignoring mediator-by-mediator interactions, while controlling for family-wise error rate or false discovery rate in multiple testing. Whereas these tests have been demonstrated effective in numerous applications, they all ignored potential paths and interactions among the mediators. Even though this strategy may be plausible in some applications, it may not hold true in others. For instance, in our brain imaging mediation analysis study in Section \ref{sec:real-data}, different brain regions are conceived to influence each other. In genetics studies, different genes are expected to interact with each other \citep{LHZ2018}. Actually, such examples are often the rule rather than the exception. Therefore, it is of great importance to develop a mediation testing method that takes directed paths and interactions among the mediators into consideration. 

Recently, \citet{LHZ2018} made an important step forward for inference of mediation effects while allowing mediator interactions. They formulated the structure of the exposure, the potential mediators, and the outcome as a directed acyclic graph (DAG). They defined the individual mediation effect of a given mediator as the summation of all the effects of the exposure on the outcome that can be attributed to that mediator. They then established the corresponding confidence interval for their interventional calculus type estimator of the mediation effect. However, the effects along different paths may cancel each other, resulting in a zero individual mediation effect in the summation. Rather than taking average and cancelling out the total effect, we argue this type of mediator is important and should be identified by the inferential test too. See Section \ref{sec:mediation} for more discussion. 

There have also been some recent proposals of penalized sparse estimation of mediation effects \citep{ZhaoLuo2016, Nandy2017}. In addition, there is a large body of literature studying penalized estimation of directed acyclic graph given observational data \citep[see, e.g.,][and the references therein]{van2013,zheng2018dags,Yuan2019}. However, estimation is an ultimately different problem from inference. Although both can in effect identify important mediators or links, estimation does not produce an explicit quantification of statistical significance, and does not explicitly control the false discovery. As such, we are targeting a completely different problem than those estimation approaches. More recently, \citet{LiShen2019} developed a constrained likelihood ratio test to infer individual links or some given directed paths of a DAG. Nevertheless, their  hypotheses are very different from our problem of inferring significant mediators. 

In this article, we propose a novel hypothesis testing procedure to evaluate individual mediation effects, which takes into account directed paths among the mediators and is equipped with statistical guarantees. A key ingredient of our proposal is to construct the test statistic using the logic of Boolean matrices, which allows us to establish the proper limiting distribution under the null hypothesis. In comparison, the asymptotic properties of the test statistic built on the usual matrix operations are extremely challenging to establish. In addition, the Boolean matrix-based test statistic can be naturally coupled with a screening procedure. This helps scale down the number of potential paths to a moderate level, and in turn reduces the variance of the test statistic, and enhances the power of the test considerably. Furthermore, we use a data splitting strategy to ensure a valid type-I error rate control for our test under minimal conditions on the screening. We employ some state-of-the-art estimator of DAG \citep{zheng2018dags} to form an initial estimator of the directed paths. We also devise a decorrelated estimator to reduce potential bias induced by high-dimensional mediators. Consequently, it ensures the resulting estimator is $\sqrt{n}$-consistent and asymptotically normal. We then employ a multiplier bootstrap method to obtain accurate critical values. Finally, we couple our test for the significance of an individual mediator with a multiple testing procedure \citep{Djor2019} to control the false discovery rate (FDR) of simultaneous testing of multivariate mediators. 

Our contributions are multi-fold. Scientifically, rigorous inference of mediation effects is a vital and long-standing problem. But nearly all existing solutions ignore potential interactions among the mediators. Our proposal thus fills a crucial gap, extends the scope of existing tests, and offers a useful inferential tool to a wide range of scientific applications. Methodologically, our proposal integrates the logic of Boolean matrices, DAG estimation, screening, data splitting, and decorrelated estimation to reduce the bias and increase the power of the test. It is ultimately different from the test of \citet{LHZ2018}, which defined the mediation effect through averaging, required the DAG selection consistency, focused on dealing with the equivalence class of DAG estimators, and did not consider multiple testing. By contrast, our method targets a different, and in our opinion, a more general definition of mediation effect, does not require the DAG selection consistency, and mostly focuses on the single DAG situation. We discuss the extension of the test to the equivalence class situation in Section \ref{sec:discussion}. We also compare our test with that of \citet{LHZ2018} numerically, and show our method is empirically more powerful while achieving a valid type-I error control. Theoretically, we systematically study the asymptotic properties of our test, while allowing the number of mediators to diverge to infinity with the sample size. We show that our test can control both the size and FDR asymptotically, and the power of the test approaches one. As a by-product, we derive an oracle inequality for the estimated DAG by the method of \citet{zheng2018dags}, which is needed to establish the consistency of our test, but is not available in \citet{zheng2018dags}. 

The rest of the article is organized as follows. We define our hypotheses in Section \ref{sec:hypotheses}, and develop the test statistics based on the logic of Boolean matrices in Section \ref{sec:test-statistics}. We propose the testing procedures in Section \ref{sec:test-proc}, and investigate the asymptotic properties in Section \ref{sec:theory}. We present the simulations in Section \ref{sec:simulations}, and a neuroimaging application in Section \ref{sec:real-data}. We conclude the paper in Section \ref{sec:discussion}, and relegate all proofs to the supplementary appendix.

\section{Hypotheses}
\label{sec:hypotheses}

In this section, we first present the Gaussian graphical model, based on which we formulate our mediation testing problem. We then formally develop the hypotheses we aim to test, and compare with the alternative formulation in \citet{LHZ2018}.

\subsection{Gaussian graphical model}
\label{sec:model}

\change{Consider an exposure variable $E$, a set of potential mediators $\bm{M}=(M_{1},\ldots,M_{d})^\top$, and an outcome variable $Y$. Let $\bm{X}=(E,\bm{M}^\top,Y)^\top$ collect all the variables,} and assume $\bm{X}$ follows the linear structural equation model, 
\vspace{-0.05in}
\begin{eqnarray}\label{eqn:md}
\bm{X} - \bm{\mu}_0 = \bm{W}_0(\bm{X}-\bm{\mu}_0)+\bm{\varepsilon},
\end{eqnarray}
where $\bm{\mu}_0 = \Mean(\bm{X})$, $\bm{W}_0$ is the $(d+2)\times (d+2)$ coefficient matrix, and $\bm{\varepsilon}=(\varepsilon_{0},\varepsilon_1,\ldots,\varepsilon_{d+1})^\top$ is the mean-zero vector of errors. The matrix $\bm{W}_0$ specifies the directional relationships among the variables in $\bm{X}$, which can be encoded by a directed graph. Let $X_j$ denote the $(j+1)$th element of $\bm{X}$, $j = 0, \ldots, d+1$. For $i,j \in \{0, 1, \ldots, d+1\}$, if $X_i$ is a direct cause of $X_j$, then an arrow is drawn from $X_i$ to $X_j$, i.e, $X_i\to X_j$, and $W_{0,j,i}\neq 0$. In this case, $X_i$ is called a parent of $X_j$, and $X_j$ a child of $X_i$. For an integer $k \ge 1$, a $k$-step directed path between $X_i$ and $X_j$ is a sequence of distinct nodes from $X_i$ to $X_j$: $X_i\to X_{i_1}\to\ldots\to X_{i_{k-1}}\to X_j$, for some $\{i_k\}_{1\le l< k}$. In this case, $X_i$ is called an ancestor of $X_j$, and $X_j$ a descendant of $X_i$. For model \eqref{eqn:md} and the associated directed graph, we impose a set of conditions. Specifically, 

\vspace{-0.05in}
\begin{enumerate}[({A}1)]
\item The directed graph is acyclic; i.e., no variable is an ancestor of itself. 
\vspace{-0.1in}

\item No potential mediator $M_i$ is a direct cause of the exposure $E$, and the outcome $Y$ is not a direct cause of neither the exposure $E$ nor any mediator $M_i$, $i = 1, \ldots, d$.
\vspace{-0.1in}

\item The errors $\varepsilon_{i}$, $i=0,1, \ldots, d+1$, are jointly normally distributed and independent. In addition, the error variances $\sigma_i^2 = \Var(\varepsilon_i)$, $i=0, \ldots, d+1$, are constant; i.e., $\sigma_0^2=\sigma_1^2=\ldots=\sigma_{d}^2=\sigma_{d+1}^2=\sigma_*^2$ for some constant $\sigma_*>0$.
\end{enumerate}
\vspace{-0.1in}

\noindent
These model assumptions are generally mild, and are often imposed in the DAG and mediation analysis literature. Specifically, Condition (A1) implies that $\bm{W}_0$ is a lower-triangular matrix, up to a permutation of the rows and columns. Condition (A2) implies that the first row of $\bm{W}_0$ and the last column of $\bm{W}_0$ are both zero vectors. Condition (A3) basically specifies that $\bm{X}$ follows a Gaussian graphical model. By Gram-Schmidt orthogonalization, any Gaussian DAG model can always be represented by \eqref{eqn:md} with independent errors. In addition, the constant variance condition in (A3) ensures that, under the Gaussian graphical model \eqref{eqn:md}, $\bm{W}_0$ is identifiable \citep[Theorem 1]{Buhlmann2014}. This avoids the situation of the equivalence class of DAG, and a similar condition has been adopted in \citet{Yuan2019} as well. \change{We note that it is possible to relax (A3) by requiring $\sigma_1^2=\ldots=\sigma_d^2$; i.e., excluding the variance requirement on the exposure and the outcome. We discuss this relaxation in more details in Section \ref{sec:weakerA3} of the appendix.} Moreover, we also discuss the extension of our method to the unequal variance case in Section \ref{sec:discussion}.

\subsection{Mediation effects and hypotheses}
\label{sec:mediation}

For a directed path $\zeta: E\to M_{i_1}\to \ldots\to M_{i_k}\to Y$ for some $\{i_t\}_{1\le t\le k}\subseteq \{1,\ldots,d\}$, we define the total effect of $E$ on $Y$ attributed to this path as,
\begin{eqnarray} \label{eqn:total-effect}
\omega_{\zeta} = W_{0,i_1,0}\left(\prod_{t=0}^{k-1} W_{0,i_{t+1},i_t}\right) W_{0,d+1,i_k},
\end{eqnarray} 
where $W_{0,i,j}$ is the $(i,j)$th entry of $\bm{W}_0$. If such a path does not exist, we have $\omega_{\zeta}=0$. This definition of total effect $\omega_{\zeta}$ plays a central role in our definition of mediation effect. 

Based on \eqref{eqn:total-effect}, we formally state our hypotheses regarding the significance of an individual mediator $M_q$, for an integer $q = 1, \ldots, d$, 
\begin{align} \label{eqn:hypo1}
\begin{split}
& H_0(q): \omega_{\zeta}=0, \;\; \textrm{ for all } \zeta \textrm{ that passes through }M_q, \;\; \quad \textrm{versus} \\
& H_1(q): \omega_{\zeta}\neq 0, \;\; \textrm{ for some } \zeta \textrm{ that passes through }M_q.
\end{split}
\end{align}
When the alternative hypothesis in \eqref{eqn:hypo1} holds, we call $M_q$ a significant mediator. 

We observe that, the hypotheses in \eqref{eqn:hypo1} can be reformulated as the following equivalent pair of hypotheses. That is, for any integer $j = 1, \ldots, d+1$, let $\hbox{ACT}(j,\bm{W}_0)$ denote the set of true ancestors of $X_j$; i.e., $i\in \hbox{ACT}(j,\bm{W}_0)$ if and only if $X_i$ is an ancestor of $X_j$. Then the pair of hypotheses \eqref{eqn:hypo1} is equivalent to the following pair of hypotheses, 
\begin{align} \label{eqn:hypo2}
\begin{split}
& H_0(q): 0 \notin \textrm{ACT}(q,\bm{W}_0) \;\; \textrm{ or } \;\; q \notin \textrm{ACT}(d+1,\bm{W}_0), \;\; \quad \textrm{versus} \\
& H_1(q): 0 \in \textrm{ACT}(q,\bm{W}_0) \;\; \textrm{ and } \;\; q \in \hbox{ACT}(d+1,\bm{W}_0).
\end{split}
\end{align}

Next, we consider a pair of hypotheses that lead to \eqref{eqn:hypo2}. For any $q_1 = 0, \ldots, d$, $q_2 = 1, \ldots, d+1$, we consider the following pair of hypotheses,
\vspace{-0.05in}
\begin{align} \label{eqn:sub-hypo}
H_0(q_1,q_2): q_1\notin \textrm{ACT}(q_2,\bm{W}_0),  \quad \textrm{versus} \quad 
H_1(q_1,q_2): q_1\in \textrm{ACT}(q_2,\bm{W}_0).
\end{align}
We observe that, the null hypothesis $H_0(q)$ in \eqref{eqn:hypo2} can be decomposed into a union of the two null hypotheses $H_0(0,q)$ and $H_0(q,d+1)$ that are defined in \eqref{eqn:sub-hypo}. Suppose $p(q_1,q_2)$ is a valid $p$-value for $H_0(q_1,q_2)$ in \eqref{eqn:sub-hypo}. According to the union-intersection principle, $\max \big\{ p(0,q), p(q,d+1) \big\}$ is a valid $p$-value for testing $H_0(q)$ in \eqref{eqn:hypo2}. Therefore, we aim at \eqref{eqn:sub-hypo} in the subsequent development of our testing procedure. 

\begin{figure}[t!]
\centering
\begin{tabular}{ccc}
\includegraphics[width=5.0cm]{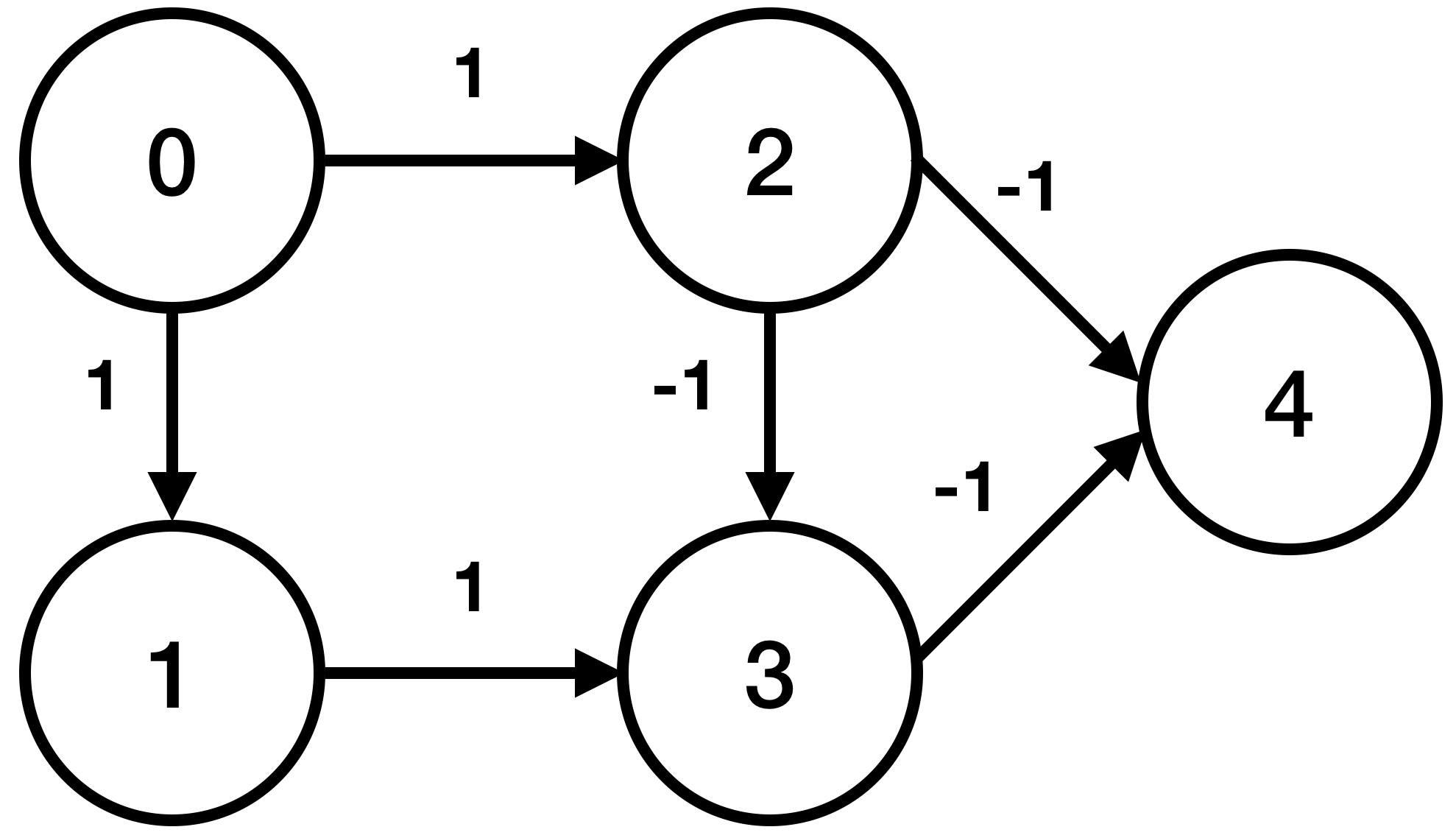} & 
\hspace{0.5in} & 
\includegraphics[width=5.0cm]{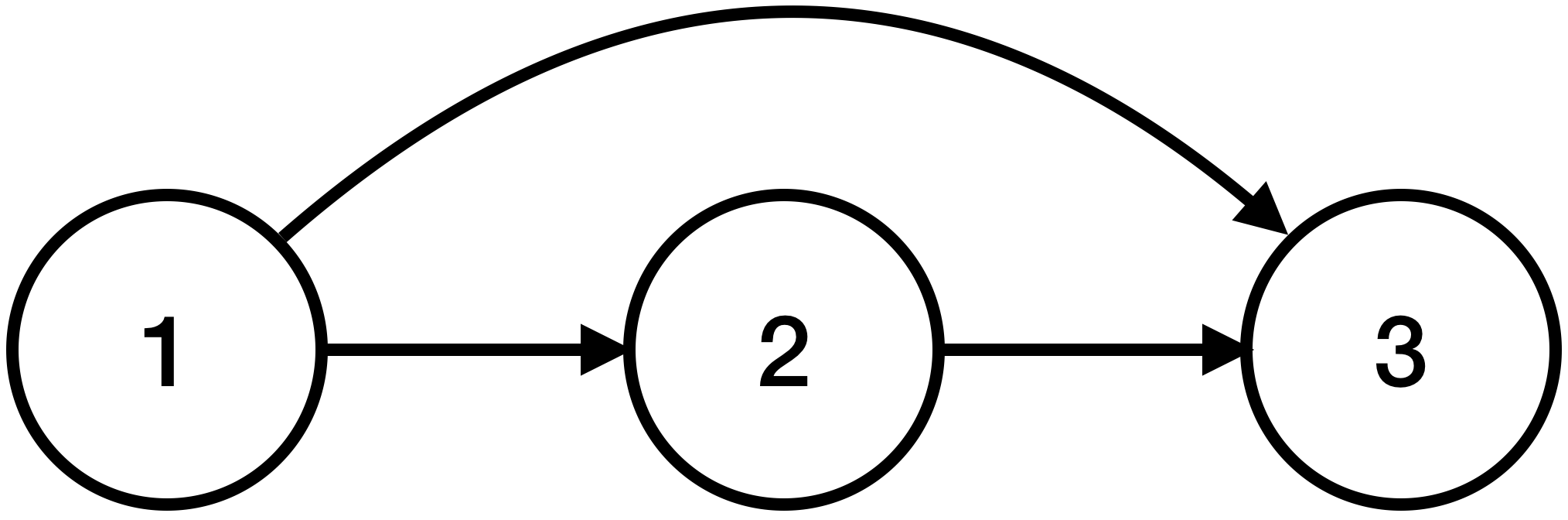} \\
(a) & & (b) 
\end{tabular}
\caption{Left panel: a DAG with five nodes, where node 0 is the exposure variable, node 4 is the outcome variable, and nodes 1 to 3 are the mediators. Right panel: a DAG with three nodes.}
\label{fig:DAG-illustration}
\end{figure}

We have defined a significant mediator through \eqref{eqn:hypo1}. There is an alternative definition employed by \citet{LHZ2018}. Specifically, they considered the hypotheses, 
\vspace{-0.05in}
\begin{eqnarray} \label{eqn:alt-hypo}
H_0^{*}(q): \sum \omega_{\zeta} = 0, \quad \textrm{versus} \quad
H_1^*(q):  \sum \omega_{\zeta} \neq 0,
\end{eqnarray}
where the summation is taken for all $\zeta$ that pass through $M_q$. \citet{LHZ2018} called $M_q$ a significant mediator when the alternative hypothesis in \eqref{eqn:alt-hypo} holds. We, however, prefer our definition of a significant mediator that is built on \eqref{eqn:hypo1} instead of \eqref{eqn:alt-hypo}. This is because the effects along the path $\zeta$ may cancel out with each other, resulting in a zero sum, even though there are significant positive and negative mediation effects along $\zeta$. As an illustration, we devise a simple example as shown in Figure \ref{fig:DAG-illustration}(a). For the mediator $M_{2}$, two paths, $E \to M_{2} \to Y$ and $E\to M_{2}\to M_{3}\to Y$, both pass through $M_{2}$. The aggregated total effect following \eqref{eqn:alt-hypo} is $\sum_{\zeta} \omega_{\zeta} = 1\times \{-1+(-1)\times (-1)\} = 0$. Similarly, we can show the aggregated total effect of $M_{3}$ is zero as well. However, both $M_{2}$ and $M_{3}$ have positive and negative mediation effects, and should be viewed as significant mediators. 

We conclude this section by computing the explicit number of potential paths that go through any mediator $M_q$. For an integer $k = 2, 3, \ldots, d+1$, the total number of $k$-step potential paths that go through $M_q$, by the combinatorial theory, is $N_k(q)={d-1 \choose k-2}(k-1)!$. Then the total number of potential paths that go through $M_q$ is $N(q)=\sum_{k=2}^{d+1} N_k(q)\ge \mathcal{N}_{d+1}(q)=d!$. As a result, it is highly nontrivial to test the significance of an individual mediator if we take into account all the potential paths among the mediators.

\section{Test Statistics}
\label{sec:test-statistics}

In this section, we first consider a potential test statistic built on the power of an estimator of the coefficient matrix $\bm{W}_0$ in model \eqref{eqn:md}, and discuss its limitation. We then present our main idea, the logic of Boolean matrices, and the test statistic built on it.

\subsection{Power of matrices}
\label{sec:power-matrix}

Matrices and vectors in this paper start the index from zero. For any matrix $\bm{A}$, let $|\bm{A}|$ denote the matrix of the same dimension whose $(i,j)$th entry is $|A_{i,j}|$. We first connect the null hypothesis $H_0(q_1,q_2)$ in \eqref{eqn:sub-hypo} with the coefficient matrix $\bm{W}_0$ in model \eqref{eqn:md}. Recall that $H_0(q_1,q_2)$ means $X_{q_1}$ is not an ancestor of $X_{q_2}$. We have the next lemma.

\begin{lemma}\label{thm:lemma1}
The null $H_0(q_1,q_2)$ holds if and only if $(|\bm{W}_0|^k)_{q_2,q_1}=0$ for any $k = 1, \ldots, d$. 
\end{lemma}

\noindent
We sketch the proof of this lemma here, which is to facilitate our understanding of the problem. The key observation is that, the $(q_2,q_1)$th entry of $|\bm{W}_0|^k$ is the sum of the absolute values of the total effects along all $k$-step paths from $X_{q_1}$ to $X_{q_2}$. For instance, for $k=2$, 
\vspace{-0.05in}
\begin{eqnarray*}
(|\bm{W}_0|^2)_{q_2,q_1}=\sum_{j=0}^{d+1} |W_{0,j,q_1}| |W_{0,q_2,j}|=\sum_{j=1}^d |W_{0,j,q_1}| |W_{0,q_2,j}|,
\end{eqnarray*}
where the last equality is due to that the first row and last column of $\bm{W}_0$ are zero vectors because of Condition (A2). If there exists a two-step path from $X_{q_1}$ to $X_{q_2}$, by definition, $W_{0,j,q_1}\neq 0$ and $W_{0,q_2,j}\neq 0$ for some $j=1, \ldots, d$, which is equivalent to $(|\bm{W}_0|^2)_{q_2,q_1}\neq 0$. Similarly, there exists a $k$-step path from $X_{0}$ to $M_q$ if and only if $(|\bm{W}_0|^k)_{q_2,q_1}\neq 0$. 

Let $\widehat{\bm{W}}$ be some consistent estimator for $\bm{W}_0$. In view of Lemma \ref{thm:lemma1}, it is natural to construct a test statistic based on $\{ ( |\widehat{\bm{W}}|^k )_{q_2,q_1} \}_{1\le k\le d}$. The major difficulty with this potential test statistic, however, is that it is unclear whether $(|\widehat{\bm{W}}|^k)_{q_2,q_1}$ has a well tabulated limiting distribution under $H_0(q_1,q_2)$. To better illustrate this, we first consider the case when $k=2$. We have $(|\widehat{\bm{W}}|^2)_{q_2,q_1}= \sum_{j=1}^{d} |\widehat{W}_{j,q_1}| |\widehat{W}_{q_2,j}|$. Under $H_0(q_1,q_2)$ and the acyclic constraint (A1), for any $j$, either $W_{0,j,q_1}$ or $W_{0,q_2,j}$ equals zero. Suppose each $\widehat{W}_{q_1,q_2}$ is root-$n$ consistent to $W_{0,q_1,q_2}$, and the mediator dimension $d$ is fixed. Then we can show $\sqrt{n}(|\widehat{\bm{W}}|^2)_{q_2,q_1}$ is asymptotically equivalent to $\sum_{\substack{ 1\le j\le d }} \sqrt{n} \left( |\widehat{W}_{j,q_1}-W_{0,j,q_1}| |W_{0,q_2,j}|+|W_{0,j,q_1}||\widehat{W}_{q_2,j}-W_{0,q_2,j}| \right)$. The limiting distribution, however, is not well-studied even in the fixed-$d$ scenario. When $k$ is large, or when the mediator dimension $d$ diverges with the sample size $n$, the derivation of the asymptotic property of $(|\widehat{\bm{W}}|^k)_{q_2,q_1}$ becomes more complicated due to the addition and multiplication operations involved in $(|\widehat{\bm{W}}|^k)_{q_2,q_1}$. Therefore, the test statistic based on $|\widehat{\bm{W}}|^k$ may not be suitable for our purpose of testing significant mediators.

\subsection{Logic of Boolean matrices}
\label{sec:boolean-matrix}

To overcome the difficulty regarding $|\widehat{\bm{W}}|^k$, and motivated by the logic of Boolean matrices, we define a new matrix multiplication operator and a new matrix addition operator to replace the usual matrix multiplication and addition operations. Specifically, for any two real-valued matrices $\bm{A}_1=\{a_{1,i,j}\}_{ij}\in \mathbb{R}^{q_1\times q_2},\bm{A}_2=\{a_{2,i,j}\}_{ij}\in \mathbb{R}^{q_2\times q_3}$, we define $\bm{A}_1\otimes \bm{A}_2$ to be a $q_1\times q_3$ matrix whose $(i,j)$th entry equals $\max_{k\in \{ 1,\ldots,q_2 \} }\min(a_{1,i,k},a_{2,k,j})$. That is, we replace the multiplication operation in the usual matrix multiplication with the minimum operator, and replace the addition operation with the maximum operator. When $\bm{A}_1$, $\bm{A}_2$ are binary matrices, the minimum and maximum operators are equivalent to the logic operators ``and" and ``or" in Boolean algebra. The defined ``$\otimes$" operator is then equivalent to the Boolean matrix multiplication operator. Moreover, for any two real-valued matrices $\bm{A}_1=\{a_{1,i,j}\}_{ij},\bm{A}_2=\{a_{2,i,j}\}_{ij}\in \mathbb{R}^{q_1\times q_2}$, we define $\bm{A}_1 \oplus \bm{A}_2$ to be a $q_1\times q_2$ matrix whose $(i,j)$th entry equals $\max(a_{1,i,j},a_{2,i,j})$. When $\bm{A}_1,\bm{A}_2$ are binary matrices, the defined ``$\oplus$" operator is equivalent to the Boolean matrix addition operator. 

Given the new definition of the multiplication and addition operators, we define $|\bm{W}|^{(k)}_0 = |\bm{W}_0|^{(k-1)} \otimes |\bm{W}_0|$ in a recursive fashion, for any $k \ge 1$. Next, we connect the null hypothesis $H_0(q_1,q_2)$ in \eqref{eqn:sub-hypo} with the newly defined $|\bm{W}_0|^{(k)}$. Its proof is given in the appendix. 

\begin{lemma}\label{thm:lemma2}
The null $H_0(q_1,q_2)$ holds if and only if $(|\bm{W}_0|^{(k)})_{q_2,q_1}=0$ for any $k = 1, \ldots, d$.
\end{lemma}

Aggregating $|\bm{W}_0|^{(k)}$ for all $k$-step paths, $k = 1, \ldots, d$, leads to the following definition, 
\vspace{-0.05in}
\begin{eqnarray*}
\bm{W}_0^*=|\bm{W}_0|\oplus |\bm{W}_0|^{(2)}\oplus \ldots \oplus |\bm{W}_0|^{(d)}.
\end{eqnarray*}
We next define two matrices $\bm{B}_0$ and $\bm{B}_0^*$ that are the binary versions of $\bm{W}_0$ and $\bm{W}_0^*$,  
\vspace{-0.05in}
\begin{eqnarray*}\label{eqn:B}
(\bm{B}_0)_{i,j} = \left\{
\begin{array}{ll}
1, & \hbox{if}~~W_{0,i,j} \neq 0,\\
0, & \hbox{otherwise},
\end{array}
\right. \quad \textrm{ and } \quad
\bm{B}_0^* = \bm{B}_0\oplus \bm{B}_0^{(2)}\oplus \ldots \oplus \bm{B}_0^{(d)}. 
\end{eqnarray*}
Then Lemma \ref{thm:lemma2} immediately implies the next result. 

\begin{coro}\label{thm:coro1}
The null $H_0(q_1,q_2)$ holds if and only if $(\bm{W}_0^*)_{q_2,q_1}=0$ and $(\bm{B}_0^*)_{q_2,q_1}=0$.  
\end{coro}

Corollary \ref{thm:coro1} suggests some natural test statistic for our hypotheses. Again, let $\widehat{\bm{W}}$ be some consistent estimator for $\bm{W}_0$, and let $\widehat{\bm{W}}^*=|\widehat{\bm{W}}|\oplus |\widehat{\bm{W}}|^2\oplus \ldots \oplus |\widehat{\bm{W}}|^d$. We further define a thresholded binary version $\widehat{\bm{B}}(c)$ and $\widehat{\bm{B}}^*(c)$, for a given thresholding value $c$, as, 
\vspace{-0.05in}
\begin{eqnarray}\label{eqn:Bc}
\{ \widehat{\bm{B}}(c) \}_{i,j} = \left\{\begin{array}{ll}
1, & \hbox{if}~~|\widehat{W}_{i,j}|>c,\\
0, & \hbox{otherwise},
\end{array}
\right. \quad \textrm{ and } \quad
\widehat{\bm{B}}^*(c)=\widehat{\bm{B}}(c)\oplus \widehat{\bm{B}}^{(2)}(c) \oplus \ldots \oplus \widehat{\bm{B}}^{(d)}(c). 
\end{eqnarray}
In view of Corollary \ref{thm:coro1}, we expect $\widehat{\bm{W}}^*_{q_2,q_1}$ to be small under $H_0(q_1,q_2)$, and we reject $H_0(q_1,q_2)$ when $(\widehat{\bm{W}}^*)_{q_2,q_1}>c$ for some thresholding value $c$, or equivalently, when $\{\widehat{\bm{B}}^*(c)\}_{q_2,q_1} = 1$. We then build a test statistic based on $\widehat{\bm{W}}^*$. 

Unlike the usual power of the matrix $|\widehat{\bm{W}}|^{k}$, the limiting distribution of $\widehat{\bm{W}}^*$ based on the logic of Boolean matrices is more tractable. Specifically, under the null hypothesis $H_0(q_1,q_2)$, for any potential path $X_{q_1}\to X_{j_1}\to \ldots\to X_{j_k}\to X_{q_2}$, such that 
\vspace{-0.05in}
\begin{eqnarray} \label{Whatpathcond}
\widehat{W}_{j_1,q_1}\neq 0, \min_{t\in \{1,\ldots,k-1\}} |\widehat{W}_{j_{t+1},j_t}|\neq 0, \;\; \textrm{ and } \; \widehat{W}_{q_2,j_k}\neq 0,
\end{eqnarray}
there exist some distinct integers $\ell_1,\ell_2\in \{q_1,j_1,\ldots,j_k,q_2\}$ as functions of $\big( q_1,\{j_t\}_{1\le t\le k},q_2 \big)$, such that $W_{0,\ell_2,\ell_1}=0$. It then follows that,
\vspace{-0.05in}
\begin{align} \label{jtsubscript}
\begin{split}
\sqrt{n}(\widehat{\bm{W}}^*)_{q_2,q_1} & \le \max_{\substack{k\in \{1,\ldots,d\}\\1\le j_1,\ldots,j_k\le d+2}} \sqrt{n} \; |\widehat{W}_{\ell_2,\ell_1}| 
\;\; = \max_{\substack{k\in \{1,\ldots,d\}\\0\le j_1,\ldots,j_k\le d+1}} \sqrt{n} \; |\widehat{W}_{\ell_2,\ell_1}-W_{0,\ell_2,\ell_1}| \\
& \le \max_{\substack{k\in \{1,\ldots,d\}\\0\le j_1,\ldots,j_k\le d+1 }} \max_{\substack{0\le t\le k\\ j_0=q_1, j_{k+1}=q_2}} \sqrt{n} \; |\widehat{W}_{j_{t+1},j_t}-W_{0,j_{t+1},j_t}|,
\end{split}
\end{align}
where the first maximum is taken over all such $k$ and $(j_1,\ldots,j_k)$ that satisfy \eqref{Whatpathcond}. When the nonzero entries of $\sqrt{n}(\widehat{\bm{W}}-\bm{W}_0)$ are asymptotically normal, the right-hand-side of \eqref{jtsubscript} converges in distribution to a maximum of some normal random variables in absolute values. Its $\alpha$th upper quantile can be consistently estimated via multiplier bootstrap. This forms the basis of our proposed testing procedure. 

On the other hand, the test outlined above has some limitations. One is that this test can be conservative when $\bm{W}_0$ is highly sparse but $\widehat{\bm{W}}$ is not. Another limitation is that it requires the support of $\widehat{\bm{W}}$ to be fixed. When this fixed support condition does not hold, it would lead to an inflated type-I error rate. To address these limitations, we next develop a testing procedure that couples such a test with screening and data sample splitting to enhance its power as well as to ensure its validity.

\section{Testing Procedure}
\label{sec:test-proc}

In this section, we first present our full testing procedure for inference of an individual mediator. We next describe in detail some major steps of this testing procedure. Finally, we present a multiple testing procedure for simultaneous inference of multivariate mediators with a proper FDR control. Given that our test is constructed based on the LOGic of booleAN matrices, we refer our testing method as LOGAN.

\subsection{The complete algorithm}
\label{sec:comalg}

Let $\bm{X}_1,\ldots,\bm{X}_n$ denote $n$ i.i.d.\ copies of $\bm{X}$, generated according to model \eqref{eqn:md}. Step 1 of our testing procedure is to randomly divide the observed data into two equal halves $\{\bm{X}_i\}_{i\in \mathcal{I}_1}\cup \{\bm{X}_i\}_{i\in \mathcal{I}_2}$, where $\mathcal{I}_{\ell}$ is the set of indices of subsamples, $\ell=1,2$. The purpose of data splitting is to ensure our test achieves a valid type-I error rate under minimal conditions when coupled with a screening step. \change{In recent years, data splitting has been commonly used in high-dimensional estimation and inference \citep[e.g.,][]{Chernozhukov2018, Newey2018, Barber2019, romano2019}. One issue with data splitting is the potential loss of power due to the usage of only a fraction of data. There have been studies showing that data splitting may improve the power in some cases \citep{Rubin2006, romano2019}. In our setting, we construct two test statistics based on both halves of data, then combine them to derive the final decision rule. We show the test constructed this way achieves a good power both asymptotically and numerically. Moreover, one may follow the idea of \citet{Mein2009} to carry out the binary split more than once, then combine the $p$-values from all splits. This strategy also helps mitigate the randomness the single data splitting introduces.  In the regression setting, \cite{Mein2009} showed empirically that this multi-split strategy improves the power than a single-split. We develop a multi-split version of our test in Section \ref{sec:multisplit} of the appendix, and show its improvement in power numerically. We also note that, one may employ the multi-split strategy of \citet[Section 4.2.1]{romano2019} for power improvement. Meanwhile, these improvements all come with a price of increased computational costs.}

Step 2 is to compute an initial estimator $\widetilde{\bm{W}}^{(\ell)}$ for $\bm{W}_0$, given each half of the data $\{\bm{X}_i\}_{i\in \mathcal{I}_{\ell}}$, $\ell = 1, 2$. Several methods can be used here, e.g., \citet{zheng2018dags, Yuan2019}. We only require $\widetilde{\bm{W}}^{(\ell)}$ to be consistent to $\bm{W}_0$. This requirement is considerably weaker than requiring $\widetilde{\bm{W}}^{(\ell)}$ to be selection consistent; i.e., $\mathbb{I}(\widetilde{W}^{(l)}_{i,j}=0)=\mathbb{I}(W_{0,i,j}=0)$ for any $i, j = 0, \ldots, d+1$, where $\mathbb{I}(\cdot)$ is the indicator function. See Section \ref{sec:init-est} for more details. 

Step 3 is to compute the binary matrix $\widehat{\bm{B}}^{(\ell)}$ for $\bm{B}_0$, given the initial estimator $\widetilde{\bm{W}}^{(\ell)}$, using   \eqref{eqn:Bc} with $c=0$. This step is straightforward, and the main purpose is to allow the subsequent decorrelated estimation step to focus only on those nonzero elements in $\widehat{\bm{B}}^{(\ell)}$. It thus acts as a screening step, and in effect reduces the number of potential paths to a moderate level. As a benefit, it reduces the variance of the Boolean matrix-based test statistic, and increases the power of the test. See Section \ref{sec:decor} for more details.

\begin{algorithm}[t!]
\caption{Testing procedure for inference of an individual mediator.}
\label{alg:full}
\begin{algorithmic}
\item
\begin{description}
\item[\textbf{Input}:] The data $\bm{X}_1,\ldots,\bm{X}_n$, $1\le q\le d$, and the significance level $0<\alpha<1$. 

\item[\textbf{Step 1}.] Randomly divide $\{1,2,\ldots,n\}$ into two disjoint subsets $\mathcal{I}_1\cup \mathcal{I}_2$ of equal sizes. 

\item[\textbf{Step 2}.] Compute an initial estimator $\widetilde{\bm{W}}^{(\ell)}$ for $\bm{W}_0$, given $\{\bm{X}_i\}_{i\in \mathcal{I}_{\ell}}$, $\ell=1,2$.

\item[\textbf{Step 3}.] Compute the binary estimator $\widehat{\bm{B}}^{(\ell)}$ for $\bm{B}_0$, given $\widetilde{\bm{W}}^{(\ell)}$, $\ell=1,2$, which is to be used for screening and also ancestor estimation in the next step.   

\item[\textbf{Step 4}.] Compute the decorrelated estimator $\widehat{\bm{W}}^{(\ell)}$ for $\bm{W}_0$, given $\widetilde{\bm{W}}^{(\ell)}$, $\widehat{\bm{B}}^{(\ell)}$ and $\{\bm{X}_i\}_{i \in \mathcal{I}_\ell}$, $\ell=1,2$.
\begin{enumerate}[({4}a)]
\item Estimate the ancestors of $M_q$, for $q =1, \ldots, d+1$, by ACT$(q, \widetilde{\bm{W}}^{(\ell)}) = \Big\{ 1 \le j \le d : \{\widehat{\bm{B}}^{*(\ell)} \}_{q,j} \neq 0 \Big\}$, where $\widehat{\bm{B}}^{*(\ell)} = |\widehat{\bm{B}}^{(\ell)}| \oplus | \widehat{\bm{B}}^{(\ell)}|^{(2)} \oplus \ldots \oplus |\widehat{\bm{B}}^{(\ell)}|^{(d)}$. 

\item Update the $j$th row of $\widetilde{\bm{W}}^{(\ell)}$, for $1 \le j \le d$, by fitting a penalized regression with $\{X_{i,j}\}_{i\in \mathcal{I}_{\ell}}$ being the response and $\{X_{i,k}\}_{i\in \mathcal{I}_{\ell}, \widetilde{W}^{(\ell)}_{j,k}\neq 0 }$ being the predictors. Denote the updated estimator as $\overline{\bm{W}}^{(\ell)}$. 

\item Compute the decorrelated estimator $\widehat{W}_{j_1,j_2}^{(\ell)}$, for any $(j_1,j_2)$ such that $\widehat{B}_{j_1,j_2}^{(\ell)}\neq 0$,  given $\{\bm{X}_i\}_{i\in \mathcal{I}_{\ell}^c}$, ACT$(j_1,\widetilde{\bm{W}}^{(\ell)})$, $\widehat{\bm{B}}^{(\ell)}$, and $\overline{\bm{W}}^{(\ell)}$. 
\end{enumerate}

\item[\textbf{Step 5}.] Compute the critical values using the bootstrap procedure, given $\widehat{\bm{W}}^{(\ell)}$, $\widehat{\bm{B}}^{(\ell)}$, and $\{\bm{X}_i\}_{i \in \mathcal{I}_\ell}$, $\ell=1,2$.
\begin{enumerate}[({5}a)]
\item Compute the critical value $\widehat{c}^{(\ell)}(0,q)$ of $\max_{(i,j)\in \mathcal{S}(0,q,\widehat{\bm{B}}^{(\ell)})} \sqrt{|\mathcal{I}_{\ell}^c|}|\widehat{W}_{i,j}^{(\ell)}-W_{0,i,j}^{(\ell)}|$ under the significance level $\alpha/2$, $\ell=1,2$.

\item Compute the critical value $\widehat{c}^{(\ell)}(q,d+1)$ of $\max_{(i,j)\in \mathcal{S}(q,d+1,\widehat{\bm{B}}^{(\ell)})} \sqrt{|\mathcal{I}_{\ell}^c|}|\widehat{W}_{i,j}^{(\ell)}-W_{0,i,j}^{(\ell)}|$ under the significance level $\alpha/2$, $\ell=1,2$. 
\end{enumerate}

\item[\textbf{Output}:] Decision. 
\begin{enumerate}[({6}a)]
\item Reject $H_0(0,q)$ if $\widehat{\bm{B}}^{*(\ell)}_{q,0}\{ n^{-1/2}\widehat{c}(0,q) \} = 1$. Denote this decision by $\mathcal{D}^{(\ell)}(0,q)$.

\item Reject $H_0(q,d+1)$ if $\widehat{\bm{B}}^{*(\ell)}_{d+1,q}\{ n^{-1/2}\widehat{c}(q,d+1) \} = 1$. Denote this decision by $\mathcal{D}^{(\ell)}(q,d+1)$. 

\item Reject $H_0(q)$ if both $\mathcal{D}^{(\ell)}(0,q)$ and $\mathcal{D}^{(\ell)}(q,d+1)$ reject, for at least one $\ell=1,2$.  
\end{enumerate}
\end{description}
\end{algorithmic}
\end{algorithm}

Step 4 is to compute a decorrelated estimator $\widehat{\bm{W}}^{(\ell)}$ using a cross-fitting procedure. We use one set of samples $\mathcal{I}_{\ell}$ to obtain the initial estimator $\widetilde{\bm{W}}^{(\ell)}$ and the binary version $\widehat{\bm{B}}^{(\ell)}$ to screen out the zero entries, then use the other set of samples $\mathcal{I}_{\ell}^c$ to compute the entries of the decorrelated estimator $\widehat{\bm{W}}^{(\ell)}$. This decorrelated estimation step is to reduce the bias of $\widetilde{\bm{W}}^{(\ell)}$ under the setting of high-dimensional mediators. Moreover, it guarantees the entry of $\widehat{\bm{W}}^{(\ell)}$, $\widehat{W}_{j_1,j_2}^{(\ell)}$, is $\sqrt{n}$-consistent and asymptotically normal. It adopts the debiasing idea that is commonly used for statistical inference of low-dimensional parameters in high-dimensional models \citep{zhang2014confidence, Ning2017}. See Section \ref{sec:decor} for more details.

Step 5 is to use a bootstrap-based procedure to compute the critical values. Let $\widehat{\bm{W}}^{*(\ell)} = |\widehat{\bm{W}}^{(\ell)}| \oplus |\widehat{\bm{W}}^{(\ell)}|^{(2)} \oplus \ldots \oplus |\widehat{\bm{W}}^{(\ell)}|^{(d)}$. Similar to \eqref{jtsubscript}, we have
\vspace{-0.1in}
\begin{eqnarray*}
\sqrt{|\mathcal{I}_{\ell}^{c}|}(\widehat{\bm{W}}^{*(\ell)})_{q_2,q_1} \le \max_{\substack{k\in \{1,\ldots,d\}\\0\le j_1,\ldots,j_k\le d+1 }} \max_{\substack{0\le t\le k\\ j_0=q_1, j_{k+1}=q_2}} \sqrt{|\mathcal{I}_{\ell}^{c}|} \; |\widehat{W}_{j_{t+1},j_t}^{(\ell)}-W_{0,j_{t+1},j_t}^{(\ell)}|.
\end{eqnarray*}	

\vspace{-0.1in}\noindent When $\widehat{W}_{j_{t+1},j_t}^{(\ell)}$ is nonzero, the mediators $j_t$ and $j_{t+1}$ satisfy that $j_t\in \hbox{ACT}(q_2,\widehat{\bm{B}}^{(\ell)})$, $j_{t+1}\in \hbox{ACT}(q_2,\widehat{\bm{B}}^{(\ell)})\cup \{q_2\}$, $q_1\in \hbox{ACT}(j_t,\widehat{\bm{B}}^{(\ell)})\cup \{j_t\}$, $q_1\in \hbox{ACT}(j_{t+1},\widehat{\bm{B}}^{(\ell)})$ and $\widehat{B}^{(\ell)}_{j_{t+1},j_t}=1$. Then,  \vspace{-0.3in}
\begin{eqnarray}\label{naivetesteq}
\sqrt{|\mathcal{I}_{\ell}^{c}|} (\widehat{\bm{W}}^{*(\ell)})_{q_1,q_2}\le \max_{\substack{(i,j)\in \mathcal{S}(q_1,q_2,\widehat{\bm{B}}^{(\ell)})}}  
\sqrt{|\mathcal{I}_{\ell}^{c}|} \; |\widehat{W}_{i,j}^{(\ell)}-W_{0,i,j}|,
\end{eqnarray}
where $\mathcal{S}(q_1,q_2,\widehat{\bm{B}}^{(\ell)}) = \big\{ (i,j):j\in \hbox{ACT}(q_2,\widehat{\bm{B}}^{(\ell)}),i \in \hbox{ACT}(q_2,\widehat{\bm{B}}^{(\ell)})\cup \{q_2\}, q_1\in \hbox{ACT}(j,\widehat{\bm{B}}^{(\ell)})$, $\textrm{ or } j=q_1, q_1\in \hbox{ACT}(i,\widehat{\bm{B}}^{(\ell)}), \{\widehat{\bm{B}}^{(\ell)}\}_{i,j}\neq 0 \big\}$. Here $\mathcal{S}(q_1,q_2,\widehat{\bm{B}}^{(\ell)})$ denotes the set of indices such that $\{ \widehat{\bm{W}}^{*(\ell)} \}_{q_1,q_2}$ depends on $\widehat{\bm{W}}^{(\ell)}$ only through its entries in $\mathcal{S}(q_1,q_2,\widehat{\bm{B}}^{(\ell)})$. Then, based on \eqref{naivetesteq}, we use bootstrap to obtain the critical values of 
\vspace{-0.05in}
\begin{eqnarray*}
\max_{(j_1,j_2)\in \mathcal{S}(0,q,\widehat{\bm{B}}^{(\ell)})} \sqrt{|\mathcal{I}_{\ell}^c|} \; | \widehat{W}_{j_1,j_2}^{(\ell)}-W_{0,j_1,j_2}^{(\ell)} | \quad \textrm{ and } 
\max_{(j_1,j_2)\in \mathcal{S}(q,d+1,\widehat{\bm{B}}^{(\ell)})} \sqrt{|\mathcal{I}_{\ell}^c|} \; | \widehat{W}_{j_1,j_2}^{(\ell)}-W_{0,j_1,j_2}^{(\ell)} |,
\end{eqnarray*}
under the significance level $\alpha/2$. Denote the two critical values by $\widehat{c}^{(\ell)}(0,q)$ and $\widehat{c}^{(\ell)}(q,d+1)$, respectively. See Section \ref{sec:boot} for more details. 

Once obtaining the critical values, we reject $H_0(0,q)$ if $\widehat{\bm{B}}^{*(\ell)}_{q,0} \left\{ |\mathcal{I}_{\ell}^c|^{-1/2}\widehat{c}^{(\ell)}(0,q) \right\} = 1$, and reject $H_0(q,d+1)$ if $\widehat{\bm{B}}^{*(\ell)}_{d+1,q} \left\{ |\mathcal{I}_{\ell}^c|^{-1/2}\widehat{c}^{(\ell)}(q,d+1) \right\} = 1$. We reject the null $H_0(q)$ when $H_0(0,q)$ and $H_0(q,d+1)$ are both rejected. Note that, for each half of the data $\ell = 1, 2$, we have made a decision $\mathcal{D}^{(\ell)}$ regarding $H_0(q)$. Finally, we reject $H_0(q)$ when either $\mathcal{D}^{(1)}$ or $\mathcal{D}^{(2)}$ decides to reject. By Bonferroni's inequality, this yields a valid $\alpha$-level test. 

We summarize the full testing procedure in Algorithm \ref{alg:full}.

\subsection{Initial DAG estimation}
\label{sec:init-est}

There are multiple estimation methods available to produce an initial estimator for $\bm{W}_0$, for instance, \citet{zheng2018dags} and \citet{Yuan2019}. We employ the method of \citet{zheng2018dags} in our implementation. Specifically, we seek $\min_{\bm{W}\in \mathbb{R}^{(d+2)\times (d+2)}} \mathcal{L}(\bm{W}) + \lambda |\mathcal{I}_{\ell}| \sum_{i,j}|W_{i,j}|$, subject to $\mathbb{G}(\bm{W})\in \hbox{DAGs}$, where $\mathcal{L}(\bm{W}) = \sum_{i\in \mathcal{I}_{\ell}}\|\widetilde{\bm{X}}_i-\bm{W} \widetilde{\bm{X}}_i\|_2^2$, $\lambda>0$ is a regularization parameter, $\mathbb{G}$ denotes the graph induced by $\bm{W}$, $\widetilde{\bm{X}}_i = \bm{X}_i-\widehat{\bm{\mu}}$ is the centered covariate, $\widehat{\bm{\mu}} =  \sum_{i=1}^n \bm{X}_i/n$, and $|\mathcal{I}_{\ell}|$ is the number of data samples in the data split $\mathcal{I}_{\ell}$. This optimization problem is challenging to solve due to the fact that the search space of DAGs scales super-exponentially with the dimension $d$. To resolve this issue, \cite{zheng2018dags} proposed a novel characterization of the acyclic constraint, by showing that the DAG constraint can be represented by trace$\{\exp(\bm{W}\circ \bm{W})\}=d+2$, where $\circ$ denotes the Hadamard product, $\exp(\bm{A})$ is the matrix exponential of $\bm{A}$, and trace($\bm{A}$) is the trace of $\bm{A}$. Then the problem becomes 
\begin{align} \label{eqn:notears-l1}
\min_{\bm{W}\in \mathbb{R}^{(d+2)\times (d+2)}}  \mathcal{L}(\bm{W}) + \lambda |\mathcal{I}_{\ell}| \sum_{i,j}|W_{i,j}|, \ \textrm{ subject to } \textrm{trace}\{\exp(\bm{W}\circ \bm{W})\} = d+2.
\end{align}
Let $\widetilde{\bm{W}}^{(\ell)}$ denote the minimizer of \eqref{eqn:notears-l1}. \citet{zheng2018dags} proposed an efficient augmented Lagrangian based algorithm to solve \eqref{eqn:notears-l1}. After obtaining $\widetilde{\bm{W}}^{(\ell)}$, we set the elements in its first row and last column to zero, following Condition (A2). 

\change{We make some remarks. First, the optimization in \eqref{eqn:notears-l1} is nonconvex, and there is no guarantee that the algorithm of \cite{zheng2018dags} can find the global minimizer. As such, $\widetilde{\bm{W}}^{(\ell)}$ may not satisfy the acyclicity condition, although a global minimizer does. To meet the acyclicity constraint, \citet{zheng2018dags} employed an additional thresholding step to truncate all the elements in the numerical solution to \eqref{eqn:notears-l1} whose absolute values are smaller than some threshold value $c_0$ to zero. We follow their implementation and adopt the same thresholding value $c_0 = 10^{-3}$. Second, to achieve the theoretical guarantees of our proposed test, we only require the estimator $\widetilde{\bm{W}}^{(\ell)}$ to be a consistent estimator of $\bm{W}_0$, which is much weaker than the requirement of the test of \citet{LHZ2018} that the DAG estimator has to be selection consistent. In Section \ref{sec:initialestimator}, we show that the global solution of \eqref{eqn:notears-l1} satisfies this consistency requirement (see Proposition \ref{prop:oracle}). Meanwhile, as long as $c_0 \ll \sqrt{n^{-1}\log n}$, we can show Proposition \ref{prop:oracle} holds for the thresholded solution of \eqref{eqn:notears-l1} as well.}

\subsection{Screening and debiasing}
\label{sec:decor}

Given the initial estimator $\widetilde{\bm{W}}^{(\ell)}$, we next compute the binary estimator $\widehat{\bm{B}}^{(\ell)}$ for $\bm{B}_0$ using  \eqref{eqn:Bc} with $c=0$. We then use the nonzero entries of $\widehat{\bm{B}}^{(\ell)}$ to determine the support of the decorrelated estimator $\widehat{\bm{W}}^{(\ell)}$ in the subsequent step of decorrelated estimation. As such, it serves as a screening step, and allows us to reduce the number of potential paths to a moderate level. As shown in \eqref{naivetesteq}, the decorrelated estimator $(\widehat{\bm{W}}^{*(\ell)})_{q_1,q_2}$ depends on $\widehat{\bm{W}}^{(\ell)}$ only through its entries in $\mathcal{S}(q_1,q_2,\widehat{\bm{B}}^{(\ell)})$. Consequently, the screening through $\widehat{\bm{B}}^{(\ell)}$ reduces the variance of $(\widehat{\bm{W}}^{*(\ell)})_{q_1,q_2}$, which in turn leads to an increased power for our test. 

Next, we employ the decorrelated estimation idea of \citet{Ning2017} to compute a decorrelated estimator $\widehat{\bm{W}}^{(\ell)}$ to reduce the bias of the initial estimator $\widetilde{\bm{W}}^{(\ell)}$ obtained from \eqref{eqn:notears-l1}. Because of the presence of the regularization term in \eqref{eqn:notears-l1} for  high-dimensional mediators, the initial estimator $\widetilde{\bm{W}}^{(\ell)}$ may suffer from a large bias and does not have a tractable limiting distribution. To address this issue, we refit $W_{0,j_1,j_2}$ for any $(j_1,j_2)$ such that $\widehat{B}^{(\ell)}_{j_1,j_2}\neq 0$, by constructing an estimating equation based on a decorrelated score function. This effectively alleviates the bias, and the resulting decorrelated estimator $\widehat{W}^{(\ell)}_{j_1,j_2}$ is both $\sqrt{n}$-consistent and asymptotically normal. 

More specifically, after some calculations, we have that, 
\begin{align}\label{eqn:decor}
\begin{split}
\Mean \left(X_{j_1}-\sum_{j\neq j_2} W_{0,j_1,j} X_j\right) \Big( X_{j_2}-\Mean \left[X_{j_2}|\{X_j\}_{j\in \scriptsize{\hbox{ACT}}(j_2,\bm{W}_0)} \right] \Big) \\
=W_{0,j_1,j_2} \Mean \left\{ X_{j_2}\Big( X_{j_2}- \Mean \left[ X_{j_2}|\{X_j\}_{j\in \scriptsize{\hbox{ACT}}(j_2,\bm{W}_0)} \right] \Big) \right\},
\end{split}
\end{align}
which is the estimating equation to construct our decorrelated estimator $\widehat{W}_{0,j_1,j_2}$. Toward that end, we need to estimate $\Mean \left[ X_{j_2}|\{X_j\}_{j\in \scriptsize{\hbox{ACT}}(j_2,\bm{W}_0)} \right]$ and $\{W_{0,j_1,j}:j\neq j_2\}$. 

To estimate $\Mean \left[ X_{j_2}|\{X_j\}_{j \in \scriptsize{\hbox{ACT}}(j_2,\bm{W}_0)} \right]$, we first estimate the set of ancestors of the $j_2$th node ${\hbox{ACT}}(j_2,\bm{W}_0)$ by ACT$(j_2, \widetilde{\bm{W}}^{(\ell)}) = \left\{ 1 \le j \le d : (\widetilde{\bm{W}}^{*(\ell)})_{j_2,j}\neq 0 \right\}$, for $j_2 = 1, \ldots, d+1$, where $\widetilde{\bm{W}}^{*(\ell)} = |\widetilde{\bm{W}}^{(\ell)}| \oplus | \widetilde{\bm{W}}^{(\ell)}|^{(2)} \oplus \ldots \oplus |\widetilde{\bm{W}}^{(\ell)}|^{(d)}$. We also note that, when estimating the ancestors, we always include the exposure variable  $E = X_0$ in the set of ancestors, and always include all mediators when estimating the ancestors of the outcome variable $Y = X_{d+1}$. Next, we approximate $\Mean \left[ X_{j_2}|\{X_j\}_{j\in \scriptsize{\hbox{ACT}}(j_2,\widetilde{\bm{W}}^{(\ell)})} \right]$ using a linear regression model, where the regression coefficients are estimated by,  
\begin{eqnarray} \label{eqn:beta}
\widehat{\bm{\beta}}^{(\ell)}_{j_1,j_2} = \argmin_{\substack{\bm{\beta}:\beta_{j_2}=0\\ \footnotesize{\hbox{supp}}(\bm{\beta}) \in \scriptsize{\hbox{ACT}}(j_1,\widetilde{\bm{W}}^{(\ell)}) } }\left\{\frac{1}{|\mathcal{I}_{\ell}^c|} \sum_{i\in \mathcal{I}_{\ell}^c} \left( \widetilde{X}_{i,j_2}-\bm{\beta}^\top \widetilde{\bm{X}}_{i} \right)^2+\sum_{\substack{k:k\neq j_2, \\ k\in \scriptsize{\hbox{ACT}}(j_1,\widetilde{\bm{W}}^{(\ell)})} } p_{\lambda}(|\beta_k|)\right\}, 
\end{eqnarray}
where $\textrm{supp}(\bm{\beta})$ denotes the support of $\bm{\beta} \in \mathbb{R}^{d+2}$, and the regression fitting is done based on the complement set of samples $\mathcal{I}_{\ell}^c$. We choose the MCP \citep{zhang09} penalty function and tune the penalty parameter by the Bayesian information criterion in our implementation. Alternatively, we can use LASSO \citep{tibsh1996}, SCAD \citep{fan2001} or Dantzig selector \citep{tao2007} in \eqref{eqn:beta}. It is crucial to note that, the resulting estimator is $\sqrt{n}$-consistent regardless of whether the linear model approximation holds or not.

To estimate $\{W_{0,j_1,j}:j\neq j_2\}$, we employ a refined version of the initial estimator $\widetilde{\bm{W}}^{(\ell)}$. That is, we update the $j$th row of $\widetilde{\bm{W}}^{(\ell)}$, $j = 1, \ldots, d$, by fitting a penalized regression with $\{X_{i,j}\}_{i\in \mathcal{I}_{\ell}}$ being the response and $\{X_{i,k}\}_{i\in \mathcal{I}_{\ell}, \widetilde{W}^{(\ell)}_{j,k}\neq 0 }$ being the predictors. We again use the MCP penalty. Denote the resulting refined estimator by $\overline{\bm{W}}^{(\ell)}$. The purpose of this refitting is to improve the estimation efficiency of the initial estimator $\widetilde{\bm{W}}^{(\ell)}$. In our numerical experiments, we find $\overline{\bm{W}}^{(\ell)}$ usually converges faster than $\widetilde{\bm{W}}^{(\ell)}$ to the truth. 

Built on the above estimators and the estimating equation \eqref{eqn:decor}, we debias $\widetilde{W}^{(\ell)}_{j_1,j_2}$ using the other half of the data $\{\bm{X}_i\}_{i\in \mathcal{I}_{\ell}^c}$, for any entry such that $\widehat{B}^{(\ell)}_{j_1,j_2}\neq 0$, by 
\begin{eqnarray}\label{eqn:decorrelatedest}
\widehat{W}^{(\ell)}_{j_1,j_2} = \frac{\displaystyle \sum_{i\in \mathcal{I}_{\ell}^c} \left( \widetilde{X}_{i,j_2}-\widehat{\bm{\beta}}^{(\ell)\top}_{j_1,j_2} \widetilde{\bm{X}}_{i } \right) \left( \widetilde{X}_{i,j_1}-\sum_{j\neq j_2} \widetilde{X}_{i,j}\overline{W}^{(\ell)}_{j_1,j} \right) }{\displaystyle \sum_{i\in \mathcal{I}_{\ell}^c} \widetilde{X}_{i,j_2} \left( \widetilde{X}_{i,j_2}-\widehat{\bm{\beta}}^{(\ell)\top}_{j_1,j_2} \widetilde{\bm{X}}_i \right)}.
\end{eqnarray}

We remark that, we have used the cross-fitting strategy in both the estimation of $\widehat{\bm{\beta}}^{(\ell)}_{j_1,j_2}$ in \eqref{eqn:beta}, and in the decorrelated estimation of $\widehat{W}^{(\ell)}_{j_1,j_2}$ in \eqref{eqn:decorrelatedest}. This strategy guarantees each entry of the decorrelated estimator $\widehat{\bm{W}}^{(\ell)}$ is asymptotically normal, regardless of whether the initial estimator $\widetilde{\bm{W}}^{(\ell)}$ is selection consistent or not.

\subsection{Bootstrap for critical values}
\label{sec:boot}

We next develop a multiplier bootstrap method to obtain the critical values, and summarize this procedure in Algorithm \ref{alg:bootstrap}. Our goal is to approximate the limiting distribution of $\widehat{S}^{(\ell)} = \max_{(j_1,j_2)\in \mathcal{S}(q_1,q_2,\widehat{\bm{B}}^{(\ell)})} \sqrt{|\mathcal{I}_{\ell}^c|} |\widehat{W}_{j_1,j_2}^{(\ell)}-W_{0,j_1,j_2}|$ on the right-hand-side of \eqref{naivetesteq}. 

We first observe that $\sqrt{|\mathcal{I}_{\ell}^c|}(\widehat{W}^{(\ell)}_{j_1,j_2}-W_{0,j_1,j_2})$ is asymptotically equivalent to
\begin{eqnarray}\label{eqn:etaequivalent}
\eta_{j_1,j_2}^{(\ell)}=\frac{\sqrt{|\mathcal{I}_{\ell}^c|}\displaystyle \sum_{i\in \mathcal{I}_{\ell}^c} \left\{ \widetilde{X}_{i,j_2}-\widehat{\bm{\beta}}^{(\ell)\top}(j_1,j_2) \widetilde{\bm{X}}_{i } \right\} \, \varepsilon_{i,j_1} }{\displaystyle \sum_{i\in \mathcal{I}_{\ell}^c} \widetilde{X}_{i,j_2} \left\{ \widetilde{X}_{i,j_2}-\widehat{\bm{\beta}}^{(\ell)\top}(j_1,j_2) \widetilde{\bm{X}}_i \right\}}.
\end{eqnarray}
Correspondingly, $\widehat{S}^{(\ell)} = \max_{(j_1,j_2)\in \mathcal{S}(q_1,q_2,\widehat{\bm{B}}^{(\ell)})} |\eta_{j_1,j_2}^{(\ell)}| + o_p(1)$, for any $q_1\in \{0,1\ldots,d\}$ and $q_2\in \{1,\ldots,d+1\}$. Conditioning on $\left\{ X_{i,j}:i\in \mathcal{I}_{\ell}^c,j\in \hbox{ACT}(j_1,\widetilde{\bm{W}}^{(\ell)}) \right\}$, $\eta_{j_1,j_2}^{(\ell)}$ corresponds to a sum of independent mean zero random variables, and is asymptotically normal. A rigorous proof is given in Step 1 of the proof of Theorem \ref{thm1} in the appendix. This implies that $\sqrt{|\mathcal{I}_{\ell}^c|}(\widehat{W}^{(\ell)}_{j_1,j_2}-W_{0,j_1,j_2})$ is asymptotically normal. Therefore, $\widehat{S}^{(\ell)}$ is to converge in distribution to a maximum of normal random variables in absolute values. Its quantile can be consistently estimated by a multiplier bootstrap method \citep{Chernozhukov2013}.

\begin{algorithm}[t!]
\caption{Bootstrap procedure to obtain the critical values.}
\label{alg:bootstrap}
\begin{algorithmic}
\item
\begin{description}
\item[\textbf{Input}:] The data $\left\{ \bm{X}_i:i\in \mathcal{I}_{\ell}^{(c)} \right\}$, the significance level $\alpha$, the variance estimator $\widehat{\sigma}_*^2$, the number of bootstrap samples $m$, the estimator $\widehat{\bm{\beta}}^{(\ell)}(j_1,j_2)$ from \eqref{eqn:beta}, and the set $\mathcal{S}(q_1,q_2,\widehat{\bm{B}}^{(\ell)})$.

\item[\textbf{Step 1}.] Generate i.i.d.\ standard normal random variables $\left\{ e_{i,j}^{(b)} \right\}_{i,j}$, $b=1,2,\ldots,m$. 

\item[\textbf{Step 2}.] Compute $\eta_{j_1,j_2}^{(\ell,b)*}$ according to \eqref{eqn:etaj1j2}, with $e_{i,j_1}$ replaced by $e_{i,j_1}^{(b)}$, and 
\begin{eqnarray*}
T^{(\ell,b)}(q_1, q_2) = \max_{(j_1,j_2)\in\mathcal{S}(q_1,q_2,\widehat{\bm{B}}^{(\ell)})} |\eta_{j_1,j_2}^{(\ell,b)*}|, \quad b=1,2,\ldots,m.
\end{eqnarray*}

\item[\textbf{Output}:] The empirical upper $\alpha$th quantile of $\{T^{(\ell, b)}(q_1, q_2) : b=1,\ldots,m\}$.
\end{description}
\end{algorithmic}
\end{algorithm}

More specifically, one can generate the bootstrap samples by replacing the residual term $\sigma_{*}^{-1}(\widetilde{X}_{i,j_1}-\sum_j \widetilde{X}_{i,j} \widetilde{W}_{i,j}^{(\ell)})$ in \eqref{eqn:decorrelatedest} with i.i.d.\ standard normal noise $\{e_{i,j}:1\le i\le n, 0\le j\le d+1\}$ that are independent of the data. That is, we approximate $\eta_{j_1,j_2}^{(\ell)}$ in \eqref{eqn:etaequivalent} by 
\begin{eqnarray}\label{eqn:etaj1j2}
\eta_{j_1,j_2}^{*(\ell)} = \frac{\sqrt{|\mathcal{I}_{\ell}^c|}\displaystyle \sum_{i\in \mathcal{I}_{\ell}^c} \left\{ \widetilde{X}_{i,j_2}-\widehat{\bm{\beta}}^{(\ell)\top}(j_1,j_2) \widetilde{\bm{X}}_{i } \right\} \, e_{i,j_1} \, \widehat{\sigma}_* }{\displaystyle \sum_{i\in \mathcal{I}_{\ell}^c} \widetilde{X}_{i,j_2} \left\{ \widetilde{X}_{i,j_2}-\widehat{\bm{\beta}}^{(\ell)\top}(j_1,j_2) \widetilde{\bm{X}}_i \right\}},
\end{eqnarray}
where $\widehat{\sigma}_*$ is some consistent estimator of $\sigma_*$. We propose to estimate $\sigma_*^{2}$ by $\widehat{\sigma}_*^{2} = \{n(d+2)\}^{-1} \sum_{\ell \in \{1,2\}} \sum_{i\in \mathcal{I}_{\ell}^c} \sum_{j=0}^{d+1} |\widetilde{X}_{i,j} - \overline{\bm{W}}^{(\ell)\top}_j \widetilde{\bm{X}}_i|_2^2$, where $\overline{\bm{W}}^{(\ell)}_j$ denotes the $j$th row of $\overline{\bm{W}}^{(\ell)}$. This estimation utilizes sample splitting again, which alleviates potential bias of the variance estimator resulting from the high correlations between the noises and the mediators in the high-dimensional setting \citep{fan2008}.  Lemma \ref{lemma:sigma} in Section \ref{sec:lemmas} of the appendix shows that $\widehat{\sigma}_*^{2}$ is indeed consistent. Then the limiting distribution of $\widehat{S}^{(\ell)}$ can be well approximated by the conditional distribution of the bootstrap samples given the data. A formal justification is given in Step 3 of the proof of Theorem \ref{thm1} in the appendix.

\subsection{False discovery rate control}
\label{sec:fdr}

We next present a multiple testing procedure for simultaneous inference of multivariate mediators with a proper FDR control. We present the full procedure in Algorithm \ref{alg:fdr}, which consists of four steps. We next detail each step. Let $\mathcal{N}$ be the set of unimportant mediators and $\mathcal{H}$ be the set of our selected mediators. The FDR is defined as the expected proportion of falsely selected mediators, i.e., $\hbox{FDR}(\mathcal{H})=\Mean \left\{ |\mathcal{N} \cup \mathcal{H}| / \max(1,|\mathcal{H}|) \right\}$. 

First, we compute the $p$-value of testing $H_0(q_1,q_2)$, $q_1 = 0, \ldots, d, q_2 = 1, \ldots, d+1$, for each half of the data. Specifically, we compute the decorrelated estimator $\widehat{\bm{W}}^{(\ell)}$ in Step 4 of Algorithm \ref{alg:full}, and $\widehat{\bm{W}}^{*(\ell)}=|\widehat{\bm{W}}^{(\ell)}|\oplus |\widehat{\bm{W}}^{(\ell)}|^{(2)}\oplus\ldots\oplus |\widehat{\bm{W}}^{(\ell)}|^{(d)}$. We next compute $T^{(\ell,b)}(q_1,q_2)$, $b=1, \ldots,m$, in Step 2 of Algorithm \ref{alg:bootstrap}. Then the $p$-value of testing $H_0(q_1,q_2)$ in \eqref{eqn:sub-hypo} is
\begin{eqnarray}\label{eqn:pvalq1q2}
\widehat{p}^{(\ell)}(q_1,q_2)=\frac{1}{m}\sum_{b=1}^m \mathbb{I}\left\{ T^{(\ell,b)}(q_1,q_2)\ge \sqrt{\mathcal{I}_{\ell}^c}(\widehat{\bm{W}}^{*(\ell)})_{q_2,q_1} \right\}.
\end{eqnarray}	
The $p$-values of testing $H_0(0,q)$ and $H_0(q,d+1)$ are $\widehat{p}^{(\ell)}(0,q)$ and $\widehat{p}^{(\ell)}(q,d+1)$, respectively.

\begin{algorithm}[t]
\caption{Multiple testing procedure for inference of multivariate mediators.}
\label{alg:fdr}
\begin{algorithmic}
\item
\begin{description}
\item[\textbf{Input}:] The significance level $\alpha$, and the thresholding values $0 < c^{(1)}, c^{(2)} < 1$. 

\item[\textbf{Step 1}.] Compute the $p$-values of testing the null hypothesis $H_0(q_1,q_2)$ for $q_1 = 0, \ldots, d, q_2 = 1, \ldots, d+1$ using \eqref{eqn:pvalq1q2} for each half of the data, $\ell = 1,2$. 

\item[\textbf{Step 2}.] Screening based on the pairwise minimum $p$-values, $\widehat{p}^{(\ell)}_{\min}(q) = \min\big\{ \widehat{p}^{(\ell)}(0,q), \widehat{p}^{(\ell)}(q,d+1) \big\}$. Let $\mathcal{H}_0^{(\ell)}=\{1\le q\le d: \widehat{p}^{(\ell)}_{\min}(q) \le c^{(\ell)}\}$ denote the set of the initially selected mediators, $\ell = 1,2$.  

\item[\textbf{Step 3}.] Order by the pairwise maximum $p$-values, $\widehat{p}^{(\ell)}_{\max}(q) = \max\big\{ \widehat{p}^{(\ell)}(0,q), \widehat{p}^{(\ell)}(q,d+1) \big\}$, for those mediators in $\mathcal{H}_0^{(\ell)}$, as $\widehat{p}^{(\ell)}_{(1)}\le \widehat{p}^{(\ell)}_{(2)}\le \ldots\le \widehat{p}^{(\ell)}_{(|\mathcal{H}_0^{(\ell)}|)}$. 

\item[\textbf{Step 4}.] Select $h^{(\ell)}$ mediators in $\mathcal{H}_0^{(\ell)}$ with the smallest $p$-values. Let $\mathcal{H}^{(\ell)}$  denote the set of selected mediators, $\ell = 1, 2$. 

\item[\textbf{Output}:] $\mathcal{H}=\mathcal{H}^{(1)}\cup \mathcal{H}^{(2)}$. 
\end{description}
\end{algorithmic}
\end{algorithm}

Next, we adopt and extend the ScreenMin procedure proposed by \cite{Djor2019} to our setting. We begin by computing the pairwise minimum $p$-values, $\widehat{p}^{(\ell)}_{\min}(q) = \min\big\{ \widehat{p}^{(\ell)}(0,q),\widehat{p}^{(\ell)}(q,d+1) \big\}$. We then screen and select those mediators whose corresponding $\widehat{p}^{(\ell)}_{\min}(q)$ is smaller than a thresholding value $c^{(\ell)}$, which is determined adaptively by $c^{(\ell)} = \max \big\{ c \in \left( \alpha/d, \ldots, \alpha/2, \alpha \right) : c |\mathcal{H}_0^{(\ell)}(c)| \le \alpha \big\}$, and $\mathcal{H}_0^{(\ell)}(c)$ denotes the set of prescreened mediators when the threshold value is $c$. \citet{djordjilovic2019optimal} showed such a thresholding value approximately maximizes the power to reject false union hypotheses. It also works well in our numerical studies. Denote the resulting set of important mediators by the ScreenMin procedure as $\mathcal{H}_0^{(\ell)}$. 

Next, we compute the pairwise maximum $p$-value, which is also the $p$-value of testing the significance of an individual mediator $H_0(q)$ in our setting, $\widehat{p}^{(\ell)}_{\max}(q) = \max\big\{ \widehat{p}^{(\ell)}(0,q), \widehat{p}^{(\ell)}(q,d$ $+1) \big\}$. We order the mediators in $\mathcal{H}_0^{(\ell)}$ according to $\widehat{p}^{(\ell)}_{\max}(q)$. 

Finally, we apply the procedure of \citet{benjamini2001control} to the ordered mediators, and select $h^{(\ell)}$ mediators with the smallest $p$-values, where $h^{(\ell)}=\max\Big[ i: \widehat{p}^{(\ell)}_{(i)} \le (i\alpha) / \{2|\mathcal{H}_0^{(\ell)}|\sum_{j=1}^{|\mathcal{H}_0^{(\ell)}|} j^{-1} \} \Big]$. Letting $\mathcal{H}^{(\ell)}$ denote the selected mediators for each half of the data, $\ell=1,2$, respectively, we set the final set of selected mediators as $\mathcal{H}=\mathcal{H}^{(1)}\cup \mathcal{H}^{(2)}$.

\section{Theory}
\label{sec:theory}

In this section, we first establish the consistency of our test for each individual mediator, by deriving the asymptotic size and power. We then show that the multiple testing procedure achieves a valid FDR control. Finally, as a by-product, we derive an oracle inequality for the estimator $\widetilde{\bm{W}}^{(\ell)}$ computed from \eqref{eqn:notears-l1} using the method of \citet{zheng2018dags}.

\subsection{Consistency and FDR control}
\label{sec:consistency}

\change{We first present a main regularity condition (A4), while we defer two additional regularity conditions (A5) and (A6) to Section \ref{sec:conditions} of the appendix in the interest of space. 

\vspace{-0.05in}
\begin{enumerate}[({A}1)]
\setcounter{enumi}{3}
\item With probability approaching one, ACT$(j,\widetilde{\bm{W}}^{(\ell)})$ contains all parents of $j$, for any $j = 0, \ldots, d+1, \ell=1,2$.
\end{enumerate}
\vspace{-0.1in}

\noindent
This condition requires an appropriate identification of the graph, in that ACT$(j,\widetilde{\bm{W}}^{(\ell)})$ contains all parents of $j$. It serves as a basis for the asymptotic properties of the proposed test. We make some remarks.  First, this condition is weaker than requiring $\widetilde{\bm{W}}_j^{(\ell)}$, the $j$th column of $\widetilde{\bm{W}}^{(\ell)}$, to satisfy the sure screening property; i.e, PA$(j) \subseteq \hbox{supp}(\widetilde{\bm{W}}_j^{(\ell)})$, where PA$(j)$ denotes the parents of node $j$. To better illustrate this, consider the DAG example in Figure \ref{fig:DAG-illustration}(b). For node $j=3$, the sure screening property of $\widetilde{\bm{W}}_3^{(\ell)}$ requires $W_{0,3,1}$ and $W_{0,3,2}$ to satisfy certain minimum-signal-strength conditions; see, e.g., \citet{fan2008}. In comparison, we require $\{1,2\} \in \hbox{ACT}(3,\widetilde{\bm{W}}^{(\ell)})$ with probability approaching one. This requires either $\{W_{0,3,1},W_{0,3,2}\}$, or $\{W_{0,2,1},W_{0,3,2}\}$, to satisfy certain minimum-signal-strength conditions. In that sense, our test is ``doubly robust". In Proposition \ref{prop:oracle}, we show (A4) is satisfied when the initial estimator is obtained using \eqref{eqn:notears-l1} of \citet{zheng2018dags}. Second, we remark that, when (A4) is not satisfied, the bias in the ancestor identification step is to affect the subsequent testing procedure. This phenomenon is similar to post-selection inference in linear regressions, where direct inference may fail when the variable selection step does not satisfy the sure screening or selection consistency property \citep{Mein2009, shi2019linear, zhu2020high}. In the linear regression setting, debiasing is an effective remedy to address the issue. However, the usual debiasing strategy may not be directly applicable in our setting; see Section \ref{sec:conditions} of the appendix for more discussion. We leave this post-selection inference problem as future research.}

We next establish the validity of our test for a single mediator in Theorem \ref{thm1}, and its local power property in Theorem \ref{thm2}. Combining the two theorems yields its consistency.  

\begin{thm}\label{thm1}
Suppose (A1) to (A5) hold. Suppose $d=O(n^{\kappa_1})$ for some constant $\kappa_1>0$, and $\|\bm{W}_0\|_2$ is bounded. Then for a significance level $0<\alpha<1$, and any mediator $q = 1, \ldots, d$, the proposed test in Algorithm \ref{alg:full} satisfies that 
\vspace{-0.05in}
\begin{eqnarray*}
\prob\Big\{ H_0(q)~\hbox{is}~\hbox{rejected} \; | \; H_0(q)~\hbox{holds} \Big\} \le \alpha + o(1).
\end{eqnarray*}
\end{thm}

Next, for any directed path $\zeta$: $E \to M_{i_1} \to \ldots \to M_{i_k} \to Y$, define $\omega_{\zeta}^*$ as the minimum signal strength along this path, 
\vspace{-0.05in}
\begin{eqnarray*}
\omega_{\zeta}^*=\min\left\{ |W_{0,i_1,0}|, \min_{j\in \{1,\ldots,k-1\}} |W_{0,i_{j+1},i_j}|,|W_{0,d+1,i_k}| \right\}. 
\end{eqnarray*}
Under the alternative hypothesis $H_1(q)$, there exists at least one path $\zeta$ that passes through $M_q$ such that $\omega_{\zeta}^*>0$. We next establish the local power property of our test. 

\begin{thm}\label{thm2}
Suppose the conditions in Theorem \ref{thm1} hold. Suppose $\max_{j\in \{0,1,\ldots,d+1\}}\|\widetilde{\bm{W}}_j^{(\ell)}-\bm{W}_{0,j}\|_2=O_p(n^{-1/2}\sqrt{\log n})$, where $\widetilde{\bm{W}}_j^{(\ell)}$ is the $j$th row of $\widetilde{\bm{W}}_j$. 
Suppose there exists one path $\zeta$:  $E\to M_{i_1}\to \ldots \to M_{i_k} \to Y$ that passes through $M_q$ such that $\omega^*_{\zeta}\gg n^{-1/2}\sqrt{\log n}$ under $H_1(q)$. Then the proposed test in Algorithm \ref{alg:full} satisfies that,  
\vspace{-0.05in}
\begin{eqnarray*}
\prob\Big\{ H_0(q)~\hbox{is}~\hbox{rejected} \; | \; H_1(q)~\hbox{holds} \Big\} \to 1, \;\; \textrm{ as } \; n \to \infty. 
\end{eqnarray*}
\end{thm}

\noindent
Note that we require $\omega^*_{\zeta}\gg n^{-1/2}\sqrt{\log n}$ for some $\zeta$ in Theorem \ref{thm2}. Consequently, our test is consistent against some local alternatives that are $\sqrt{n}$-consistent to the null up to some logarithmic term. Let $s_0 = \max_j |\hbox{supp}(\bm{W}_{0,j})|$ denotes the maximum sparsity size where $\bm{W}_{0,j}$ stands for the $j$th row of $\bm{W}_0$. In Proposition \ref{prop:oracle}, we show that $\max_{j\in \{0,\ldots,d+1\}}\|\widetilde{\bm{W}}_{j}^{(\ell)}-\bm{W}_{j}\|_2=O_p(n^{-1/2}\sqrt{\log n})$ when the maximum sparsity size $s_0$ is bounded. 

Next, we show that our multiple testing procedure achieves a valid FDR control. Note that we use a union-intersection principle to construct the $p$-value for $H_0(q)$. The key idea of the ScreenMin procedure of \cite{Djor2019} lies in exploiting the independence between the two $p$-values $\widehat{p}^{(\ell)}(0,q)$ and $\widehat{p}^{(\ell)}(q,d+1)$. In our setting, these $p$-values are actually asymptotically independent. We thus have the following result. 

\begin{thm}\label{thm3}
Suppose the conditions in Theorem \ref{thm1} hold. Then the set of selected mediators $\mathcal{H}$ in Algorithm \ref{alg:fdr} satisfies that  $\textrm{FDR}(\mathcal{H}) \le \alpha+o(1)$.
\end{thm}

\subsection{Oracle inequality for the initial DAG estimator}
\label{sec:initialestimator}

As a by-product, we establish the oracle inequality for the estimator of \citet{zheng2018dags}. We first introduce the oracle estimator. For a given ordering $\pi = \{ \pi_0,\pi_1,\ldots,\pi_{d+1} \}$, consider the estimator $\widetilde{\bm{W}}^{(\ell)}(\pi) = \left\{ \widetilde{\bm{W}}_0^{(\ell)}(\pi), \widetilde{\bm{W}}^{(\ell)}_1(\pi),\ldots,\widetilde{\bm{W}}^{(\ell)}_{d+1}(\pi) \right\}^\top$ where 
\begin{eqnarray*}
	\widetilde{\bm{W}}_{\pi_{j}}^{(\ell)}(\pi)=\argmin_{\bm{\beta}:\hbox{\footnotesize{supp}}(\bm{\beta}) \in \{\pi_{0},\pi_{1},\ldots,\pi_{j-1}\} } \sum_{i\in \mathcal{I}_{\ell}} (X_{i,\pi_{j}}-\bm{\beta}^\top \bm{X}_i)^2 + \lambda |\mathcal{I}_{\ell}| \|\bm{\beta}\|_1,
\end{eqnarray*}
for $j \in \{0,1,\ldots,d+1\}$. That is, $\widetilde{\bm{W}}^{(\ell)}(\pi)$ is computed as if the ordering of $\pi$ were known. Let $\Pi^*$ denote the set of all true orderings, while a more rigorous definition is given in Section \ref{sec:conditions} of the supplementary appendix. Then the oracle estimator $\widetilde{\bm{W}}_{\pi_{j}}^{(\ell)}(\pi^*)$, for some $\pi^* \in \Pi^*$, is computed as if the true ordering $\pi^*$ were known. With a proper choice of $\lambda$, it follows from the oracle inequality for LASSO \citep{Bickel2009} that, 
\begin{eqnarray*}
	\max_{j\in \{0,\ldots,d+1\}} \|\widetilde{\bm{W}}^{(\ell)}_j(\pi^*)-\bm{W}_{0,j}\|_2\le O(1) n^{-1/2}\sqrt{s_0\log n}.
\end{eqnarray*}

The next proposition establishes the convergence rate of $\widetilde{\bm{W}}^{(\ell)}$ obtained from \eqref{eqn:notears-l1}. 

\begin{prop}\label{prop:oracle}
Suppose (A1), (A2), (A3) and (A6) hold. Suppose $d=O(n^{\kappa_1})$ for some constant $\kappa_1<1$, $\|\bm{W}_0\|_2$ is bounded, and $\lambda=\kappa_2 n^{-1/2}\sqrt{\log n}$ for some sufficiently large constant $\kappa_2>0$. Then with probability tending to $1$, the initial estimator $\widetilde{\bm{W}}^{(\ell)}$ obtained from \eqref{eqn:notears-l1} satisfies that, 
\begin{eqnarray*}
\widetilde{\bm{W}}^{(\ell)}=\widetilde{\bm{W}}^{(\ell)}(\pi^*) \textrm{ for some } \pi^* \in \Pi^*, \textrm{ and } 
\max_{j\in \{0,\ldots,d+1\}} \|\widetilde{\bm{W}}^{(\ell)}_j-\bm{W}_{0,j}\|_2\le O(1) n^{-1/2}\sqrt{s_0\log n}.
\end{eqnarray*}
\end{prop}

\noindent
Proposition \ref{prop:oracle} shows that the convergence rate of $\widetilde{\bm{W}}^{(\ell)}$ is the same as that of the oracle estimator. Moreover, the true ordering $\pi^*$ can be inferred from $\widetilde{\bm{W}}^{(\ell)}$. \change{If we produce the initial estimator from \eqref{eqn:notears-l1}, it further implies that (A4) holds. We again make some remarks. First, in this proposition, we require the dimension $d$ to grow at a slower rate than $n$. We note that this condition is not needed in Theorems \ref{thm1}-\ref{thm3}. Moreover, we can further relax this requirement on $d$ by imposing some sparsity conditions on $\widetilde{\bm{W}}^{(\ell)}$ and the population limit of $\widetilde{\bm{W}}^{(\ell)}(\pi)$; see the final remark of the proof of Proposition 1 in the appendix. Second, this proposition is for the global minimizer of \eqref{eqn:notears-l1}. Of course, 
as we have commented, there is no guarantee that the algorithm of \cite{zheng2018dags} can find the global minimizer. This is a universal problem for almost all nonconvex optimizations. Nevertheless, \citet[Theorem 1]{zhong2014proximal} showed that the actually minimizer to \eqref{eqn:notears-l1}, which is obtained through the proximal quasi-Newton method, converges to the local minimizer of the augmented Lagrange problem, while \citet[Table 1]{zheng2018dags} showed that numerically the difference between the actual minimizer and the global minimizer is much smaller than that between the global minimizer and the ground truth. As such, we expect Proposition \ref{prop:oracle} to hold for the actually minimizer as well. Meanwhile, we acknowledge that this local minimizer problem is challenging and is warranted for future research.}

\section{Simulations}
\label{sec:simulations}

We simulate the data following model \eqref{eqn:md}. We set $\bm{\mu}_0$ to a vector of ones, and $\sigma_*^2=1$. We generate the adjacency matrix $\bm{W}_0$ as follows: We begin with a zero matrix, then replace every entry $W_{0,j_1,j_2}$ in the lower off-diagonals by the product of two random variables $R_{j_1,j_2}^{(1)} R_{j_1,j_2}^{(2)}$. Here $R_{j_1,j_2}^{(1)} \sim \textrm{Bernoulli}(p_1)$, if $j_2=0$, or $j_1=d+1$, and $R_{j_1,j_2}^{(1)} \sim \textrm{Bernoulli}(p_2)$, otherwise, and $R_{j_1,j_2}^{(2)}$ is uniformly distributed on $[-2,-0.5]\cup [0.5,2]$. All these variables are independently generated. We consider three scenarios of the total number of mediators $d$, with varying binary probabilities $p_1, p_2$, each under two sample sizes $n$; i.e., $(d,p_1,p_2)=(50,0.05,0.15)$ with $n=100,200$, $(d,p_1,p_2)=(100,0.03,0.1)$ with $n=250,500$, and $(d,p_1,p_2)=(150,0.02,0.05)$ with $n=250,500$.  Table \ref{tab:strength} in Section \ref{sec:add-numerical} of the appendix reports the corresponding mediators with nonzero mediation effects, and their associated $\delta(q)$, where $\delta(q) = (\bm{W}_0^*)_{d+1,q}(\bm{W}_0^*)_{q,0}$, and $\bm{W}_0^*$ is constructed based on $\bm{W}_0$, $q = 1, \ldots, d$. By Lemma \ref{thm:lemma2}, $\delta(q)$ measures the size of the mediation effect. When $\delta(q)=0$, $H_0(q)$ holds; otherwise, $H_1(q)$ holds. A larger $\delta(q)$ indicates a stronger mediation effect. The percentage of nonzero mediators for the three scenarios is $0.12, 0.09$ and $0.06$, respectively. 

We first evaluate the empirical performance of our test for a single mediator in Algorithm \ref{alg:full}. We also compare it with that of \cite{LHZ2018}, which they named as Mediation Interventional calculus when the DAG is Absent (MIDA). We use the same initial estimator as ours for MIDA. We construct the $100(1-\alpha)\%$ confidence interval for the total effect of each mediator, following the procedure as described in \cite{LHZ2018}. We reject the null hypothesis if zero is not covered by the confidence interval.

\begin{figure}[t!]
\centering
\includegraphics[width=15cm]{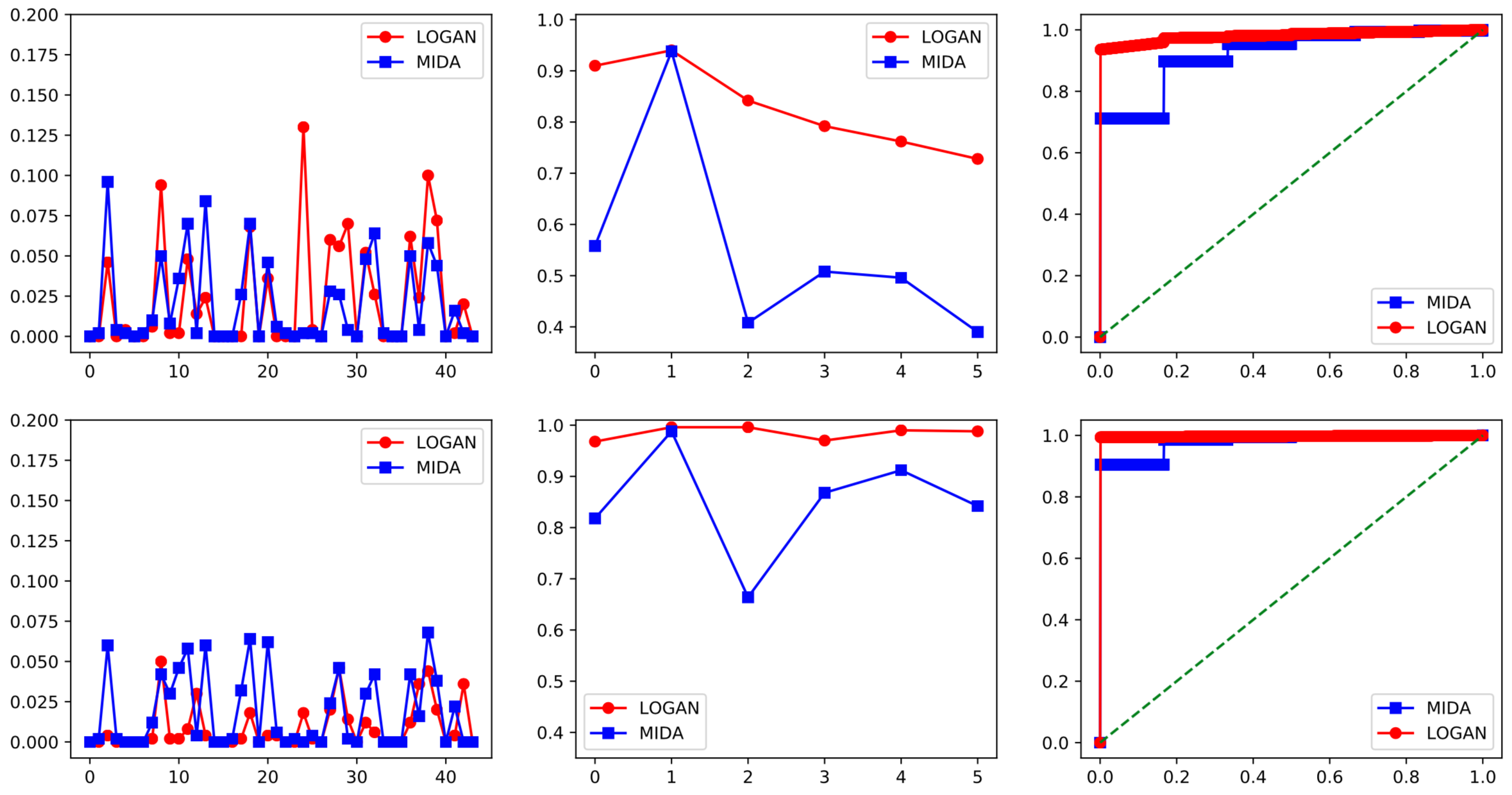}
\caption{Empirical rejection rate and ROC curve of the proposed test, LOGAN, and the test of \cite{LHZ2018}, MIDA, when $d=50$. The upper panels: $n=100$, and the bottom panels: $n=200$. The left panels: under $H_0$, the middles panels: under $H_1$, where the horizontal axis is the mediator index, \change{and the right panels: the average ROC curve.}}
\label{figAB}
\end{figure}

We evaluate each testing method by the empirical rejection rate, in percentage, out of 500 data replications at the significance level $\alpha = 5\%$. This rate reflects the size of the test when the null hypothesis holds, and reflects the power otherwise. We also compute the average receiver operating characteristic (ROC) curves, aggregated over 500 replications, when the significance level $\alpha$ varies. Figure \ref{figAB} reports the results when $d=50$. The results for $d=100$ and $d=150$ show a similar qualitative pattern, and are reported in Section \ref{sec:add-numerical} of the appendix. We make a few observations. First, our test achieves a valid size under the null hypothesis. The empirical rejection rate is close to or below the nominal level for most cases. When the sample size $n$ is small, our test has a few inflated type-I errors. As $n$ increases, all the rejection rates are below the nominal level. By contrast, the test of \cite{LHZ2018} still has a good number of inflated type-I errors even when $n$ is large.  Such inflated errors may be due to the fact that MIDA relies on the selection consistency of the estimated DAG, which may not hold under the finite samples. Second, our test consistently achieves a larger empirical power over MIDA under the alternative hypothesis. This may be due to that the effects calculated by MIDA along different paths may cancel each other, leading to a decreased power. Combined with the results on the empirical size, the power of our test is not gained at the cost of the inflated Type-I errors. Moreover, the empirical power of our test increases along with the sample size, demonstrating the consistency of the test. Finally, we observe that the ROC curve of our test lies above that of MIDA in all settings as $\alpha$ varies, which clearly demonstrates the advantage of our test over MIDA.

\begin{figure}[t!]
\centering
\includegraphics[width=16cm]{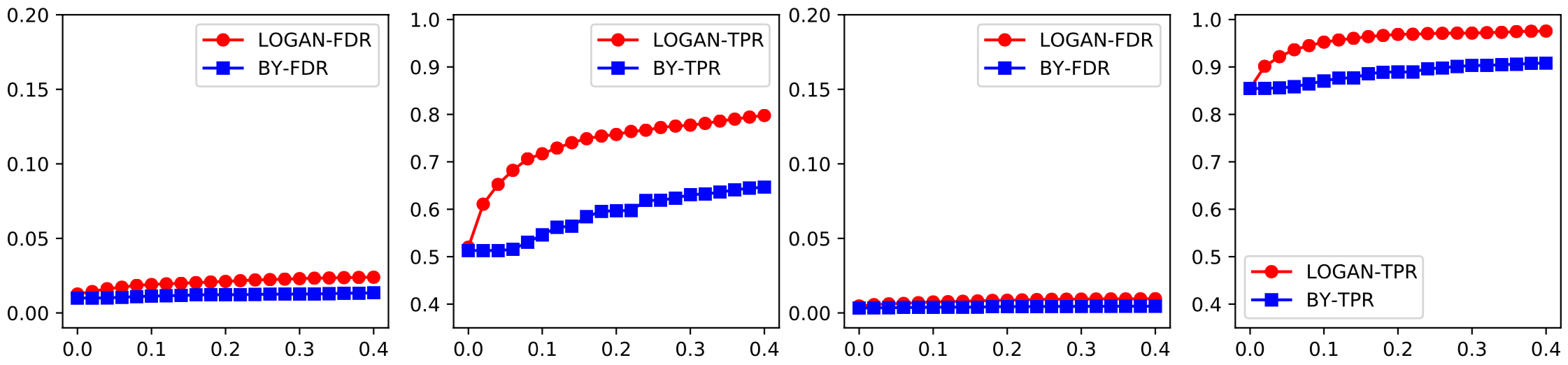}
\caption{\change{False discover rate and true positive rate of the proposed method and the  Benjamini-Yekutieli procedure when $d=50$. The horizontal axis corresponds to the significance level $\alpha$. The left two panels: $n=100$, and the right two panels: $n=200$.}}
\label{fig:fdr}
\end{figure}

We next evaluate the empirical performance of our multiple testing procedure in Algorithm \ref{alg:fdr}. We also compare it with the standard Benjamini-Yekutieli (BY) procedure. For the latter, in Step 2 of Algorithm \ref{alg:fdr}, instead of applying ScreenMin to determine the set $\mathcal{H}_0^{(\ell)}$, one simply sets $\mathcal{H}_0^{(\ell)}=\{1,2,\ldots,d\}$, i.e., the set of all mediators. We evaluate each testing procedure by the false discovery rate (FDR) and the true positive rate (TPR), over 500 data replications. Figure \ref{fig:fdr} reports the results under the varying significance level $\alpha$ from 0 to 0.4 when $d= 50$. The results for $d=100$ and $d=150$ are similar, and are reported in Section \ref{sec:add-numerical} of the appendix. It is seen that both methods achieve a valid false discovery control, in that the FDRs are all below the nominal level. However, our method is more powerful than BY, as reflected by a larger TPR in all cases. 

\section{Application}
\label{sec:real-data}

In this section, we illustrate our testing method with an application to a neuroimaging study of Alzheimer's disease (AD). AD is an irreversible neurodegenerative disorder, and is characterized by progressive impairment of cognitive and memory functions. It is the leading form of dementia, and the sixth leading cause of death in the U.S \citep{AD2020}.  The data we analyze is part of the ongoing Berkeley Aging Cohort Study. It consists of 698 participants aging between 55.3 and 94.1 years old. For each participant, the well established PACC composite score was recorded, which combines tests that assess episodic memory, timed executive function, and global cognition  \citep{Donohue2014}. Moreover, for each participant, a 1.5T structural magnetic resonance imaging (MRI) scan and a positron emission tomography (PET) scan using 18-F florbetaben tracer were acquired. All imaging data were preprocessed following the established protocols. Particularly, for MRI, all T1 images were bias-corrected, segmented, then warped and normalized to a common template space. Then the volumes were examined quantitatively by a cortical surface-based analysis and turned into cortical thickness measures. Cortical thickness is an important biomarker that reflects AD severity. We employ the FreeSurfer brain atlas and summarize each MRI image by a 68-dimensional vector, whose entries measure cortical thickness of 68 brain regions of interest. For PET, native-space images were realigned and coregistered to each participant's MRI scan, and centiloid analysis was performed to transform the standardized uptake value ratio to centiloid units. The PET scan provides a measure of deposition of amyloid-beta, a hallmark pathological protein of AD that is  commonly found in the brains of AD and elderly subjecs. The total amount of amyloid-beta deposition was extracted from PET for each subject. There are well validated methods for thresholding the subjects based on the total deposition as amyloid positive and amyloid negative groups, which are known to behave differently in AD progression \citep{Landau2013}. For our data, 309 subjects were classified as amyloid positive, and 389 as amyloid negative. Since age is a well known risk factor for AD, in our study, we aim to understand how age mediates cortical thickness of different brain regions then the PACC score. We carry out the mediation analysis for the amyloid positive and amyloid negative groups separately.

\begin{table}[b!]
\centering
\caption{Identified significant mediators for the amyloid positive and amyloid negative groups.}
\label{tab:bac}
\begin{tabular}{cccc}\toprule
\multicolumn{1}{c}{Amyloid positive group} & \multicolumn{3}{c}{Amyloid negative group} \\ \hline
r-entorhinal & l-entorhinal & l-precuneus & l-superiortemporal  \\
                   & r-inferiorparietal & r-superiorfrontal  & r-superiortemporal \\                      
\bottomrule
\end{tabular}
\end{table}

We apply the proposed multiple testing procedure in Algorithm \ref{alg:fdr} to this data, with age as the exposure, the cortical thickness of 68 brain regions as the potential mediators, and the PACC score as the outcome. We set the FDR level at 10\%. For the amyloid positive group, we find one significant mediator, and for the amyloid negative group, we found six significant mediators. Table \ref{tab:bac} reports the results. These findings agree well with the neuroscience literature. In particular, the entorhinal cortex functions as a hub in a widespread network for memory, navigation and the perception of time. It is found implicated in the early stages of AD, and is one of the most heavily damaged cortices in AD \citep{Hoesen1991}. The precuneus is involved with episodic memory, visuospatial processing, reflections upon self, and aspects of consciousness, and is found to be an AD-signature region \citep{Bakkour2013}. Moreover, the superior temporal gyrus is involved in auditory processing, and also has been implicated as a critical structure in social cognition. The superior frontal gyrus is involved in self-awareness, and the inferior parietal lobule is involved in the perception of emotions. Numerous studies have found involvement of these brain regions in the development of AD \citep{Du2007, Bakkour2013}.

\section{Discussion}
\label{sec:discussion}

In this article, we have primarily focused on the case when there is only a single DAG associated with our model. Now, we briefly discuss the extension to the case when there is an equivalence class of DAGs. Specifically, when the error variances $\sigma_i^2$, $i=0, \ldots, d+1$, in (A3) are not all equal, there exist an equivalence class of DAGs, denoted by $\mathcal{G}$, that could generate the same joint distribution of the variables. Such a class can be uniquely represented by a completed partially directed acyclic graph. For each DAG $\mathbb{G} \in \mathcal{G}$, we define $\omega_{\zeta}(\mathbb{G})$ as the total effect of $E$ on $Y$ attributed to a given path $\zeta$ following \eqref{eqn:total-effect}. Then, our hypotheses of interest become, 
\begin{align} \label{eqn:hypo-equiv}
\begin{split}
& H_0(q): \omega_{\zeta}(\mathbb{G})=0, \;\; \textrm{ for all } \zeta \textrm{ that passes through }M_{q} \;\; \textrm{and all $\mathbb{G}\in \mathcal{G}$} \;\; \quad \textrm{versus} \\
& H_1(q): \omega_{\zeta}(\mathbb{G})\neq 0, \;\; \textrm{ for some } \zeta \textrm{ that passes through }M_{q} \;\; \textrm{and some $\mathbb{G}\in \mathcal{G}$}.
\end{split}
\end{align}
To test \eqref{eqn:hypo-equiv}, we begin by estimating the equivalence class $\mathcal{G}$ based on each half of the dataset. This can be done by applying the structural learning algorithm such as \citet{chickering2003}. Let $\widehat{\mathcal{G}}$ denote the resulting estimator. For each $\widehat{\mathbb{G}} \in \widehat{\mathcal{G}}$, we employ the procedure in Section \ref{sec:test-proc} to construct a test statistic. We then take the supremum of these test statistics over all $\widehat{\mathbb{G}}$, and obtain its critical value via bootstrap. 

Finally, we comment that our proposed testing procedures can be extended to more scenarios, e.g., when there are sequentially ordered multiple sets of mediators, or when there are multiple exposure variables. We can also speed up the computation of the Boolean matrices using some transition closure algorithm \citep{chakradhar1993transitive} when the dimension of the DAG is large. We leave those pursuits as our future research. 

\section*{Acknowledgements}
Li's research was partially supported by NIH grants R01AG061303, R01AG062542, and R01AG034570. 
The authors thank the AE, and the reviewers for their constructive comments, which have led to a significant improvement of the earlier version of
this article.

\appendix
\section*{Appendix}
We first outline a multi-split version of our test. We then introduce a DAG learning procedure under a weaker constant variance condition. We next present some additional regularity conditions, followed by two supporting lemmas, then the proofs of the main theoretical results in the paper. Finally, we present some additional numerical results. 

We employ the following notation. For any sequence $\{a_n:n\ge 1\}$, $a_n=O(1)$ means $|a_n|\le C$ for some constant $C>0$, and $a_n=o(1)$ means $\lim_n a_n=o(1)$. For a sequence of random variables $\{Z_n:n\ge 1\}$, $Z_n=O_p(1)$ means, for any sufficiently small $\varepsilon>0$, there exists some constant $M>0$ such that $\prob(|Z_n|\le M)\ge 1-\varepsilon$, and $Z_n=o_p(1)$ means $\{Z_n:n\ge 1\}$ converges in probability to zero. Without loss of generality, we assume $\bm{\mu}_0=0$. 
To simplify the presentation, we only consider the case where $\widehat{\bm{\mu}}=0$ and hence $\widetilde{\bm{X}}_i=\bm{X}_i$ for $i=1,\ldots,n$. In the case where $\widehat{\bm{\mu}} \neq 0$, the theories can be similarly proved.

\begin{figure}[!t]
	\centering
	\includegraphics[width=14.5cm]{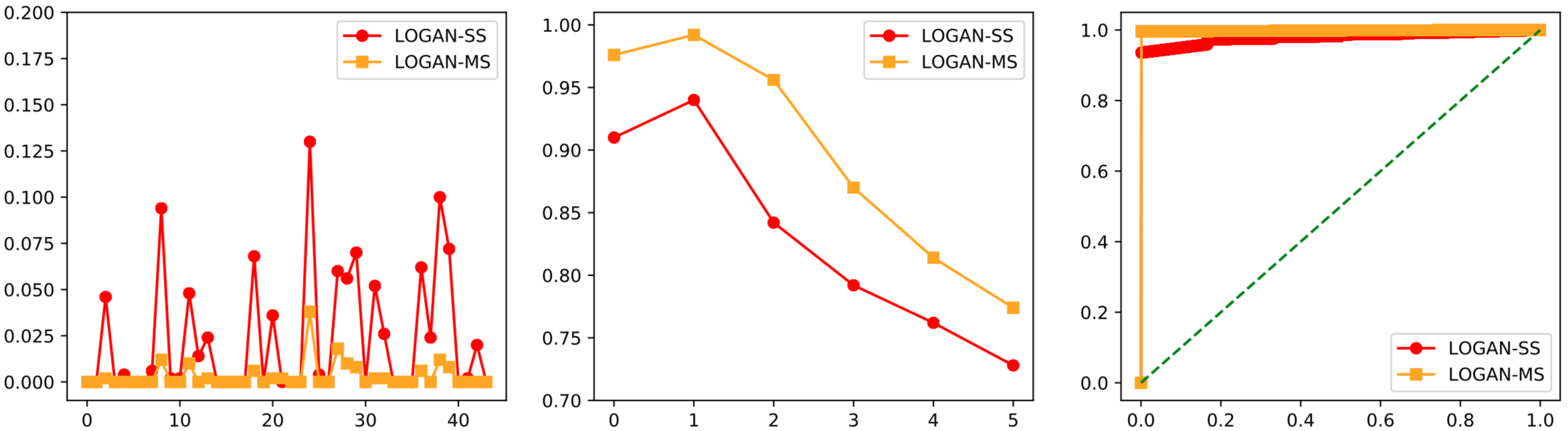}
	\includegraphics[width=14.5cm]{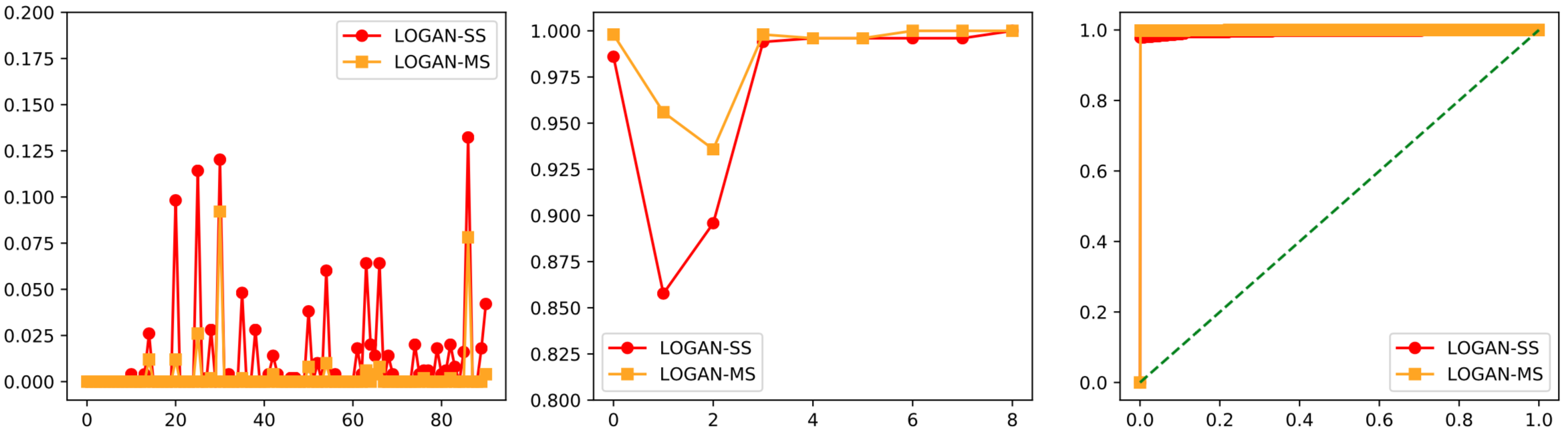}
	\includegraphics[width=14.5cm]{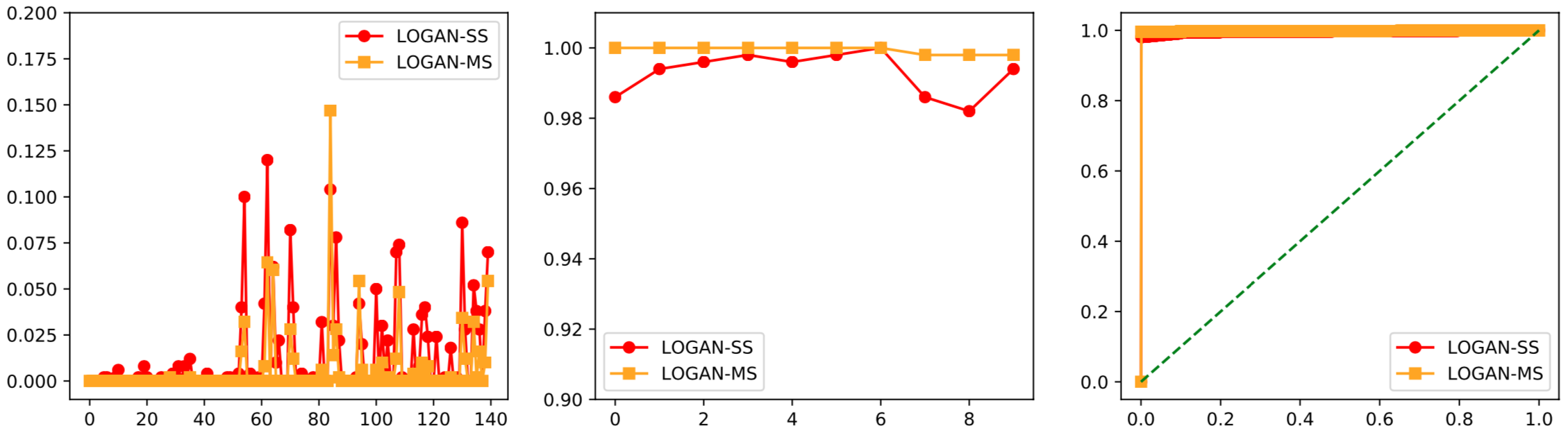}
	\caption{Empirical rejection rate of the single-split (LOGAN-SS) and the multi-split (LOGAN-MS) method. The upper panels: $(d,n)=(50,100)$, the middle panels: $(d,n)=(100,250)$, and the bottom panels: $(d,n)=(150,250)$. The left panels: under $H_0$, the middle panels: under $H_1$, where the horizontal axis is the mediator index, and the right panels: the average ROC curve.}
	\vspace{-0.1in}
	\label{fig:multisplit}
\end{figure}

{\color{black}
	\subsection{A multi-split version of the test}
	\label{sec:multisplit}
	
	We first develop a version of our individual mediator test based on multiple binary splits. This helps improve the power when the sample size is limited, and also helps mitigate the randomization arising from a single binary split. The main idea is to apply the single-split method in Algorithm \ref{alg:full} multiple times, then combine the $p$-values from all splits. Specifically, we carry out the binary split $S$ times. For the $s$th binary split, we divide $\{1,\ldots,n\}$ into two disjoint subsets $\mathcal{I}_{s,1}\cup \mathcal{I}_{s,2}$ of equal sizes. We then apply Algorithm \ref{alg:full}  to compute the $p$-values for $H_0(0,q)$ and $H_0(q,d+1)$ for each half of the data. Denote the obtained $p$-values by $\widehat{p}^{(s,1)}(0,q)$, $\widehat{p}^{(s,1)}(q,d+1)$, $\widehat{p}^{(s,2)}(0,q)$ and $\widehat{p}^{(s,2)}(q,d+1)$, respectively. We next combine these $p$-values following the idea of \cite{Mein2009}, by defining
	\vspace{-0.05in}
	\begin{eqnarray*}
		\widehat{p}(0,q) & = & \min\Big(1,q_{\gamma}\left[ \left\{\gamma^{-1}\widehat{p}^{(s,\ell)}(0,q), s=1,\ldots,S, \ell=1,2 \right\} \right] \Big), \\
		\widehat{p}(q,d+1) & = & \min\Big(1,q_{\gamma}\left[ \{\gamma^{-1}\widehat{p}^{(s,\ell)}(q,d+1), s=1,\ldots,S, \ell=1,2\} \right] \Big),
	\end{eqnarray*}
	where $\gamma$ is some constant between $0$ and $1$, and $q_{\gamma}$ is the empirical $\gamma$-quantile. In our simulations, we have experimented with a range of values of $\gamma$ between $0.1$ and $0.2$, and the results are similar, and thus we set $\gamma=0.15$ in our implementation. The corresponding $p$-value for $H_0(q)$ is given by $\max\{\widehat{p}(0,q), \widehat{p}(q,d+1)\}$, following the union-intersection principle. 
	
	We apply this multi-split method to the simulation examples in Section \ref{sec:simulations}, and compare with the single-split method. Figure \ref{fig:multisplit} shows the empirical rejection rates of the two methods. It is seen that the multi-split method in general improves over the single-split method, by achieving smaller type-I errors and larger powers.} 

{\color{black}
	\subsection{Learning DAG under a weaker constant variance condition}
	\label{sec:weakerA3}
	
	We note that the constant variance condition in (A3) can be relaxed as follows. 
	
	\vspace{-0.05in}
	\begin{enumerate}[({A}1$^*$)]
		\setcounter{enumi}{2}
		\item The errors $\varepsilon_{i}$, $i=0,1, \ldots, d+1$, are jointly normally distributed and independent. In addition, the error variances $\sigma_i^2 = \Var(\varepsilon_i)$, $i=1, \ldots, d$, are constant; i.e., $\sigma_1^2=\cdots=\sigma_{d}^2=\sigma_*^2$ for some constant $\sigma_*>0$.
	\end{enumerate}
	\vspace{-0.05in}
	
	\noindent
	In other words, the constant variance requirement does not have to be imposed on the exposure and outcome variables. Under this weaker requirement, we first need to ensure $\bm{W}_0$ remains identifiable, since \citet[Theorem 1]{Buhlmann2014} is no longer directly applicable. In addition, we need to modify the initial estimator of $\bm{W}_0$ accordingly. 
	
	The next lemma shows that $\bm{W}_0$ remains identifiable under (A3$^*$). 
	\begin{lemma}\label{lemma:iden}
		Suppose (A1), (A2) and (A3$^*$) hold. Then $\bm{W}_0$ is identifiable from the joint distribution function of $\bm{X}$. 
	\end{lemma}
	
	\noindent
	\textbf{Proof}: By (A2), we decompose $\bm{W}_0$ as,
	\begin{eqnarray}\label{eqn:W0}
	\bm{W}_0=
	\left(
	\begin{array}{ccc}
	0 & \bm{0}_d^\top & 0\\
	\bm{W}_{0,1} & \bm{W}_{1,1} & \bm{0}_d \\
	W_{0,2} & \bm{W}_{1,2}^\top & 0
	\end{array}
	\right),
	\end{eqnarray}
	where $W_{0,2}\in \mathbb{R}$, $\bm{W}_{0,1},\bm{W}_{1,2}\in \mathbb{R}^d$, $\bm{W}_{1,1}\in \mathbb{R}^{d\times d}$, and the matrix $\bm{W}_{1,1}$ is acyclic under (A1). 
	
	We first note that $W_{0,2}$ and $\bm{W}_{1,2}$ correspond to the regression coefficients of $(E,\bm{M})^\top$ on $Y$. Under the given model, the covariance matrix of $(E,\bm{M}^\top)$ is non-degenerated. As such, $W_{0,2}$ and $\bm{W}_{1,2}$ are uniquely determined by the distribution function of $\bm{X}$. 
	
	We next show that $\bm{W}_{1,1}$ is also uniquely determined by the distribution function of $\bm{X}$. For each $j=1,\ldots,d$, let $\widetilde{M}_j$ denote the population residual adjusted by the exposure, i.e., $\widetilde{M}_j=(M_j-\mu_{0,j})-\Corr(M_j,E)(E-\mu_{0,0})$. It follows that the set of residuals $\widetilde{\bm{M}}$ satisfy that $\widetilde{\bm{M}}=\bm{W}_{1,1}\widetilde{\bm{M}}+(\varepsilon_1,\ldots,\varepsilon_d)^\top$. By (A1), $\bm{W}_{1,1}$ is acyclic. As such, $\widetilde{\bm{M}}$ forms a structural linear equation with the coefficient matrix $\bm{W}_{1,1}$. Under (A3*), all the residuals have the constant variance. It then follows from Theorem 1 of \cite{Buhlmann2014} that $\bm{W}_{1,1}$ is identifiable. 
	
	Finally, we note that $\bm{W}_{0,1}$ satisfies $\bm{W}_{0,1}=\Cov(E,\bm{M}-\bm{W}_{1,1}\bm{M})/\Var(E)$. It follows from the identifiability of $\bm{W}_{1,1}$ that $\bm{W}_{0,1}$ is identifiable as well. 
	
	This completes the proof. 
	\eop
	\bigskip

	We next outline a modified initial DAG estimation procedure under the new constant variance condition (A3$^*$). Similar to Proposition \ref{prop:oracle}, we can show that this new estimator satisfies the oracle inequality as well.
	
	Following the decomposition of $\bm{W}_0$ in \eqref{eqn:W0}, we first estimate $\bm{W}_{0,2}$ and $\bm{W}_{1,2}$ using penalized regressions of $(E,\bm{M})^\top$ on $Y$ such as MCP, LASSO, SCAD, and Dantzig selector. Denote the corresponding estimators as $\widetilde{W}^{(\ell)}_{0,2}$ and $\widetilde{\bm{W}}^{(\ell)}_{1,2}$.
	
	We next estimate $\bm{W}_{1,1}$, by first regressing $\bm{M}_i$ on $E_i$, $i=1,\ldots,n$, to obtain the estimated residual $\widehat{\widetilde{\bm{M}}}_i$, then employing the method of \citet{zheng2018dags} to solve
	\begin{align*} 
	\begin{split}
	\widetilde{\bm{W}}^{(\ell)}_{1,1} = \argmin_{\bm{W}\in \mathbb{R}^{d\times d}} \sum_{i\in \mathcal{I}_{\ell}}\|\widehat{\widetilde{\bm{M}}}_i-\bm{W} \widehat{\widetilde{\bm{M}}}_i\|_2^2  + \lambda |\mathcal{I}_{\ell}| \sum_{i,j}|W_{i,j}| 
	\textrm{ subject to } \textrm{trace}\{\exp(\bm{W}\circ \bm{W})\} = d.
	\end{split}
	\end{align*}
	
	After obtaining $\widetilde{\bm{W}}^{(\ell)}_{1,1}$, we set $\widetilde{\bm{W}}_{0,1}^{(\ell)}=\widetilde{\bm{W}}^{(\ell)}_{1,1} \widehat{\Cov}(\bm{M},E)\widehat{\Var}^{-1}(E)$, where $\widehat{\Cov}(\bm{M},E)$ is the sampling covariance estimator of $\bm{M}$, and $\widehat{\Var}(E)$ is the sampling variance estimator of $E$. 
	
	Finally, we put together $\widetilde{\bm{W}}_{0,1}^{(\ell)}$, $\widetilde{\bm{W}}_{1,1}^{(\ell)}$, $\widetilde{W}_{0,2}^{(\ell)}$ and $\widetilde{\bm{W}}_{1,2}^{(\ell)}$ according to \eqref{eqn:W0} to form the modified initial estimator $\widetilde{\bm{W}}^{(\ell)}$ for $\bm{W}_0$. 
}

\subsection{Additional regularity conditions}
\label{sec:conditions}

\begin{figure}[t!]
	\centering
	\includegraphics[width=3.25cm]{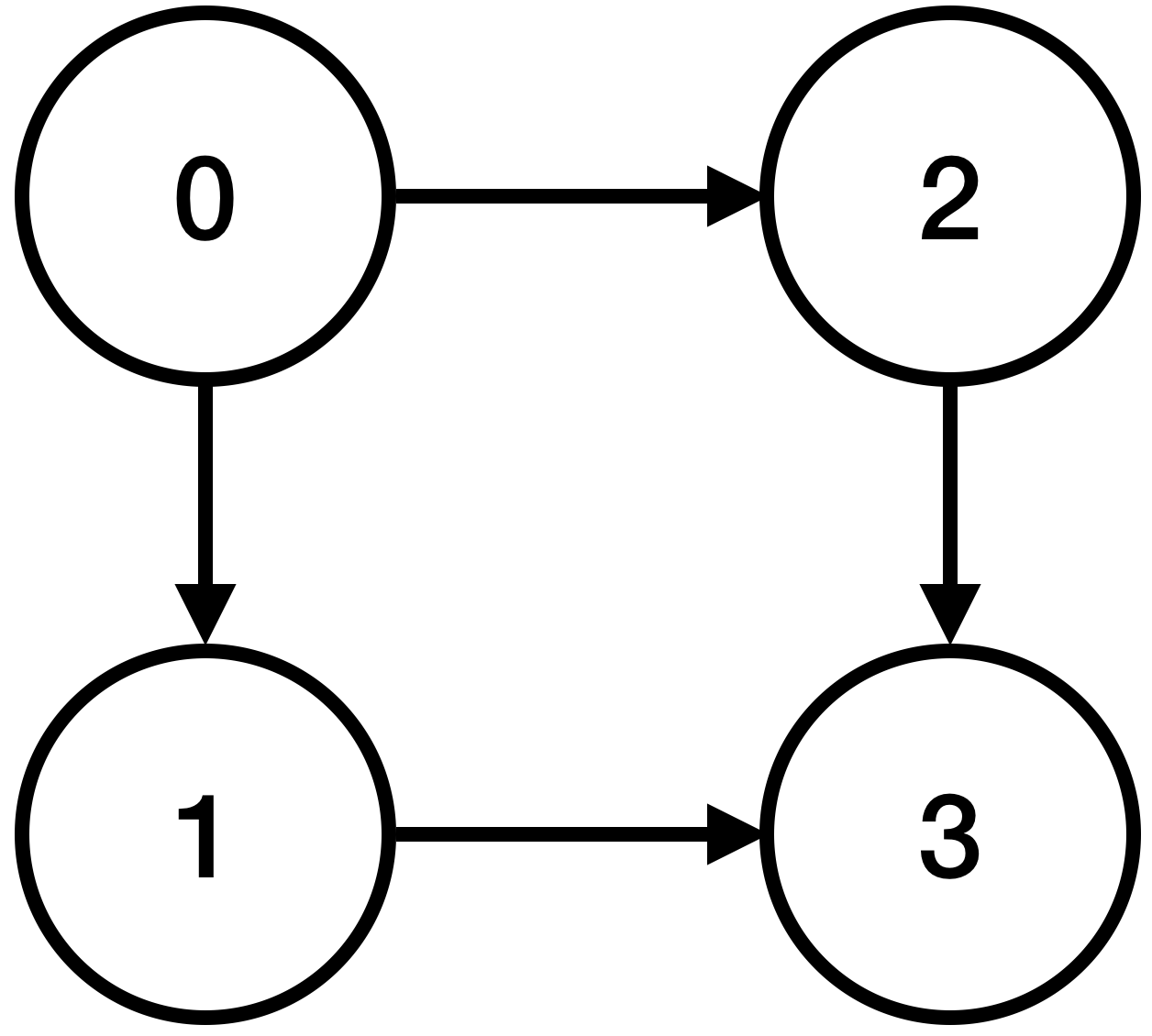} 
	\caption{An illustrative DAG with four nodes, where both $(0,1,2,3)$ and $(0,2,1,3)$ correspond to the true orderings.}
	\label{DAG-illustration}
\end{figure}

We introduce two additional regularity conditions for the theoretical guarantees of the proposed test.  We begin with some notation. Define the limit of the estimator $\widehat{\bm{\beta}}^{(\ell)}(j_1,j_2)$ 
\begin{eqnarray*}
	\bm{\beta}^{(\ell)}_0(j_1,j_2) = \argmin_{\substack{\bm{\beta}:\beta_{j_2}=0, \footnotesize{\hbox{supp}}(\bm{\beta})\in \hbox{ACT}(j_1,\widetilde{\bm{W}}^{(\ell)}) } } \Mean \left( X_{j_2}-\bm{\beta}^\top \bm{X}_i \right)^2.
\end{eqnarray*}
Any permutation $\pi=(\pi_0,\pi_1,\ldots,\pi_d,\pi_{d+1})^\top$ of $\{0,1,\ldots,d,d+1\}$ determines an order of the mediators $\{M_{j}\}_{1\le j\le d}$. Define
\begin{eqnarray*}
	\bm{W}_{\pi_j}(\pi) = \argmin_{\bm{\beta}:\hbox{\footnotesize{supp}}(\bm{\beta}) \in \{\pi_0,\pi_1,\ldots,\pi_{j-1},\pi_j\} } \Mean \left( M_{\pi_j}-\bm{\beta}^\top \bm{M} \right)^2, \quad \textrm{ for } \; j = 1, \ldots, d.
\end{eqnarray*}
Let $\bm{W}(\pi) = \left\{ \bm{W}_{0}(\pi),\bm{W}_{1}(\pi),\ldots,\bm{W}_{d+1}(\pi) \right\}^\top$. It corresponds to the coefficient matrix obtained by doing a Gram-Schmidt orthogonalization, starting with $X_{\pi_0}$, and finishing by projecting $X_{\pi_{d+1}}$ on $X_{\pi_0},X_{\pi_1},\ldots,X_{\pi_d}$. Let $\bm{\Omega}(\pi)$ be a diagonal matrix where the diagonal elements $\omega_0^2(\pi),\omega_1^2(\pi),\ldots,\omega_{d+1}^2(\pi)$ correspond to the error variances, $\Var\{X_{0}-\bm{X}^\top \bm{W}_{0}(\pi)\}, \Var\{X_{1}-\bm{X}^\top \bm{W}_{1}(\pi)\}, \ldots, \Var\{X_{d+1}-\bm{X}^\top \bm{W}_{d+1}(\pi)\}$, respectively. Let $\Pi^*$ denote the set consisting of all true orderings $\pi^*$ such that $\bm{W}(\pi^*)=\bm{W}_0$. Note that $\pi^*$ may not be unique. As an illustration, consider the DAG in Figure \ref{DAG-illustration}, where both $(0,1,2,3)$ and $(0,2,1,3)$ correspond to the true orderings of the four nodes.

We impose the following additional regularity conditions. In particular, (A5) is required to establish the consistency and FDR control of the proposed test, while (A6) is to establish the oracle inequality for the initial estimator. 

\vspace{-0.05in}
\begin{enumerate}[({A}1)]
	\setcounter{enumi}{4}
	
	\item There exist some constants $\kappa_3, \kappa_4, \kappa_5,\kappa_6>0$, with $\kappa_4+\kappa_5>1/2$, such that, $\|\widehat{\bm{\beta}}^{(\ell)}(j_1,j_2)-\bm{\beta}_0^{(\ell)}(j_1,j_2)\|_2\le \kappa_3 n^{-\kappa_4}$, $\|\overline{\bm{W}}_{j_1}^{(\ell)}-\bm{W}_{0,j_1}\|_2\le \kappa_3 n^{-\kappa_5}$ and $\|\widehat{\bm{\beta}}^{(\ell)}(j_1,j_2)-\bm{\beta}_0^{(\ell)}(j_1,j_2)\|_1\le \kappa_3 n^{-\kappa_6}$, $\|\overline{\bm{W}}_{j_1}^{(\ell)}-\bm{W}_{0,j_1}\|_1\le \kappa_3 n^{-\kappa_6}$, for any $0 \le j_1,j_2 \le d+1, \ell=1,2$.
	
	\item There exists a constant $\omega>0$ such that for all $\pi \notin \Pi^*$, $\frac{1}{d}\sum_{j=0}^{d+1} \left\{ \omega_j^2(\pi) -\sigma_*^2 \right\}^2 >\omega$.
	
\end{enumerate}
\vspace{-0.05in}

Condition (A5) is mild. This is because, when (A4) holds and $\bm{\beta}_0^{(\ell)}$ is estimated via the MCP, LASSO, SCAD, or Dantzig selector, we have $\max_{j_1,j_2,\ell} \|\widehat{\bm{\beta}}^{(\ell)}(j_1,j_2)-\bm{\beta}_0^{(\ell)}(j_1,j_2)\|_2\le O(1) n^{-1/2} \sqrt{s^*\log n}$ and $\max_{j_1,j_2,\ell} \|\widehat{\bm{\beta}}^{(\ell)}(j_1,j_2)-\bm{\beta}_0^{(\ell)}(j_1,j_2)\|_1\le O(1) n^{-1/2} s^*\sqrt{\log n}$, with probability approaching one, where $s^* = \max_{j_1,j_2,\ell} |\mathcal{M}^{(\ell)}(j_1,j_2)|$ denotes the maximum sparsity size, $\mathcal{M}^{(\ell)}(j_1,j_2)=\hbox{supp}\left\{\bm{\beta}_0^{(\ell)}(j_1,j_2)\right\}$, and $O(1)$ denotes some positive constant. Similarly, we have $\max_{j,\ell} \|\overline{\bm{W}}^{(\ell)}_j-\bm{W}_{0,j}\|_2\le O(1) n^{-1/2} \sqrt{s_0\log n}$ and $\max_{j,\ell} \|\overline{\bm{W}}^{(\ell)}_j-\bm{W}_{0,j}\|_1\le O(1) n^{-1/2} s_0\sqrt{\log n}$, with probability approaching one. Therefore, Condition (A5) holds as long as $s_0,s^* =O(n^{\kappa_7})$ for some $\kappa_7<1/2$. 

Condition (A6) is referred to as the ``omega-min" condition in \citet{van2013}. It essentially guarantees that the true ordering of the mediators can be consistently estimated, which is needed to establish the oracle inequality for the estimator from \eqref{eqn:notears-l1} of \citet{zheng2018dags}. When the number of mediators $d$ is fixed and the error variances are equal as in (A3), this condition automatically holds. 

\change{Finally, we make some remark on why the usual debiasing strategy may not be directly applicable to relax the regularity condition (A4) in Section \ref{sec:consistency} of the paper. Specifically, if (A4) does not hold and the true ordering cannot not be recovered, then no matter whether we debias the estimated coefficient matrix or not, the resulting estimator for $\bm{W}_0$ may not be consistent. We illustrate with a simple example. Consider a DAG with two variables, where $X_1=\varepsilon_1$, $X_2=aX_1+\varepsilon_2$ for some $a\neq 0$, and $\varepsilon_1$ and $\varepsilon_2$ are independent mean-zero random errors. Then the corresponding coefficient matrix is
	\begin{eqnarray*}
		\bm{W}_0 = 
		\left(\begin{array}{cc}
			0 & 0 \\
			a & 0
		\end{array}\right).
	\end{eqnarray*}
	Meanwhile, we note that the linear structure equation can be rewritten as $X_2=\varepsilon^*_2$ and $X_1=a(a^2+1)^{-1} X_2+\varepsilon^*_1$, for some mean-zero random errors $\varepsilon^*_1$ and $\varepsilon^*_2$. In addition, with some calculation, we have that
	\vspace{-0.05in}
	\begin{eqnarray*}
		\varepsilon^*_2&=&a\varepsilon_1+\varepsilon_2,\\
		\varepsilon^*_1&=&\frac{1}{a^2+1}\varepsilon_1-\frac{a}{a^2+1}\varepsilon_2,
	\end{eqnarray*}
	and thus $\varepsilon^*_1$ and $\varepsilon^*_2$ are independent. Then the corresponding coefficient matrix $\bm{W}_0^*$ becomes 
	\begin{eqnarray*}
		\left(\begin{array}{cc}
			0 & a(a^2+1)^{-1} \\
			0 & 0
		\end{array}\right).
	\end{eqnarray*}
	If the ordering is not correctly specified and (A4) does not hold, then debiasing or not, the second element on the first row will be close to $a(a^2+1)^{-1}$, rather than the true value $0$. As such, the debiased estimator is not consistent when (A4) is violated. This simple example reflects the challenge of post-selection inference in our setting.}

\subsection{Supporting lemmas}
\label{sec:lemmas}

Next, we present two supporting lemmas. Lemma \ref{lemma:sigma} establishes the convergence rate of the variance estimator $\widehat{\sigma}_*^2$, whereas Lemma \ref{lemma:actdesc} is needed for the limiting distribution. 

\begin{lemma}\label{lemma:sigma}
	Suppose (A5) holds, and $\|\bm{W}_0\|_2$ is bounded. Then $|\widehat{\sigma}_*^2-\sigma_*^2|=O(n^{-\kappa_7})$ for some constant $\kappa_7$ satisfying $0<\kappa_7 \le \min(2 \kappa_5,1/2)$, with probability tending to $1$. 
\end{lemma}

\noindent
\textbf{Proof}: 
It suffices to show that, for $\ell = 1,2$ and any $\kappa_7$ satisfying that $\kappa_7\le 2\kappa_5$, $\kappa_7<1/2$, 
\begin{eqnarray}\label{prooflemma4eq1}
\frac{1}{(d+2)|\mathcal{I}_{\ell}^c|} \sum_{j=0}^{d+1}\sum_{i\in \mathcal{I}_{\ell}^c} |X_{i,j}-\overline{\bm{W}}^{(\ell)\top}_j \bm{X}_i|_2^2-\sigma_*^2=O(n^{-\kappa_7}),
\end{eqnarray}
with probability tending to $1$. In turn, it suffices to show that, 
\begin{eqnarray}\label{prooflemma4eq2}
\prob\left(\bigg| \frac{1}{(d+2)|\mathcal{I}_{\ell}^c|} \sum_{i\in \mathcal{I}_{\ell}^c} \sum_{j=0}^{d+1} |X_{i,j}-\overline{\bm{W}}^{(\ell)\top}_j \bm{X}_i|_2^2-\sigma_*^2 \bigg| > 2\kappa_3^2n^{-\kappa_7} \; \bigg | \; \overline{\bm{W}}^{(\ell)}\right) = o_p(1).
\end{eqnarray} 
This is because, if \eqref{prooflemma4eq2} holds, by bounded convergence theorem, we have
\begin{eqnarray*}
	\prob\left(\bigg| \frac{1}{(d+2)|\mathcal{I}_{\ell}^c|} \sum_{i\in \mathcal{I}_{\ell}^c} \sum_{j=0}^{d+1} |X_{i,j}-\overline{\bm{W}}^{(\ell)\top}_j \bm{X}_i|_2^2-\sigma_*^2 \bigg| > 2\kappa_3^2n^{-\kappa_7}\right) = o(1), 
\end{eqnarray*}
which in turn yields \eqref{prooflemma4eq1}. 

For the conditional mean of the left-hand-side of \eqref{prooflemma4eq1} given $\{\bm{X}_i:i\in \mathcal{I}_{\ell}\}$,
\begin{eqnarray*}
	&&\frac{1}{(d+2)}\Mean \left( \sum_{j=0}^{d+1} |X_j-\overline{\bm{W}}^{(\ell)\top}_j \bm{X}|_2^2 \; | \; \overline{\bm{W}}^{(\ell)} \right) - \sigma_*^2 \\&=&  \frac{1}{(d+2)}\sum_{j=0}^{d+1}\Mean \|X_j-\bm{W}_{0,j}^\top \bm{X}\|_2^2-\sigma_*^2+\frac{1}{(d+2)} \|\bm{W}_0-\overline{\bm{W}}^{(\ell)}\|_2^2 \\
	&=& \frac{1}{(d+2)} \|\bm{W}_0-\overline{\bm{W}}^{(\ell)}\|_2^2=\frac{1}{(d+2)}\sum_{j=0}^{d+1} \|\bm{W}_{0,j}-\overline{\bm{W}}_{j}^{(\ell)}\|_2^2 
	\; \le \; \kappa_3^2 n^{-2\kappa_5} \; \le \; \kappa_3^2 n^{-\kappa_7},
\end{eqnarray*} 
where the first equality is due to the fact that $\Mean \bm{X}=0$ and the second-to-last inequality is due to Condition (A5). The event defined in \eqref{prooflemma4eq2} occurs only when 
\begin{eqnarray*}
	\bigg| \frac{1}{(d+2)|\mathcal{I}_{\ell}^c|} \sum_{i\in \mathcal{I}_{\ell}^c} \sum_{j=0}^{d+1} |X_{i,j}-\overline{\bm{W}}^{(\ell)\top}_j \bm{X}_i|_2^2-\frac{1}{(d+2)}\Mean \left( \sum_{j=0}^{d+1} |X_j-\overline{\bm{W}}^{(\ell)\top}_j \bm{X}|_2^2 \; | \; \overline{\bm{W}}^{(\ell)} \right) \bigg| > \kappa_3^2n^{-\kappa_7}.
\end{eqnarray*}
Thus, to prove \eqref{prooflemma4eq2}, it suffices to show that 
\begin{align*}
\prob\Bigg( \bigg| \frac{1}{(d+2)|\mathcal{I}_{\ell}^c|} \sum_{i\in \mathcal{I}_{\ell}^c} \sum_{j=0}^{d+1} |X_{i,j}-\overline{\bm{W}}^{(\ell)\top}_j \bm{X}_i|_2^2-\frac{1}{(d+2)}\Mean \left(\sum_{j=0}^{d+1}|X_j-\overline{\bm{W}}^{(\ell)\top}_j \bm{X}|_2^2 \; | \; \overline{\bm{W}}^{(\ell)} \right) \bigg| \\
> \kappa_3^2n^{-\kappa_7}|\overline{\bm{W}}^{(\ell)} \Bigg) = o_p(1).
\end{align*} 
By Chebyshev's inequality, this probability is bounded from above by
\begin{eqnarray*}
	\frac{n^{{2\kappa_7-1}}}{2\kappa_3^4(d+2)^2} \Var\left( \sum_{j=0}^{d+1}|X_j-\overline{\bm{W}}^{(\ell)\top}_j \bm{X}|^2 \; | \; \overline{\bm{W}}^{(\ell)} \right) \le \frac{n^{{2\kappa_7-1}}}{\kappa_3^4(d+2)^2} \Mean\left( \left\{ \sum_{j=0}^{d+1}\|X_j-\overline{\bm{W}}^{(\ell)\top}_j \bm{X}|_2^2 \right\}^2 \; \bigg| \; \overline{\bm{W}}^{(\ell)} \right).
\end{eqnarray*}
Since $\kappa_7<1/2$, it suffices to show that, with probability approaching one, 
\begin{eqnarray}\label{meanX0W}
\frac{1}{(d+2)^2} \Mean\left( \left\{ \sum_{j=0}^{d+1} |X_j-\overline{\bm{W}}^{(\ell)\top}_j \bm{X}|_2^2 \right\}^2 \; \bigg | \; \overline{\bm{W}}^{(\ell)} \right) = O(1),
\end{eqnarray}
By Cauchy-Schwarz inequality, the left-hand-side of \eqref{meanX0W} is bounded from above by
\vspace{-0.1in}
\begin{eqnarray}\label{meanX0W1}
\begin{split}
\frac{1}{d+2} \sum_{j=0}^{d+1} \Mean \left\{ |X_{j}-\overline{\bm{W}}^{(\ell)\top}_{j} \bm{X}|^4\; | \; \overline{\bm{W}}^{(\ell)} \right\}
& \le  \\
\frac{16}{d+2} \sum_{j=0}^{d+1} & \left( \Mean \left\{ |X_{j}|^4\; | \; \overline{\bm{W}}^{(\ell)} \right\} + \Mean \left\{ |\overline{\bm{W}}^{(\ell)\top}_{j} \bm{X}|^4\; | \; \overline{\bm{W}}^{(\ell)} \right\} \right).
\end{split}
\end{eqnarray}

For any random variable $Z$ and any constant $\kappa > 0$, by Taylor's theorem, we have
\begin{eqnarray*}
	\Mean \left\{ \exp\left(\frac{Z^2}{\kappa^2}\right) \right\} = \sum_{k=0}^{+\infty} \frac{\Mean \left(Z^{2k}\right)}{\kappa^{2k} k!}.
\end{eqnarray*}
It follows that $\Mean \{ Z^4/(2\kappa^4) \} \le \Mean \{ \exp(Z^2/\kappa^2) \} -1$. Let $\|Z\|_{\psi_p}$ denote its Orlicz norm, i.e., 
\begin{eqnarray*}
	\|Z\|_{\psi_p}=\inf \left[C>0: \Mean \left\{ \exp\left(\frac{|Z|^p}{|C|^p}\le 2\right) \right\} \right].
\end{eqnarray*}
By definition, we have $\Mean \{Z^4/(2\|Z\|_{\psi_2}^4) \} \le 1$, and henceforth, $\Mean (Z^4) \le 2\|Z\|_{\psi_2}^4$. 

Under our model assumptions, the covariance matrix $\bm{\Sigma}_0=\Cov(\bm{X})$ is given by $ \sigma_*^2(\bm{I}_{d+2}-\bm{W}_{0})^{-1} \{(\bm{I}_{d+2}-\bm{W}_{0})^{-1 }\}^\top$, where $\bm{I}_{d+2}$ is a $(d+2)\times (d+2)$ identity matrix. For any $\bm{a}\in \mathbb{R}^{d+2}$, by Cauchy-Schwarz inequality, 
\begin{eqnarray*}
	\bm{a}^\top \bm{\Sigma}_0^{-1}\bm{a} \le \kappa_* \|\bm{a}^\top (\bm{I}_{d+2}-\bm{W}_{0})\|_2^2\le \kappa_* \|\bm{a}\|_2^2 \|\bm{I}_{d+2}-\bm{W}_{0}\|_2^2\le 2\kappa_* \|\bm{a}\|_2^2\left( \|\bm{I}_{d+2}\|_2^2+\|\bm{W}_{0}\|_2^2 \right),
\end{eqnarray*}
for some constant $\kappa_*>0$. 
Since $\|\bm{W}_0\|_2$ is bounded, it implies that the maximum eigenvalues of $\bm{\Sigma}_0^{-1}$  is bounded. Thus, the minimum eigenvalue of $\bm{\Sigma}_0$ is bounded away from zero. Also, note that $(\bm{I}_{d+2}-\bm{W}_{0})^{-1}=\bm{I}_{d+2}+\bm{W}_{0}$. Following similar arguments, the maximum eigenvalue of $\bm{\Sigma}_0$ is bounded as well. Since $\bm{X}$ is jointly normal with bounded $\lambda_{\max}(\Sigma_0)$, we have, for some constant $\kappa>0$ and any vector $\bm{a} \in \mathbb{R}^{d+2}$, 
\vspace{-0.1in}
\begin{eqnarray}\label{orlicznorm}
\|\bm{a}^\top \bm{X}\|_{\psi_2}\le \kappa \|\bm{a}\|_2.
\end{eqnarray} 

It then follows from \eqref{orlicznorm} that conditional on $\overline{\bm{W}}^{(\ell)}$,
\begin{eqnarray*}
	\Mean\left( |X_{j}|^4 \right) \le 2\|X_{j}\|_{\psi_2}^4\le 2 \kappa^4, \; \textrm{ and } \; \Mean \left( |\overline{\bm{W}}^{(\ell)\top}_{j} \bm{X}|^4 \right) \le 2\|\overline{\bm{W}}^{(\ell)\top}_{j} \bm{X}\|_{\psi_2}^4\le 2\kappa^4\|\overline{\bm{W}}^{(\ell)}_{j}\|_2^4,
\end{eqnarray*}
for any $j = 1, \ldots, d$. It follows from \eqref{meanX0W1} that conditional on $\overline{\bm{W}}^{(\ell)}$,
\begin{eqnarray*}
	\frac{1}{(d+2)} \sum_{j=1}^{d} \Mean \left( |X_{j}-\overline{\bm{W}}^{(\ell)\top}_{j} \bm{X}|^4 \right) \le \frac{32\kappa^4}{d+2}\sum_{j=0}^{d+1} \left( 1+\|\overline{\bm{W}}^{(\ell)}_{j}\|_2^4 \right) \le 32\kappa^4 \left( 1+\max_j  \|\overline{\bm{W}}^{(\ell)}_{j}\|_2^4 \right).
\end{eqnarray*}
By (A5), we have $\max_j  \|\overline{\bm{W}}^{(\ell)}_{j}\|_2\le \max_j  \|\bm{W}_{0,j}\|_2+\max_j \|\bm{W}_{0,j}-\overline{\bm{W}}^{(\ell)}_{j}\|_2\le \max_j  \|\bm{W}_{0,j}\|_2+\kappa_3 n^{-\kappa_5}$. Since $\|\bm{W}_0\|_2$ is bounded, we have $\|\bm{e}_j^\top \bm{W}_0\|_2=O(1)$, and hence $\|\bm{W}_{0,j}\|_2=O(1)$, where $\bm{e}_j$ denotes a $(d+2)$-dimensional vector with the $j$th element equal to one and the rest equal to zero. It follows that $\sum_{j=0}^{d+1} \Mean \left\{\left. |X_{j}-\overline{\bm{W}}^{(\ell)\top}_{j} \bm{X}|^4/(d+2) \right| \overline{\bm{W}}^{(\ell)}\right\} = O(1)$, with probability approaching one. Then \eqref{meanX0W} is proven. This completes the proof. 
\eop
\medskip

\begin{lemma}\label{lemma:actdesc}
	Suppose (A1), (A4) hold. Then $\hbox{ACT}(j,\widetilde{\bm{W}}^{(\ell)})$ contains no descendants of $j$. 
\end{lemma}

\noindent
\textbf{Proof}: 
Suppose there exists some $j'\in \hbox{ACT}(j,\widetilde{\bm{W}}^{(\ell)})$, such that $j'$ is a descendant of $j$. By definition, there exists a directed path from $X_{j}$ to $X_{j'}$: $X_{j}\to X_{i_1}\to \ldots X_{i_K}\to X_{j'}$. By Condition (A4), we have $j\in \hbox{ACT}(i_1,\widetilde{\bm{W}}^{(\ell)})$, $i_k\in \hbox{ACT}(i_{k+1},\widetilde{\bm{W}}^{(\ell)})$, for $k=1,\ldots,K-1$ and $i_K\in \hbox{ACT}(j',\widetilde{\bm{W}}^{(\ell)})$. This, together with  $j'\in \hbox{ACT}(j,\widetilde{\bm{W}}^{(\ell)})$, implies that there exists a directed path from $X_{j}$ to $X_{j}$ on the DAG generated by $\widetilde{\bm{W}}^{(\ell)}$. Then the acyclic constraint of $\widetilde{\bm{W}}^{(\ell)}$ is violated. This completes the proof. 
\eop

\subsection{Proof of Lemma \ref{thm:lemma2}}

To prove Lemma \ref{thm:lemma2}, it suffices to show that 
\vspace{-0.1in}
\begin{eqnarray}\label{lemma2keyequation}
(|\bm{W}_0|^{(k)})_{q_2,q_1}=\max_{0\le j_1,\ldots,j_{k-1}\le d+1} \min\left(|W_{0,j_1,q_1}|, \min_{l\in \{1,\ldots,k-2\}} |W_{0,j_{l+1},j_l}|,|W_{0,q_2,j_{k-1}}|\right).
\end{eqnarray}
Lemma \ref{thm:lemma2} can then be similarly proven as Lemma \ref{thm:lemma1}. We use induction to prove \eqref{lemma2keyequation} for any $q_1 = 0, \ldots, d$, and $q_2 = 1, \ldots, d+1$. When $k=2$, by the definition of $\otimes$, we have $(|\bm{W}_0|^{(2)})_{q_2,q_1}=\max_{0\le j\le d+1} \min(|W_{0,j,q_1}|, |W_{0,q_2,j}|)$. Thus, \eqref{lemma2keyequation} holds with $k=2$. 

Suppose \eqref{lemma2keyequation} holds with $k=t$ for some $t \ge 2$, i.e,
\vspace{-0.1in}
\begin{eqnarray}\label{prooflemma2keyequation}
(|\bm{W}_0|^{(t)})_{q_2,q_1}=\max_{0\le j_1,\ldots,j_{t-1}\le d+1} \min\left(|W_{0,j_1,q_1}|, \min_{l\in \{1,\ldots,t-2\}} |W_{0,j_{l+1},j_l}|,|W_{0,q_2,j_{t-1}}|\right).
\end{eqnarray} 
Therefore,
\vspace{-0.1in}
\begin{eqnarray*}
	&&(|\bm{W}_0|^{(t+1)})_{q_2,q_1}=(|\bm{W}_0|^{(t)} \circ |\bm{W}_0|)_{q_2,q_1}=\max_{j\in \{0,\ldots,d+1\}} \min \{(|\bm{W}_0|^{(t)})_{j,q_1}, |W_{0,q_2,j}|\}\\
	&=&\max_{j\in \{0,\ldots,d+1\}} \min\left\{\max_{0\le j_1,\ldots,j_{t-1}\le d+1} \min\left(|W_{0,j_1,q_1}|, \min_{l\in \{1,\ldots,t-2\}} |W_{0,j_{l+1},j_l}|,|W_{0,j,j_{t-1}}|\right),  |W_{0,q_2,j}|\right\}\\
	&=&\max_{j\in \{0,\ldots,d+1\}} \max_{0\le j_1,\ldots,j_{t-1}\le d+1} \min\left(|W_{0,j_1,q_1}|, \min_{l\in \{1,\ldots,t-2\}} |W_{0,j_{l+1},j_l}|,|W_{0,j,j_{t-1}}|, |W_{0,q_2,j}|\right)\\
	&=&\max_{0\le j_1,\ldots,j_{t-1},j_t\le d+1} \min\left(|W_{0,j_1,q_1}|, \min_{l\in \{1,\ldots,t-1\}} |W_{0,j_{l+1},j_l}|, |W_{0,q_2,j_t}|\right).
\end{eqnarray*}
Thus, \eqref{prooflemma2keyequation} holds with $k=t+1$. This completes the proof of Lemma \ref{thm:lemma2}. 
\eop

\subsection{Proof of Theorem \ref{thm1}}
\label{sec:proofthm1}

\textbf{Outline of the proof}: 
Our goal is to prove
\begin{eqnarray}\label{assertion}
\prob\left\{ \max_{(j_1,j_2)\in \mathcal{S}(q_1,q_2,\widehat{\bm{B}}^{(\ell)})} \sqrt{|\mathcal{I}_{\ell}^c|} |\widehat{W}_{i,j}^{(\ell)}-W_{0,i,j}|>\widehat{c}^{(\ell)}(q_1,q_2) \right\} = \frac{\alpha}{2}+o(1),
\end{eqnarray}
for any $q_1 = 0, \ldots, d, q_2 = 1, \ldots, d+1$ and $\ell = 1,2$. Then the validity of our test follows by the union-intersection principle. We begin with an outline of our proof, which relies on the high-dimensional central limit theorem that was recently developed by \cite{Cherno2013} and \cite{Cherno2014}. We divide the proof into three steps. 

In Step 1, we show that, 
\vspace{-0.1in}
\begin{eqnarray}\label{prooffirststep}
\max_{(j_1,j_2) \in \mathcal{S}(q_1,q_2,\widehat{\bm{B}}^{(\ell)})} \left| \sqrt{|\mathcal{I}_{\ell}^c|} \left( \widehat{W}_{j_1,j_2}^{(\ell)}-W_{0,j_1,j_2} \right) - \eta_{0,j_1,j_2}^{(\ell)} \right| = o_p(\log^{-1/2} n),
\end{eqnarray}
where
\vspace{-0.15in}
\begin{eqnarray} \label{eta0j1j2}
\eta_{0,j_1,j_2}^{(\ell)}=\frac{ |\mathcal{I}_{\ell}^c|^{-1/2}\displaystyle \sum_{i\in \mathcal{I}_{\ell}^c} \left\{ X_{i,j_2}-\bm{\beta}^{(\ell)\top}_0(j_1,j_2) \bm{X}_{i } \right\} \; \varepsilon_{i,j_1} }{ \Mean \left[ X_{j_1} \left\{ X_{j_2}-\bm{\beta}_0^{(\ell)\top}(j_1,j_2) \bm{X} \right\} \; \big | \;\widetilde{\bm{W}}^{(\ell)} \right] }.
\end{eqnarray}
This further implies that
\begin{eqnarray}\label{prooffirststep1}
\underbrace{\max_{(j_1,j_2)\in \mathcal{S}(q_1,q_2,\widehat{\bm{B}}^{(\ell)})} \sqrt{|\mathcal{I}_{\ell}^c|} | \widehat{W}_{j_1,j_2}^{(\ell)}-W_{0,j_1,j_2} |}_{\widehat{S}^{(\ell)}}=\underbrace{\max_{(j_1,j_2)\in \mathcal{S}(q_1,q_2,\widehat{\bm{B}}^{(\ell)})} |\eta_{0,j_1,j_2}^{(\ell)}|}_{S_0^{(\ell)}} \; + \; o_p(\log^{-1/2} n). 
\end{eqnarray}
Let $\mathcal{M}^{(\ell)}(j_1,j_2)$ denote the support of $\bm{\beta}^{(\ell)}_0(j_1,j_2)$. 
By definition, for any $j\in \mathcal{M}^{(\ell)}(j_1,j_2)$, we have $j,j_2\in \hbox{ACT}(j_1, \widetilde{\bm{W}}^{(\ell)})$. Since $\widetilde{\bm{W}}^{(1)}$ and $\widetilde{\bm{W}}^{(2)}$ satisfy the acyclic constraint in (A1), by Lemma \ref{lemma:actdesc}, neither $j$ nor $j_2$ is a descendant of $j_1$. As a result, $X_{j_1}$ is conditionally independent of $X_{j_2}$ and $X_{j}$ given its parents. As such, the numerator of $\eta_{0,j_1,j_2}^{(\ell)}$ forms a sum of i.i.d.\ mean zero random variables.

In Step 2, we show that, 
\begin{eqnarray} \label{proofsecondstep}
\sup_{z\in \mathbb{R}} \left| \prob\left( S_0^{(\ell)}\le z|\widetilde{\bm{W}}^{(\ell)} \right) - \prob\left( \|N(0,V_0)\|_{\infty}\le z|\widetilde{\bm{W}}^{(\ell)} \right) \right| = o(1),
\end{eqnarray}
where $V_0$ is a matrix involving the covariance of $\eta_{0,j_1,j_2}^{(\ell)}$ in \eqref{eta0j1j2} and is defined in Step 2. 

In Step 3, we show that, for some constant $\kappa^*>0$, 
\begin{eqnarray} \label{proofthirdstep}
\|V_0-\widehat{V}\|_{\infty,\infty}=O_p(n^{-\kappa^*}), 
\end{eqnarray}
where $\|\cdot\|_{\infty,\infty}$ denotes the elementwise max-norm, and $\widehat{V}$ is a matrix involving the covariance of $\eta_{j_1,j_2}^{*(\ell)}$ in \eqref{eqn:etaj1j2} of the paper, and is defined later in Step 3. 

Following \eqref{proofsecondstep} and bounded convergence theorem, we have that, 
\vspace{-0.1in}
\begin{eqnarray*}
	\sup_{z\in \mathbb{R}}\left|\prob\left( S_0^{(\ell)}\le z \right)- \prob\left( \|N(0,V_0)\|_{\infty}\le z \right) \right| = o(1).
\end{eqnarray*}
This, together with \eqref{prooffirststep1}, yields that
\vspace{-0.1in}
\begin{align}\label{eqstep2}
\begin{split}
\prob\left( \widehat{S}^{(\ell)}\le z \right) & \ge \prob\left( \|N(0,V_0)\|_{\infty} \le z-\varepsilon\log^{-1/2} n \right) - o(1), \\
\prob\left( \widehat{S}^{(\ell)}\le z \right) & \le \prob\left( \|N(0,V_0)\|_{\infty}\le z+\varepsilon\log^{-1/2} n \right) + o(1),
\end{split}
\end{align}
for any sufficiently small $\varepsilon>0$, where the little-o term is uniform in $z$. Using similar arguments for \eqref{eqstep2} and also Lemma 3.1 of \citet{Cherno2015}, we have by \eqref{proofthirdstep}
\vspace{-0.1in}
\begin{align*}
\begin{split}
\prob\left( \widehat{S}^{(\ell)}\le z \right) & \ge \prob\left( \|N(0,\widehat{V})\|_{\infty}\le z-2\varepsilon\log^{-1/2} n|\widehat{V} \right) - o(1), \\
\prob\left( \widehat{S}^{(\ell)}\le z \right) & \le \prob\left( \|N(0,\widehat{V})\|_{\infty}\le z+2\varepsilon\log^{-1/2} n|\widehat{V} \right) + o(1), 
\end{split}
\end{align*}
for any sufficiently small $\varepsilon>0$. Set $z=\widehat{c}^{(\ell)}(q_1,q_2)$. Since the little-o term is uniform in $z\in \mathbb{R}$, we have
\begin{align}\label{eqstep3}
\begin{split}
\prob\left\{ \widehat{S}^{(\ell)}\le \widehat{c}^{(\ell)}(q_1,q_2) \right\} & \ge \prob\left\{ \|N(0,\widehat{V})\|_{\infty}\le \widehat{c}^{(\ell)}(q_1,q_2)-2\varepsilon\log^{-1/2} n|\widehat{V} \right\} - o(1), \\
\prob\left\{ \widehat{S}^{(\ell)}\le \widehat{c}^{(\ell)}(q_1,q_2) \right\} & \le \prob\left\{ \|N(0,\widehat{V})\|_{\infty}\le \widehat{c}^{(\ell)}(q_1,q_2)+2\varepsilon\log^{-1/2} n|\widehat{V} \right\} + o(1).
\end{split}
\end{align}
We show in Step 2 that all diagonal elements in $V_0$ are well bounded away from zero. Henceforth, with probability approaching $1$, all diagonal elements in $\widehat{V}$ are well bounded away from zero as well. It follows from Theorem 1 of \cite{chernozhukov2017detailed} that 
\begin{eqnarray*}
	\prob\left\{ \|N(0,\widehat{V})\|_{\infty}\le \widehat{c}^{(\ell)}(q_1,q_2)+2\varepsilon\log^{-1/2} n|\widehat{V} \right\} - \prob\left\{ \|N(0,\widehat{V})\|_{\infty}\le \widehat{c}^{(\ell)}(q_1,q_2) \right. \\ 
	\left. - 2\varepsilon\log^{-1/2} n|\widehat{V} \right\} \le O(1) \varepsilon \log^{1/2}d \log^{-1/2} n,
\end{eqnarray*}
where $O(1)$ denotes some positive constant. Under the given conditions, we have $\log d=O(\log n)$. The right-hand-side is bounded by $\kappa \varepsilon$ for some constant $\kappa>0$. This, together with \eqref{eqstep3}, yields that 
\begin{eqnarray*}
	\left| \prob\left\{ \widehat{S}^{(\ell)}\le \widehat{c}^{(\ell)}(q_1,q_2) \right\} - \prob\left\{ \|N(0,\widehat{V})\|_{\infty}\le \widehat{c}^{(\ell)}(q_1,q_2)|\widehat{V} \right\} \right| \le \kappa \varepsilon + o(1).
\end{eqnarray*}
Since $\varepsilon$ can be made arbitrarily small, \eqref{assertion} follows, which completes the proof of Theorem \ref{thm1}. Next, we give detailed proofs for each step.

\bigskip
\noindent
\textbf{Step 1}:
The proof of this step relies on the arguments developed to establish the limiting distribution of the debiased LASSO \citep{van2014} and the decorrelated score statistic \citep{Ning2017}. We aim to establish the upper bound for $\max_{(j_1,j_2)\in \mathcal{S}(q_1,q_2,\widehat{\bm{B}}^{(\ell)})}$ $| \sqrt{|\mathcal{I}_{\ell}^c|}(\widehat{W}_{j_1,j_2}-W_{0,j_1,j_2})-\eta_{j_1,j_2}^{(\ell)} |$, and for $\max_{(j_1,j_2)\in \mathcal{S}(q_1,q_2,\widehat{\bm{B}}^{(\ell)})} |\eta_{j_1,j_2}^{(\ell)}-\eta_{0,j_1,j_2}^{(\ell)}|$. Together, these two upper bounds would lead to \eqref{prooffirststep}.

To obtain the first upper bound, we have that, 
\vspace{-0.1in}
\begin{eqnarray*}
	& & \max_{(j_1,j_2)\in \mathcal{S}(q_1,q_2,\widehat{\bm{B}}^{(\ell)})} \left| \sqrt{|\mathcal{I}_{\ell}^c|} \left( \widehat{W}_{j_1,j_2}-W_{0,j_1,j_2} \right) - \eta_{j_1,j_2}^{(\ell)} \right|  \\
	& \le & \left|\frac{|\mathcal{I}_{\ell}^c|^{-1/2}\displaystyle \sum_{i\in \mathcal{I}_{\ell}^c} \left\{ X_{i,j_2}-\widehat{\bm{\beta}}^{(\ell)\top}(j_1,j_2) \bm{X}_{i } \right\} \left\{ \sum_{j\neq j_2} X_{i,j}(\overline{W}_{j_1,j}^{(\ell)}-W_{0,j_1,j}) \right\} }{\displaystyle |\mathcal{I}_{\ell}^c|^{-1} \sum_{i\in \mathcal{I}_{\ell}^c} X_{i,j_2} \left\{ X_{i,j_2}-\widehat{\bm{\beta}}^{(\ell)\top}(j_1,j_2) \bm{X}_i \right\}}\right|=\frac{I_1(j_1,j_2, \ell)}{I_2(j_1,j_2,\ell)}.
\end{eqnarray*}
We next bound $I_1(j_1,j_2, \ell)$ and $I_2(j_1,j_2, \ell)$, respectively. 

For $I_1(j_1,j_2, \ell)$, we have that, $I_1(j_1,j_2, \ell) \le I_{1,1}(j_1,j_2,\ell)+I_{1,2}(j_1,j_2,\ell)$, where
\begin{eqnarray*}
	I_{1,1}(j_1,j_2,\ell) & = & \bigg| |\mathcal{I}_{\ell}^c|^{-1/2}\displaystyle \sum_{i\in \mathcal{I}_{\ell}^c} \left\{ \bm{\beta}_0^{(\ell)\top}(j_1,j_2)\bm{X}_i-\widehat{\bm{\beta}}^{(\ell)\top}(j_1,j_2) \bm{X}_{i } \right\} \bigg\{ \sum_{j\neq j_2} X_{i,j}(\overline{W}_{j_1,j}^{(\ell)}-W_{0,j_1,j})  \bigg\} \bigg|,\\
	I_{1,2}(j_1,j_2,\ell) & = & \bigg| |\mathcal{I}_{\ell}^c|^{-1/2}\displaystyle \sum_{i\in \mathcal{I}_{\ell}^c} \left\{ X_{i,j_1}-\bm{\beta}_0^{(\ell)\top}(j_1,j_2)\bm{X}_i \right\} \bigg\{ \sum_{j\neq j_2} X_{i,j}(\overline{W}_{j_1,j}^{(\ell)}-W_{0,j_1,j}) \bigg\} \bigg|.
\end{eqnarray*}

For $I_{1,1}(j_1,j_2)$, by Cauchy-Schwarz inequality, it can be upper bounded by
\begin{eqnarray*}
	&&|\mathcal{I}_{\ell}^c|^{1/2}\|\widehat{\bm{\beta}}^{(\ell)}(j_1,j_2)-\bm{\beta}_0^{\ell}(j_1,j_2)\|_2 \|\overline{\bm{W}}_{j_1}^{(\ell)}-\bm{W}_{0,j_1}\|_2 \| \bm{\Sigma}_0 \|_2\\
	&+&|\mathcal{I}_{\ell}^c|^{1/2}\|\widehat{\bm{\beta}}^{(\ell)}(j_1,j_2)-\bm{\beta}_0^{\ell}(j_1,j_2)\|_1 \|\overline{\bm{W}}_{j_1}^{(\ell)}-\bm{W}_{0,j_1}\|_1\Big\|\mathcal{I}_{\ell}^c|^{-1} \sum_{i\in \mathcal{I}_{\ell}^c} \bm{X}_i \bm{X}_i^{\top}-\bm{\Sigma}_0\Big\|_{\infty,\infty},
\end{eqnarray*}
where $\|\cdot\|_{\infty,\infty}$ denotes the elementwise maximum norm in absolute value. In the proof of Lemma \ref{lemma:sigma}, we have shown that $\| \bm{\Sigma}_0 \|_2$ is bounded. By Condition (A5), the first line is upper bounded by $O(1)\sqrt{n}  n^{-(\kappa_4+\kappa_5)}$ 
where $O(1)$ denotes some positive constant. 

In addition, using similar arguments as in Equation (A.70) of \cite{Rose}, we have
\begin{eqnarray*}
	\Big\| |\mathcal{I}_{\ell}^c|^{-1} \sum_{i\in \mathcal{I}_{\ell}^c} \bm{X}_i \bm{X}_i^{\top}-\bm{\Sigma}_0 \Big\|_{\infty,\infty} = O(n^{-1/2}\sqrt{\log d}),
\end{eqnarray*}
with probability approaching $1$. This together with the condition on the number of mediators $d$ and Condition (A5) implies that the second line is $O(n^{-\kappa_6}\sqrt{\log n})$, with probability approaching $1$. Consequently, 
\vspace{-0.05in}
\begin{eqnarray}\label{IN1eq2}
\max_{j_1,j_2,\ell}I_{1,1}(j_1,j_2,\ell)\le O(1) n^{\max(1/2-(\kappa_4+\kappa_5),-\kappa_6)}\sqrt{\log n},
\end{eqnarray}
where $O(1)$ denotes some positive constant.

For $I_{1,2}(j_1,j_2)$, note that $W_{0,j_1,j}\neq 0$ only when $j$ is a parent of $j_1$. By (A4), we have $W_{0,j_1,j}\neq 0$ only when $j\in \hbox{ACT}(j_1,\widetilde{\bm{W}}^{(\ell)})$. It follows that,
\begin{eqnarray*}
	\sum_{j\neq j_2} X_{i,j} \left( \overline{W}_{j_1,j}^{(\ell)}-W_{0,j_1,j} \right) = \sum_{\substack{j\neq j_2,j\in \scriptsize{\hbox{ACT}} ( j_1,\widetilde{\bm{W}}^{(\ell)} )}} X_{i,j} \left( \overline{W}_{j_1,j}^{(\ell)}-W_{0,j_1,j} \right).
\end{eqnarray*}
By the definition of $\bm{\beta}^{(\ell)}_0(j_1,j_2)$, we have $\Mean \left[\left\{ X_{j_2}-\bm{X}^\top \bm{\beta}_0^{(\ell)}(j_1,j_2) \right\}X_{j} | \widetilde{\bm{W}}^{(\ell)},\overline{\bm{W}}^{(\ell)} \right] = 0$ for any $j\in \hbox{ACT}(j_1,\widetilde{\bm{W}}^{(\ell)})-\{j_2\}$. Note that $\widetilde{\bm{W}}^{(\ell)}$ and $\overline{\bm{W}}^{(\ell)}$ are constructed by samples in $\{X_{i,j}\}_{i\in \mathcal{I}_{\ell}}$. It follows that $\Mean \left\{ \varphi_i^{(\ell)}(j_1,j_2)|\widetilde{\bm{W}}^{(\ell)},\overline{\bm{W}}^{(\ell)} \right\} = 0$ for any $i\in \mathcal{I}_{\ell}$, where
\begin{eqnarray*}
	\varphi_i^{(\ell)}(j_1,j_2) = \left\{ X_{i,j_2}-\bm{X}_i^\top \bm{\beta}_0^{(\ell)}(j_1,j_2) \right\} \left\{\sum_{\substack{j\neq j_2,j\in \scriptsize{\hbox{ACT}}(j_1,\widetilde{\bm{W}}^{(\ell)})}} X_{i,j} (\overline{W}_{j_1,j}^{(\ell)}-W_{0,j_1,j})\right\}.
\end{eqnarray*}
By the definition of the Orlicz norm and Cauchy-Schwarz inequality, we have
\begin{eqnarray}\label{Orlicz1}
\begin{split}
&\|\varphi_i^{(\ell)}(j_1,j_2)\|_{\psi_1|\widetilde{\bm{W}}^{(\ell)},\overline{\bm{W}}^{(\ell)}}=\|\varphi_i^{(\ell)}(j_1,j_2)\|_{\psi_2|\widetilde{\bm{W}}^{(\ell)},\overline{\bm{W}}^{(\ell)}}^2\\
\le& \frac{1}{2\tau} \|X_{i,j_2}-\bm{X}_i^\top \bm{\beta}_0^{(\ell)}(j_1,j_2)\|_{\psi_2|\widetilde{\bm{W}}^{(\ell)},\overline{\bm{W}}^{(\ell)}}^2
+\frac{\tau}{2}\|\bm{X}_i^\top(\overline{\bm{W}}^{(\ell)}_{j_1}-\bm{W}_{0,j_1}) \|_{\psi_2|\widetilde{\bm{W}}^{(\ell)},\overline{\bm{W}}^{(\ell)}}^2,
\end{split}
\end{eqnarray}
where $\|\cdot\|_{\psi_p|\widetilde{\bm{W}}^{(\ell)},\overline{\bm{W}}^{(\ell)}}$ denotes the Orlicz norm conditional on $\widetilde{\bm{W}}^{(\ell)}$ and $\overline{\bm{W}}^{(\ell)}$. Since $j_2$ does not belong to the support of $\bm{\beta}_0^{(\ell)}(j_1,j_2)$, we have by \eqref{orlicznorm} that
\begin{eqnarray*}
	\|X_{i,j_2}-\bm{X}_i^\top \bm{\beta}_0^{(\ell)}(j_1,j_2)\|_{\psi_2|\widetilde{\bm{W}}^{(\ell)},\overline{\bm{W}}^{(\ell)}} \le c_0 \left\{1+\|\bm{\beta}_0^{(\ell)}(j_1,j_2)\|_2 \right\}.
\end{eqnarray*} 
Using similar arguments in Lemma A.1 of \cite{Rose}, we have $\max_{j_1,j_2,\ell}\|\bm{\beta}_0^{(\ell)}(j_1,j_2)\|_2=O(1)$. Henceforth, $\max_{i,j_1,j_2,\ell}\|X_{i,j_2}-\bm{X}_i^\top \bm{\beta}_0^{(\ell)}(j_1,j_2)\|_{\psi_2|\widetilde{\bm{W}}^{(\ell)},\overline{\bm{W}}^{(\ell)}}=O(1)$. Similarly, we have,\vspace{-0.1in}
\begin{eqnarray*}
	\max_{i,j_1,j_2,\ell} \|\bm{X}_i^\top(\overline{\bm{W}}^{(\ell)}_{j_1}-\bm{W}_{0,j_1}) \|_{\psi_2|\widetilde{\bm{W}}^{(\ell)},\overline{\bm{W}}^{(\ell)}}\le c_0 \max_{j_1} \|\overline{\bm{W}}^{(\ell)}_{j_1}-\bm{W}_{0,j_1}\|_2\le c_0 \kappa_3 n^{-\kappa_5},
\end{eqnarray*}
by Condition (A5). Setting $\tau=n^{-\kappa_5}$, it follows from \eqref{Orlicz1} that
\begin{eqnarray}\label{Orlicz2}
\max_{i,\ell} \|\varphi_i^{(\ell)}(j_1,j_2)\|_{\psi_1|\widetilde{\bm{W}}^{(\ell)},\overline{\bm{W}}^{(\ell)}}=O(n^{-\kappa_5}),
\end{eqnarray}
by (A4) and (A5). It follows from Lemma G.3 of \cite{PAL} that
\vspace{-0.1in}
\begin{eqnarray*}
	\prob\left(\bigg| \sum_{i\in \mathcal{I}_{\ell}^c} \varphi_i^{(\ell)}(j_1,j_2) \bigg| > t|\max_j  \|\overline{\bm{W}}^{(\ell)}_{j}-\bm{W}_{0,j}\|_2 \le \kappa_3n^{-\kappa_5},\widetilde{\bm{W}}^{(\ell)} \right) 
	\le 2\exp\left\{-\kappa \min\left(\frac{t^2}{n^{1-2\kappa_5}},\frac{t}{n^{-\kappa_5}}\right) \right\},
\end{eqnarray*}
for some constant $\kappa>0$. By Bonferroni's inequality,  we have that, 
\begin{eqnarray*}
	&&\prob\left(\max_{j_1,j_2,\ell} \bigg| \sum_{i\in \mathcal{I}_{\ell}^c} \varphi_i^{(\ell)}(j_1,j_2) \bigg| > t|\max_j  \|\overline{\bm{W}}^{(\ell)}_{j}-\bm{W}_{0,j}\|_2\le \kappa_3n^{-\kappa_5}, \widetilde{\bm{W}}^{(\ell)} \right)\\
	&\le& 4(d+2)^2\exp\left\{-c\min\left(\frac{t^2}{n^{1-2\kappa_5}},\frac{t}{n^{-\kappa_5}}\right) \right\} 
	\le 4\exp\left\{-c\min\left(\frac{t^2}{n^{1-2\kappa_5}},\frac{t}{n^{-\kappa_5}}\right)+2\log (d+2) \right\}.
\end{eqnarray*}
Setting $t=3\kappa_1c^{-1} n^{1/2-\kappa_5}\sqrt{\log n}$, for a sufficiently large $n$, we have that,
\begin{eqnarray*}
	4\exp\left\{-c\min\left(\frac{t^2}{n^{1-2\kappa_5}},\frac{t}{n^{-\kappa_5}}\right)+2\log (d+2) \right\} & = & 4\exp\left\{-3\kappa_1\log n+2\log (d+2) \right\}\\
	&=& \frac{4(d+2)^2}{n^{3\kappa_1}}=O(n^{-\kappa_1})=o(1).
\end{eqnarray*}
Therefore, conditional on the events in (A4) and (A5), we have that, with probability approaching one, $\max_{j_1,j_2,\ell}|\sum_{i\in \mathcal{I}^{c}_{\ell}} \varphi_i^{(\ell)}(j_1,j_2)|=O(n^{1/2-\kappa_5}\sqrt{\log n})$, or equivalently,
\begin{eqnarray}\label{Bernstein}
\max_{j_1,j_2,\ell}I_{1,2}(j_1,j_2,\ell)=O(n^{-\kappa_5}\sqrt{\log n}).
\end{eqnarray}
Combining \eqref{IN1eq2} and \eqref{Bernstein} together yields, for some constant $\kappa>0$, 
\vspace{-0.1in} 
\begin{eqnarray}\label{IN}
\max_{j_1,j_2,\ell}I_1(j_1,j_2,\ell)=O(n^{-\kappa}),
\end{eqnarray}

For $I_2(j_1,j_2,\ell)$, we first define 
\vspace{-0.1in}
\begin{eqnarray*}
	I_2^*(j_1,j_2,\ell) & = & |\mathcal{I}_{\ell}^c|^{-1}\sum_{i\in \mathcal{I}_{\ell}^c} \left\{ X_{i,j_2}-\bm{\beta}_0^{(\ell)\top}(j_1,j_2)\bm{X}_i \right\}X_{i,j_2}, \\
	I_2^{**}(j_1,j_2,\ell) & = & \Mean \left\{ I_2^*(j_1,j_2,\ell)|\widetilde{\bm{W}}^{(\ell)} \right\} = \Mean\left[\left\{ X_{i,j_2}-\bm{\beta}_0^{(\ell)\top}(j_1,j_2) \bm{X}_i \right\}^2 \; | \; \widetilde{\bm{W}}^{(\ell)}\right].
\end{eqnarray*}
Following the proof of Corollaries 4.1 and 4.2 of \cite{Ning2017}, we have that,
\begin{eqnarray}\nonumber
&&|I_2(j_1,j_2,\ell)-I_2^*(j_1,j_2,\ell)|\le|\mathcal{I}_{\ell}^c|^{-1} \bigg| \sum_{i\in \mathcal{I}_{\ell}^c} X_{i,j_2} \{\bm{\beta}_0^{(\ell)}(j_1,j_2)-\widehat{\bm{\beta}}^{(\ell)}(j_1,j_2)\}^\top \bm{X}_i \bigg| \\ \nonumber
&\le&\|\bm{\beta}_0^{(\ell)}(j_1,j_2)-\widehat{\bm{\beta}}^{(\ell)}(j_1,j_2)\|_1\left(\frac{2}{n} \Big\| \sum_{i\in \mathcal{I}_{\ell}^c} X_{i,j_2}\bm{X}_i-|\mathcal{I}_{\ell}^c|\Mean X_{j_2}\bm{X} \Big\|_{\infty}\right) \\\label{denominatoreq1}
&+& \|\bm{\beta}_0^{(\ell)}(j_1,j_2)-\widehat{\bm{\beta}}^{(\ell)}(j_1,j_2)\|_2 \lambda_{\max}(\bm{\Sigma}_0)
=O(n^{-\kappa_4})+O(n^{-\kappa_6-1/2}\sqrt{\log n}),
\end{eqnarray}
where the big-O term is uniform in $(j_1,j_2,\ell)$. In addition, note that
\begin{eqnarray*}
	\Mean |\mathcal{I}_{\ell}^c|^{-1} \sum_{i\in \mathcal{I}_{\ell}^c} \left[ X_{i,j_2} \left\{ X_{i,j_2}-\bm{\beta}_0^{(\ell)\top}(j_1,j_2)\bm{X}_i \right\} | \widetilde{\bm{W}}^{(\ell)} \right] 
	= \Mean \left[ X_{j_2} \left\{ X_{j_2}-\bm{\beta}_0^{(\ell)\top}(j_1,j_2)\bm{X} \right\} | \widetilde{\bm{W}}^{(\ell)} \right] \\
	= \Mean \left[ \left\{X_{j_2}-\bm{\beta}_0^{(\ell)\top}(j_1,j_2)\bm{X} \right\}^2|\widetilde{\bm{W}}^{(\ell)} \right] 
	\ge \|\{1,\bm{\beta}_0^{(\ell)\top}(j_1,j_2)\}\|_2^2 \lambda_{\min}(\Sigma_0)\ge \lambda_{\min}(\Sigma_0).
\end{eqnarray*}
Since the minimum eigenvalue of $\Sigma_0$ is bounded away from zero, we have
\vspace{-0.05in}
\begin{eqnarray}\label{denominatoreq2}
\min_{j_1,j_2,\ell} I_2^{**}(j_1,j_2,\ell) = \min_{j_1,j_2,\ell}\Mean \left\{ I_2^*(j_1,j_2,\ell)|\widetilde{\bm{W}}^{(\ell)} \right\} \ge 2\varepsilon,
\end{eqnarray}
for some $\varepsilon> 0$. Similar to \eqref{Bernstein}, we have,
\vspace{-0.05in}
\begin{eqnarray}\label{denominatoreq2.5}
\max_{_1,j_2,\ell} \Big\| \sum_{i\in \mathcal{I}_{\ell}^c} \{X_{i,j_2}-\bm{\beta}_0^{(\ell)\top}(j_1,j_2)\bm{X}_i\}X_{i,j_2}-|\mathcal{I}_{\ell}^c| I_2^{**}(j_1,j_2,\ell) \Big\|_{\infty}=O(\sqrt{n\log n}).
\end{eqnarray}
This together with \eqref{denominatoreq2} yields,
\begin{eqnarray*}
	\min_{j_1,j_2,\ell} I_2^*(j_1,j_2,\ell) =\min_{j_1,j_2,\ell} |\mathcal{I}_{\ell}^c|^{-1}\sum_{i\in \mathcal{I}_{\ell}^c} \{X_{i,j_2}-\bm{\beta}_0^{(\ell)\top}(j_1,j_2)\bm{X}_i\}X_{i,j_2}\ge 2\varepsilon.
\end{eqnarray*}
Combining this together with \eqref{denominatoreq1}, we have,
\vspace{-0.1in}
\begin{align}\label{eq1}
\begin{split}
\min_{j_1,j_2,\ell} I_2(j_1,j_2,\ell) \ge & \min_{j_1,j_2,\ell} \left[|\mathcal{I}_{\ell}^c|^{-1}\sum_{i\in \mathcal{I}_{\ell}^c} \left\{ X_{i,j_2}-\bm{\beta}_0^{(\ell)\top}(j_1,j_2)\bm{X}_i \right\}X_{i,j_2}\right.\\ 
& \left.+ \; |\mathcal{I}_{\ell}^c|^{-1} \sum_{i\in \mathcal{I}_{\ell}^c} X_{i,j_2} \left\{ \bm{\beta}_0^{(\ell)}(j_1,j_2)-\widehat{\bm{\beta}}^{(\ell)}(j_1,j_2) \right\}^\top \bm{X}_i \right] \ge \varepsilon.
\end{split}
\end{align}

Combining \eqref{IN} and \eqref{eq1} together yields,  
\begin{eqnarray}\label{eq0}
\max_{(j_1,j_2)\in \mathcal{S}(q_1,q_2,\widehat{\bm{B}}^{(\ell)})}|	\sqrt{|\mathcal{I}_{\ell}^c|}(\widehat{W}_{j_1,j_2}-W_{0,j_1,j_2})-\eta_{j_1,j_2}^{(\ell)} |\le O(1) n^{-\kappa},
\end{eqnarray}
for some constant $\kappa>0$. This gives the first desired upper bound. 

To obtain the second upper bound, we have that,
\begin{eqnarray*}
	\max_{(j_1,j_2)\in \mathcal{S}(q_1,q_2,\widehat{\bm{B}}^{(\ell)})} |\eta_{j_1,j_2}^{(\ell)}-\eta_{0,j_1,j_2}^{(\ell)}| 
	\le \max_{\substack{(j_1,j_2)\in \mathcal{S}(q_1,q_2,\widehat{\bm{B}}^{(\ell)})}} \frac{|\mathcal{I}_{\ell}^c|^{-1/2} |\sum_{i\in \mathcal{I}_{\ell}^c} \{ \widehat{\bm{\beta}}^{(\ell)}(j_1,j_2)-\bm{\beta}_0^{(\ell)}(j_1,j_2) \}^\top\bm{X}_i\varepsilon_{i,j_1}| }{I_2(j_1,j_2,\ell)} \\
	+ \max_{\substack{(j_1,j_2)\in \mathcal{S}(q_1,q_2,\widehat{\bm{B}}^{(\ell)})}} \frac{|I_2(j_1,j_2,\ell)-I_2^{**}(j_1,j_2,\ell)| |\mathcal{I}_{\ell}^c|^{-1/2} |\sum_{i\in \mathcal{I}_{\ell}^c} \{\widehat{\bm{\beta}}^{(\ell)}(j_1,j_2)-\bm{\beta}_0^{(\ell)}(j_1,j_2)\}^\top\bm{X}_i\varepsilon_{i,j_1}| }{I_2(j_1,j_2,\ell)I_2^{**}(j_1,j_2,\ell)}. \nonumber \\
\end{eqnarray*}
By \eqref{denominatoreq1}, \eqref{denominatoreq2}, \eqref{denominatoreq2.5} and \eqref{eq1}, we have,
\begin{align}\label{eq2}
\begin{split}
& \max_{(j_1,j_2)\in \mathcal{S}(q_1,q_2,\widehat{\bm{B}}^{(\ell)})} |\eta_{j_1,j_2}^{(\ell)}-\eta_{0,j_1,j_2}^{(\ell)}| \\
\le & \; \frac{2}{\varepsilon}\max_{\substack{(j_1,j_2)\in \mathcal{S}(q_1,q_2,\widehat{\bm{B}}^{(\ell)})}} \Big| |\mathcal{I}_{\ell}^c|^{-1/2} \sum_{i\in \mathcal{I}_{\ell}^c} \{\widehat{\bm{\beta}}^{(\ell)}(j_1,j_2)-\bm{\beta}_0^{(\ell)\top}(j_1,j_2)\}^\top \bm{X}_i\varepsilon_{i,j_1} \Big|.
\end{split}
\end{align}
Note that the supports of $\widehat{\bm{\beta}}^{(\ell)}(j_1,j_2)$ and $\bm{\beta}_0^{(\ell)\top}(j_1,j_2)$ belong to $\hbox{ACT}(j_1,\widetilde{\bm{W}}^{(\ell)})$. Since $\varepsilon_{i,j_1}$ is independent of $\{X_{i,j}:j\in \hbox{ACT}(j_1,\widetilde{\bm{W}}^{(\ell)})\}$ conditional on $\widetilde{\bm{W}}^{(\ell)}$, we have,
\begin{eqnarray*}
	&&\max_{\substack{(j_1,j_2)\in \mathcal{S}(q_1,q_2,\widehat{\bm{B}}^{(\ell)})}} \Big| |\mathcal{I}_{\ell}^c|^{-1/2} \sum_{i\in \mathcal{I}_{\ell}^c} \{\widehat{\bm{\beta}}^{(\ell)}(j_1,j_2)-\bm{\beta}_0^{(\ell)\top}(j_1,j_2)\}^\top \bm{X}_i\varepsilon_{i,j_1} \Big| \\ 
	&\le& \max_{\substack{(j_1,j_2)\in \mathcal{S}(q_1,q_2,\widehat{\bm{B}}^{(\ell)})}} \|\widehat{\bm{\beta}}^{(\ell)}(j_1,j_2)-\bm{\beta}_0^{(\ell)\top}(j_1,j_2)\|_1 \max_{j\in \scriptsize{\hbox{ACT}}(j_1,\widehat{\bm{B}}^{(\ell)}) } \Big| |\mathcal{I}_{\ell}^c|^{-1/2} \sum_{i\in \mathcal{I}_{\ell}^c} X_{i,j}\varepsilon_{i,j_1} \Big|.
\end{eqnarray*}
By (A4), $\varepsilon_{i,j_1}$ is conditionally independent of $X_{j}$ for any $j\in \hbox{ACT}(j_1,\widetilde{\bm{W}}^{(\ell)})$. Similar to \eqref{Bernstein}, by Bernstein's inequality, we have that,
\vspace{-0.1in}
\begin{eqnarray*}
	\max_{\substack{(j_1,j_2)\in \mathcal{S}(q_1,q_2,\widehat{\bm{B}}^{(\ell)})}}\max_{j\in \scriptsize{\hbox{ACT}}(j_1,\widetilde{\bm{W}}^{(\ell)})} \Big| |\mathcal{I}_{\ell}^c|^{-1/2} \sum_{i\in \mathcal{I}_{\ell}^c} X_{i,j}\varepsilon_{i,j_1} \Big| = O(\sqrt{\log n}).
\end{eqnarray*}
It follows from Condition (A5) that,
\vspace{-0.1in}
\begin{eqnarray*}
	\max_{\substack{(j_1,j_2)\in \mathcal{S}(q_1,q_2,\widehat{\bm{B}}^{(\ell)})}} \Big| |\mathcal{I}_{\ell}^c|^{-1/2} \sum_{i\in \mathcal{I}_{\ell}^c} \{\widehat{\bm{\beta}}^{(\ell)}(j_1,j_2)-\bm{\beta}_0^{(\ell)\top}(j_1,j_2)\}^\top \bm{X}_i\varepsilon_{i,j_1} \Big| = O(n^{-\kappa}),
\end{eqnarray*}
for some constant $\kappa>0$. In view of \eqref{eq2}, we obtain the second desired upper bound, 
\vspace{-0.1in}
\begin{eqnarray}\label{eqn:2ndbound}
\max_{\substack{(j_1,j_2)\in \mathcal{S}(q_1,q_2,\widehat{\bm{B}}^{(\ell)})}} |\eta_{j_1,j_2}^{(\ell)}-\eta_{0,j_1,j_2}^{(\ell)}|=O(n^{-\kappa}),
\end{eqnarray}
for some constant $\kappa>0$. 

Finally, combining \eqref{eq0} and \eqref{eqn:2ndbound} together yields \eqref{prooffirststep}. This completes Step 1.

\bigskip
\noindent
\textbf{Step 2}:
Recall that, each $\varepsilon_{i,j_1}$ is uncorrelated with $X_{i,j_2}-\bm{\beta}_0^{(\ell)\top} (j_1,j_2)\bm{X}_i$. Since both $\varepsilon_{i,j_1}$ and $X_{i,j_2}-\bm{\beta}_0^{(\ell)\top}(j_1,j_2) \bm{X}_i$ are normally distributed given $\widetilde{\bm{W}}^{(\ell)}$, they are independent as well. As a result, we have,
\vspace{-0.1in}
\begin{eqnarray*}
	\Var\left[ \left\{ X_{i,j_2}-\bm{\beta}_0^{(\ell)\top}(j_1,j_2) \bm{X}_i \right\} \varepsilon_{i,j_1}| \widetilde{\bm{W}}^{(\ell)} \right] = \sigma_*^2 \Mean\left[ \left\{ X_{i,j_2}-\bm{\beta}_0^{(\ell)\top}(j_1,j_2) \bm{X}_i \right\}^2 \; | \; \widetilde{\bm{W}}^{(\ell)} \right],
\end{eqnarray*}
for $1\le j_1\le d+1$. Recall that $I_2^{**}(j_1,j_2,\ell) = \Mean\left[\left\{ X_{i,j_2}-\bm{\beta}_0^{(\ell)\top}(j_1,j_2) \bm{X}_i \right\}^2 \; | \; \widetilde{\bm{W}}^{(\ell)}\right]$. Then, we have for $1\le j_1\le d+1$ that \vspace{-0.1in}
\begin{align}\label{vareta}
\begin{split} 
\Var\left( \eta_{0,j_1,j_2}^{(\ell)}|\widetilde{\bm{W}}^{(\ell)} \right) & = \frac{1}{\{I_2^{**}(j_1,j_2,\ell)\}^2 }\Var\left[ \left\{ X_{i,j_2}-\bm{\beta}_0^{(\ell)\top}(j_1,j_2) \bm{X}_i \right\} \varepsilon_{i,j_1} \; | \; \widetilde{\bm{W}}^{(\ell)} \right] \\
& = \frac{1}{\sigma_*^2 \Mean\left[ \left\{ X_{i,j_2}-\bm{\beta}_0^{(\ell)\top}(j_1,j_2) \bm{X}_i \right\}^2 \; | \; \widetilde{\bm{W}}^{(\ell)} \right]}.
\end{split}
\end{align}

Since $j_2\notin \mathcal{M}^{(\ell)}(j_1,j_2)$, we have
\begin{eqnarray*}
	\lambda_{\min}(\bm{\Sigma}_0) \big\| \{ 1,\bm{\beta}_0^{(\ell)\top}(j_1,j_2) \}^\top \big\|_2^2 
	& \le & \Mean\left[\left\{ X_{i,j_2}-\bm{\beta}_0^{(\ell)\top}(j_1,j_2) \bm{X}_i \right\}^2 \; | \; \widetilde{\bm{W}}^{(\ell)} \right] \\
	& \le & \lambda_{\max}(\bm{\Sigma}_0) \big\| (1,\bm{\beta}_0^{(\ell)\top}(j_1,j_2))^\top \big\|_2^2.
\end{eqnarray*}
Following the proof of Lemma A.1 of \cite{Rose}, we have $\max_{j_1,j_2,\ell}\|\bm{\beta}_0^{(\ell)}(j_1,j_2)\|_2=O(1)$. Since $\lambda_{\min}(\bm{\Sigma}_0)$ and $\lambda_{\max}(\bm{\Sigma}_0)$ are uniformly bounded away from zero and infinity, there exists some constant $\kappa \ge 1$ such that, for any $j_1,j_2$, 
\begin{eqnarray}\label{eq3}
\kappa^{-1} \le \Mean[\{X_{i,j_2}-\bm{\beta}_0^{(\ell)\top}(j_1,j_2) \bm{X}_i\}^2|\widetilde{\bm{W}}^{(\ell)}] \le \kappa.
\end{eqnarray}
It follows from \eqref{vareta} that, for any $j_1,j_2$, 
\begin{eqnarray}\label{vareta1}
\kappa^{-1} \le \Var(\eta_{0,j_1,j_2}^{(\ell)}|\widetilde{\bm{W}}^{(\ell)}) \le \kappa.
\end{eqnarray}

We index all pairs of indices $(j_1,j_2)$ in $\mathcal{S}(q_1,q_2,\widehat{\bm{B}}^{(\ell)})$ by $\{j_1(0),j_2(0)\}, \{j_1(1),j_2(1)\}, \ldots,$ $\{j_1(L-1),j_2(L-1)\}$, where $L = |\mathcal{S}(q_1,q_2,\widehat{\bm{B}}^{(\ell)})|$. Next, define a covariance matrix $V_0 \in \mathbb{R}^{L \times L}$, such that its $(l_1,l_2)$th entry is the covariance of $\eta_{0,j_1(l_1),j_2(l_1)}^{(\ell)}$ and $\eta_{0,j_1(l_2),j_2(l_2)}^{(\ell)}$ conditional on $\widetilde{\bm{W}}^{(\ell)}$. By \eqref{vareta1}, the diagonal elements of $V_0$ are uniformly bounded away from zero and infinity. We also comment that $\eta_{j_1,j_2}^{(\ell)}$ defined in \eqref{eqn:etaequivalent} in Section \ref{sec:boot} of the paper can be viewed as a more intuitive version of $\eta_{0,j_1,j_2}^{(\ell)}$ defined in \eqref{eta0j1j2}.

When $L$ is finite, by the classical Lindeberg-Feller central limit theorem and Condition (A4), we have \eqref{proofsecondstep} holds. When $L$ diverges, we have $L \le d^2$ that grows at an polynomial order of $n$. By \eqref{vareta1}, we have $\min_{j_1,j_2,\ell} \Var\left( \eta_{0,j_1,j_2}^{(\ell)}|\widetilde{\bm{W}}^{(\ell)} \right)$ is uniformly bounded away from zero. Moreover, following similar arguments in proving \eqref{Orlicz2}, we have $\max_{j_1,j_2,\ell}\|\eta_{0,j_1,j_2}^{(\ell)}\|_{\psi_1|\widetilde{\bm{W}}^{(\ell)}}=O(1)$. Then, by Corollary 2.1 of \cite{Cherno2013}, \eqref{proofsecondstep} holds as well. This completes Step 2.

\bigskip
\noindent
\textbf{Step 3}:
Define a covariance matrix $\widehat{V} \in \mathbb{R}^{L \times L}$, such that its $(l_1,l_2)$th entry is the covariance of $\eta_{j_1(l_1),j_2(l_1)}^{*(\ell)}$ and $\eta_{j_1(l_2),j_2(l_2)}^{*(\ell)}$  conditional on the data. To bound $\|\widehat{V}-V_0\|_{\infty,\infty}$ in \eqref{proofthirdstep}, it suffices to bound
\vspace{-0.1in}
\begin{eqnarray*}
	\max_{\substack{(j_1,j_2),(j_3,j_4) \\ \in \mathcal{S}(q_1,q_2,\widehat{\bm{B}}^{(\ell)})}}  \left|\Cov\left[\eta_{j_1,j_2}^{*(\ell)},\eta_{j_3,j_4}^{*(\ell)}|\{X_{i,j}\}_{1\le i\le n,0\le j\le d+1}\right] - \Cov\left[\eta_{0,j_1,j_2}^{(\ell)},\eta_{0,j_3,j_4}^{(\ell)}|\{X_{i,j}\}_{i\in \mathcal{I}_{\ell}^c,0\le j\le d+1}\right]\right|.
\end{eqnarray*}
Recall in Section \ref{sec:boot}, when $j_1\neq j_3$, we have $\Cov(\eta_{j_1,j_2}^{*(\ell)},\eta_{j_3,j_4}^{*(\ell)}|\{X_{i,j}\}_{1\le i\le n,0\le j\le d+1})=0$. Similarly, $\Cov(\eta_{0,j_1,j_2}^{(\ell)},\eta_{0,j_3,j_4}^{(\ell)}|\{X_{i,j}\}_{i\in \mathcal{I}_{\ell}^c, 0\le j\le d+1 })=0$ when $j_1\neq j_3$. As a result, we have
\begin{eqnarray*}
	\|\widehat{V}-V_0\|_{\infty,\infty}=\max_{\substack{(j_1,j_2),(j_1,j_3) \\ \in \mathcal{S}(q_1,q_2,\widehat{\bm{B}}^{(\ell)})}} \left|\Cov\left[ \eta_{j_1,j_2}^{*(\ell)},\eta_{j_1,j_3}^{*(\ell)}|\{X_{i,j}\}_{1\le i\le n,0\le j\le d+1} \right] \right.\\
	\left.-\Cov\left[ \eta_{0,j_1,j_2}^{(\ell)},\eta_{0,j_1,j_3}^{(\ell)}|\{X_{i,j}\}_{i\in \mathcal{I}_{\ell}^c,0\le j\le d+1} \right]\right|.
\end{eqnarray*}
After some calculations, we have, for $1\le j_1\le d+1$, 
\begin{eqnarray*}
	\Cov\left[ \eta_{0,j_1,j_2}^{(\ell)},\eta_{0,j_1,j_3}^{(\ell)}|\{X_{i,j}\}_{i\in \mathcal{I}_{\ell}^c, 0\le j\le d+1} \right] & = & \frac{\Mean\left[\prod_{k\in \{j_2,j_3\}} \left\{ X_{k}-\bm{\beta}_0^{(\ell)\top}(j_1,k) \bm{X}_{0} \right\}|\widetilde{\bm{W}}^{(\ell)}\right]\sigma_*^2}{I^{**}_2(j_1,j_2,\ell)I_2^{**}(j_1,j_3,\ell)}, \\
	\Cov\left[ \eta_{j_1,j_2}^{*(\ell)},\eta_{j_1,j_3}^{*(\ell)}|\{X_{i,j}\}_{1\le i\le n,0\le j\le d+1} \right] & = & \frac{|\mathcal{I}_{\ell}^c|^{-1}\sum_{i\in \mathcal{I}_{\ell}^c}\prod_{k\in \{j_2,j_3\}}\left[ \left\{ X_{i,k}-\widehat{\bm{\beta}}^{(\ell)\top}(j_1,k) \bm{X}_{i } \right\}\right]\widehat{\sigma}_*^2}{ I_2(j_1,j_2,\ell)I_2(j_1,j_3,\ell)}.
\end{eqnarray*}

It follows that 
\begin{eqnarray*}
	\|\widehat{V}-V_0\|_{\infty,\infty} & = & \max_{\substack{(j_1,j_2),(j_1,j_3) \\ \in \mathcal{S}(q_1,q_2,\widehat{\bm{B}}^{(\ell)})}} I_{3,1}(j_1,j_2,j_3,\ell) 
	+ \max_{\substack{(j_1,j_2),(j_1,j_3) \\ \in \mathcal{S}(q_1,q_2,\widehat{\bm{B}}^{(\ell)})}} I_{3,2}(j_1,j_2,j_3,\ell) \\
	& & \; + \max_{\substack{(j_1,j_2),(j_1,j_3) \\ \in \mathcal{S}(q_1,q_2,\widehat{\bm{B}}^{(\ell)})}} I_{3,3}(j_1,j_2,j_3,\ell),
\end{eqnarray*}
where
\begin{eqnarray*}
	I_{3,1}(j_1,j_2,j_3,\ell) &=& \frac{\left|\Mean\left[\prod_{k\in \{j_2,j_3\}} \left\{X_{k}-\bm{\beta}_0^{(\ell)\top}(j_1,k) \bm{X}_{0} \right\} \; | \; \widetilde{\bm{W}}^{(\ell)}\right]\right|}{I^{**}_2(j_1,j_2,\ell)I_2^{**}(j_1,j_3,\ell)} \; |\widehat{\sigma}_*^2-\sigma_*^2|,\\
	I_{3,2}(j_1,j_2,j_3,\ell)&=&\frac{\widehat{\sigma}_*^2}{I^{**}_2(j_1,j_2,\ell)I_2^{**}(j_1,j_3,\ell)}\left|\frac{1}{|\mathcal{I}_{\ell}^c|}\sum_{i\in \mathcal{I}_{\ell}^c}\prod_{k\in \{j_2,j_3\}}\left[\left\{ X_{i,k}-\widehat{\bm{\beta}}^{(\ell)\top}(j_1,k) \bm{X}_{i } \right\}\right]\right. \\
	&& \; - \left.\Mean\prod_{k\in \{j_2,j_3\}} \left[\left\{ X_{k}-\bm{\beta}_0^{(\ell)\top}(j_1,k) \bm{X}_{0} \right\}\right]\right|,\\
	I_{3,3}(j_1,j_2,j_3,\ell)&=&\left|\frac{1}{|\mathcal{I}_{\ell}^c|}\sum_{i\in \mathcal{I}_{\ell}^c}\prod_{k\in \{j_2,j_3\}}\left[\left\{ X_{i,k}-\widehat{\bm{\beta}}^{(\ell)\top}(j_1,k) \bm{X}_{i } \right\}\right]\widehat{\sigma}_*^2\right| \\
	&& \; \times \frac{|{I^{**}_2(j_1,j_2,\ell)I_2^{**}(j_1,j_3,\ell)}-{I_2(j_1,j_2,\ell)I_2(j_1,j_3,\ell)}|}{{I^{**}_2(j_1,j_2,\ell)I_2^{**}(j_1,j_3,\ell)I_2(j_1,j_2,\ell)I_2(j_1,j_3,\ell)}}.
\end{eqnarray*}
We next bound $\max_{j_1,j_2,j_3,\ell}I_{3,q}(j_1,j_2,j_3,\ell)$, $q=1,2,3$, respectively.

For $\max_{j_1,j_2,j_3,\ell}I_{3,1}(j_1,j_2,j_3,\ell)$, by Cauchy-Schwarz inequality, we have
\begin{eqnarray*}
	\max_{j_1,j_2,j_3,\ell}\left|\Mean\left[\prod_{k\in \{j_2,j_3\}} \left\{ X_{k}-\bm{\beta}_0^{(\ell)\top}(j_1,k) \bm{X}_{0} \right\} \; | \; \widetilde{\bm{W}}^{(\ell)}\right]\right| \\
	\le \max_{j_1,j_2,j_3,\ell} \prod_{k\in \{j_2,j_3\}} \Mean \left[\left\{ X_{k}-\bm{\beta}_0^{(\ell)\top}(j_1,k) \bm{X}_{0} \right\}^2 \; | \; \widetilde{\bm{W}}^{(\ell)} \right]. 
\end{eqnarray*}
By \eqref{eq3}, we have,
\begin{eqnarray}\label{eq4}
\max_{j_1,j_2,j_3,\ell}\left|\Mean\left[\prod_{k\in \{j_2,j_3\}} \left\{ X_{k}-\bm{\beta}_0^{(\ell)\top}(j_1,k) \bm{X}_{0} \right\} \; | \; \widetilde{\bm{W}}^{(\ell)}\right]\right|=O(1). 
\end{eqnarray}
Combining \eqref{eq4} with \eqref{denominatoreq2} and Lemma \ref{lemma:sigma} yields,
\begin{eqnarray}\label{I1}
\max_{j_1,j_2,j_3,\ell} I_{3,1}(j_1,j_2,j_3,\ell)=O(n^{-\kappa_7}).
\end{eqnarray}

For $\max_{j_1,j_2,j_3,\ell}I_{3,2}(j_1,j_2,j_3,\ell)$, by Lemma \ref{lemma:sigma}, we have $\widehat{\sigma}_*^2=O(1)$. Define 
\begin{eqnarray*}
	I_3^*(j_1,j_2,j_3,\ell) & = & \left|\frac{1}{|\mathcal{I}_{\ell}^c|}\sum_{i\in \mathcal{I}_{\ell}^c}\prod_{k\in \{j_2,j_3\}}\left[\left\{ X_{i,k}-\widehat{\bm{\beta}}^{(\ell)\top}(j_1,k) \bm{X}_{i } \right\}\right] \right. \\ 
	& & \left. - \; \Mean\left[\prod_{k\in \{j_2,j_3\}} \left\{ X_{k}-\bm{\beta}_0^{(\ell)\top}(j_1,k) \bm{X}_{0} \right\} \; | \; \widetilde{\bm{W}}^{(\ell)}\right]\right|. \\
	I_3^{**}(j_1,j_2,j_3) & = & \left|\frac{1}{|\mathcal{I}_{\ell}^c|}\sum_{i\in \mathcal{I}_{\ell}^c}\left[\prod_{k\in \{j_2,j_3\}} \left\{X_{i,k}-\widehat{\bm{\beta}}^{(\ell)\top}(j_1,k) \bm{X}_{i } \right\} \right.\right. \\
	& & \left.\left. - \; \prod_{k\in \{j_2,j_3\}} \left\{ X_{i,k}-\bm{\beta}_0^{(\ell)\top}(j_1,k) \bm{X}_{i } \right\}\right]\right|.
\end{eqnarray*}
Similar to \eqref{Bernstein}, we have, 
\begin{eqnarray*}
	\max_{j_1,j_2,j_3,\ell}\left|\frac{1}{|\mathcal{I}_{\ell}^c|}\sum_{i\in \mathcal{I}_{\ell}^c}\prod_{k\in \{j_2,j_3\}} \left\{ X_{i,k}-\bm{\beta}_0^{\top}(j_1,k) \bm{X}_{i } \right\} 
	- \Mean\left[\prod_{k\in \{j_2,j_3\}} \left\{ X_{k}-\bm{\beta}_0^{(\ell)\top}(j_1,k) \bm{X}_{0} \right\} \; | \; \widetilde{\bm{W}}^{(\ell)}\right]\right|\\
	=O(n^{-1/2}\sqrt{\log n}),
\end{eqnarray*}
Next, $\max_{j_1,j_2,j_3,\ell} I_3^{**}(j_1,j_2,j_3,\ell) \le I_{4,1}(j_1,j_2,j_3,\ell)+I_{4,2}(j_1,j_2,j_3,\ell)+I_{4,3}(j_1,j_2,j_3,\ell)$, where
\vspace{-0.1in}
\begin{eqnarray*}
	I_{4,1}(j_1,j_2,j_3,\ell) & = & \bigg| \frac{1}{|\mathcal{I}_{\ell}^c|}\sum_{i\in \mathcal{I}_{\ell}^c} \left\{ \bm{\beta}_0^{(\ell)}(j_1,j_2)-\widehat{\bm{\beta}}^{(\ell)}(j_1,j_2) \right\}^\top\bm{X}_{i } \left\{ X_{i,j_3}-\bm{\beta}_0^{(\ell)\top}(j_1,j_3) \bm{X}_{i } \right\} \bigg|,\\
	I_{4,2}(j_1,j_2,j_3,\ell) & = & \bigg| \frac{1}{|\mathcal{I}_{\ell}^c|}\sum_{i\in \mathcal{I}_{\ell}^c} \left\{ \bm{\beta}_0^{(\ell)}(j_1,j_3)-\widehat{\bm{\beta}}^{(\ell)}(j_1,j_3) \right\}^\top\bm{X}_{i } \left\{ X_{i,j_2}-\bm{\beta}_0^{(\ell)\top}(j_1,j_2) \bm{X}_{i } \right\} \bigg|,\\
	I_{4,3}(j_1,j_2,j_3,\ell) & = & \bigg| \frac{1}{|\mathcal{I}_{\ell}^c|}\sum_{i\in \mathcal{I}_{\ell}^c} \left\{ \bm{\beta}_0^{(\ell)}(j_1,j_2)-\widehat{\bm{\beta}}^{(\ell)}(j_1,j_2) \right\}^\top\bm{X}_{i }\bm{X}_i^\top \left\{ \bm{\beta}_0^{(\ell)}(j_1,j_2)-\widehat{\bm{\beta}}^{(\ell)}(j_1,j_2) \right\} \bigg|.
\end{eqnarray*}
Using similar arguments in proving \eqref{IN1eq2} and \eqref{denominatoreq1}, we have 
\vspace{-0.05in}
\begin{eqnarray*}
	\max_{j_1,j_2,j_3,\ell} I_{4,1}(j_1,j_2,j_3,\ell) & = & O(n^{-\kappa_5}), \\
	\max_{j_1,j_2,j_3,\ell} I_{4,2}(j_1,j_2,j_3,\ell) & = & O(n^{-\kappa_5}), \\
	\max_{j_1,j_2,j_3,\ell} I_{4,3}(j_1,j_2,j_3,\ell) & = & O(n^{-2\kappa_5}). 
\end{eqnarray*}
Therefore, we have, for some constant $\kappa > 0$, 
\vspace{-0.1in}
\begin{eqnarray}\label{I2star}
\max_{j_1,j_2,j_3,\ell} I_2^*(j_1,j_2,j_3,\ell) = O(n^{-\kappa}). 
\end{eqnarray}
It then follows from \eqref{denominatoreq2} that, for some constant $\kappa> 0$, 
\vspace{-0.1in}
\begin{eqnarray}\label{I2}
\max_{j_1,j_2,j_3,\ell} I_2(j_1,j_2,j_3,\ell)=O(n^{-\kappa}). 
\end{eqnarray}

For $\max_{j_1,j_2,j_3,\ell}I_{3,3}(j_1,j_2,j_3,\ell)$, combining \eqref{I2star} with \eqref{eq4}, we have,
\begin{eqnarray}\label{eq5}
\left|\frac{1}{|\mathcal{I}_{\ell}^c|}\sum_{i\in \mathcal{I}_{\ell}^c}\prod_{k\in \{j_2,j_3\}}\left[ \{X_{i,k}-\widehat{\bm{\beta}}^{(\ell)\top}(j_1,k) \bm{X}_{i }\}\right]\right|=O(1).
\end{eqnarray}
Using similar arguments in bounding $\max_{j_1,j_2,\ell} |I_2(j_1,j_2,\ell)-I_2^{**}(j_1,j_2,\ell)|$ in Step 2, we get
\begin{eqnarray*}
	\max_{j_1,j_2,j_3,\ell} |{I^{**}_2(j_1,j_2,\ell)I_2^{**}(j_1,j_3,\ell)}-{I_2(j_1,j_2,\ell)I_2(j_1,j_3,\ell)}|=O(n^{-\kappa}),
\end{eqnarray*}
for some constant $\kappa>0$. Combining this together with \eqref{eq5}, \eqref{denominatoreq2}, \eqref{eq1} and that $\widehat{\sigma}_*^2=O(1)$, we have, for some constant $\kappa > 0$, 
\begin{eqnarray} \label{I3}
\max_{j_1,j_2,j_3,\ell}I_{3,3}(j_1,j_2,j_3,\ell)=O(n^{-\kappa}).
\end{eqnarray}

Combining \eqref{I1}, \eqref{I2} and \eqref{I3} together yields the bound for $\|\widehat{V}-V_0\|_{\infty,\infty}$ in \eqref{proofthirdstep}. This completes Step 3.
\eop

\subsection{Proof of Theorem \ref{thm2}}

Recall the proposed bootstrap procedure repeatedly generate random variables from $N(0,\widehat{V})$, and the critical value $\widehat{c}^{(\ell)}(q_1,q_2)$ is the upper $(\alpha/2)$th quantile of $\|N(0,\widehat{V})\|_{\infty}$. That is, 
\begin{eqnarray}\label{quantile}
\prob\left( \|N(0,\widehat{V})\|_{\infty}\le \widehat{c}^{(\ell)}(q_1,q_2)|\widehat{V} \right) = \frac{\alpha}{2}.
\end{eqnarray}
We have shown in the proof of Theorem \ref{thm1} that the diagonal elements of $\widehat{V}$ are uniformly bounded by some constant $\kappa>0$. It follows from Bonferroni's inequality that
\begin{eqnarray*}
	\prob\left( \|N(0,\widehat{V})\|_{\infty}>t \sqrt{\kappa \log n} \right) & \le & (d+2)\max_{j\in \{0,\ldots,d+1\}} \prob\left( |N(0,\widehat{V}_{j,j})|>t\sqrt{\kappa \log n} \right) \\
	& \le & (d+2) \left\{1-\Phi\left( t\sqrt{\log n} \right)\right\},
\end{eqnarray*}
where $\widehat{V}_{j,j}$ is the $(j,j)$th entry of $\widehat{V}$. For $t\ge 1$ and $n\ge 3$, we have $t\sqrt{\log n}\ge 1$, and hence
\begin{eqnarray*}
	1-\Phi\left( t\sqrt{\log n} \right) & = & \frac{1}{\sqrt{2\pi}}\int_{t\sqrt{\log n}}^{+\infty} \exp\left(-\frac{x^2}{2}\right)dx\le \int_{t\sqrt{\log n}}^{+\infty} x\exp\left(-\frac{x^2}{2}\right)dx \\
	& = & \exp\left(-\frac{t^2 \log n}{2}\right) = n^{-t^2/2}. 
\end{eqnarray*}
Setting $t=2\sqrt{\kappa_1+1}$, it follows from the condition $d=O(n^{\kappa_1})$ that,
\begin{eqnarray*}
	\prob\left( \|N(0,\widehat{V})\|_{\infty}> \sqrt{2(\kappa_1+1)c\log n} \right) = O(d/n^{\kappa_1+1})=o(1).
\end{eqnarray*}
In view of \eqref{quantile}, we obtain $\widehat{c}^{(\ell)}(q_1,q_2)\le \sqrt{2(\kappa_1+1)c\log n}$. 

According to our test procedure, we reject the null if $\sqrt{n}(\widehat{\bm{W}}^{*(\ell)})_{q,0}>\widehat{c}(0,q)$ and $\sqrt{n}(\widehat{\bm{W}}^{*(\ell)})_{d+1,q}>\widehat{c}(q,d+1)$, for some $\ell = 1, 2$, where $\widehat{\bm{W}}^{*(\ell)}=|\widehat{\bm{W}}^{(\ell)}|\oplus |\widehat{\bm{W}}^{(\ell)}|^{(2)} \oplus \ldots \oplus |\widehat{\bm{W}}^{(\ell)}|^{(d)}$. To prove Theorem \ref{thm2}, it suffices to show 
\begin{eqnarray}\label{prove2sufficient}
\begin{split}
&&\prob(\sqrt{n}(\widehat{\bm{W}}^{*(\ell)})_{q,0}\ge \sqrt{2(\kappa_1+1)c\log n})\to 1,\\ &&\prob(\sqrt{n}(\widehat{\bm{W}}^{*(\ell)})_{d+1,q}\ge \sqrt{2(\kappa_1+1)c\log n})\to 1,
\end{split}
\end{eqnarray}
under the alternative hypothesis $H_1$. We next prove \eqref{prove2sufficient}. 

Under the given conditions of the theorem, there exists a path, $X_{0}\to X_{i_1}\to \ldots \to X_{i_k}\to X_{q}$, such that $|W_{0,0,i_1}|,|W_{0,i_1,i_2}|,\ldots,|W_{0,i_{k-1},i_k}|,|W_{0,i_k,q}|\gg n^{-1/2}\sqrt{\log n}$. Under the given conditions, we have with probability tending to $1$ that, 
\begin{eqnarray*}
	\max\left(|\widehat{W}_{0,i_1}^{(\ell)}-W_{0,0,i_1}|,\max_{1\le j\le k-1} |\widehat{W}_{i_j,i_{j+1}}^{(\ell)}-W_{0,i_j,i_{j+1}}|,|\widehat{W}_{i_k,q}^{(\ell)}-W_{0,i_k,q}| \right)=O(n^{-1/2}\sqrt{\log n}).
\end{eqnarray*}
It follows that
\begin{eqnarray}\label{What}
\min\left( |\widehat{W}_{0,i_1}^{(\ell)}|, \min_{1\le j\le k-1} |\widehat{W}_{i_j,i_{j+1}}^{(\ell)}|, |\widehat{W}_{i_k,q}^{(\ell)}| \right)\gg n^{-1/2} \sqrt{\log n}.
\end{eqnarray}
By definition, we have that, 
\begin{eqnarray*}
	\sqrt{n}(\widehat{\bm{W}}^{*(\ell)})_{q,0}\ge \sqrt{n}\min\left( |\widehat{W}_{0,i_1}^{(\ell)}|, \min_{1\le j\le k-1} |\widehat{W}_{i_j,i_{j+1}}^{(\ell)}|, |\widehat{W}_{i_k,q}^{(\ell)}| \right).
\end{eqnarray*}
It follows from \eqref{What} that $\sqrt{n}(\widehat{\bm{W}}^{*(\ell)})_{q,0} \gg \sqrt{\log n}$ under Condition (A5). Therefore, $\prob\left( \sqrt{n}(\widehat{\bm{W}}^{*(\ell)})_{q,0}\ge \sqrt{2(\kappa_1+1)c\log n} \right) \to 1$, which proves the first result in \eqref{prove2sufficient}. Similarly, we can prove the second result in \eqref{prove2sufficient} that $\prob\left( \sqrt{n}(\widehat{\bm{W}}^{*(\ell)})_{d+1,q}\ge \sqrt{2(\kappa_1+1)c\log n} \right)$ $\to 1$. This completes the proof of Theorem \ref{thm2}. 
\eop

\subsection{Proof of Theorem \ref{thm3}}
\label{sec:sketchproofthm3}

First, we show that the $p$-values $\widehat{p}^{(\ell)}(0,q)$ and $\widehat{p}^{(\ell)}(q,d+1)$ are asymptotically independent. Let $\hbox{ACT}(q)$ and $\hbox{DES}(q)$ denote the  set of ancestors and descendants of $X_{q}$, respectively. By Condition (A4), the two test statistics $\sqrt{\mathcal{I}_{\ell}^c} (\widehat{\bm{W}}^{*(\ell)})_{q,0}$ and $\sqrt{\mathcal{I}_{\ell}^c} (\widehat{\bm{W}}^{*(\ell)})_{d+1,q}$ are constructed based on the set of estimators $\left\{ \widehat{W}^{(\ell)}_{j_1,j_2}: j_1 \in \hbox{ACT}(q)\cup \{q\},j_2\in \hbox{ACT}(q) \right\}$ and $\left\{ \widehat{W}^{(\ell)}_{j_1,j_2}: j_1 \in \hbox{DES}(q),j_2\in \hbox{DES}(q)\cup \{q\} \right\}$, respectively. These two sets of estimators are asymptotically independent conditional on $\{X_i\}_{i\in \mathcal{I}_{\ell}}$. To better illustrate this, note that for any $(j_1,j_2,j_3,j_4)$ such that $j_1 \in \hbox{ACT}(q)\cup \{q\}, j_2\in \hbox{ACT}(q), j_3 \in \hbox{DES}(q), j_4\in \hbox{DES}(q)\cup \{q\}$, we have $j_1\neq j_3$. As discussed in Section \ref{sec:boot}, $\sqrt{n}(\widehat{W}^{(\ell)}_{j_1,j_2}-W_{0,j_1,j_2})$ and $\sqrt{n}(\widehat{W}^{(\ell)}_{j_3,j_4}-W_{0,j_3,j_4})$ are asymptotically uncorrelated. Since the two variables are jointly normal, they are asymptotically independent as well. As a result, the two test statistics $\sqrt{\mathcal{I}_{\ell}^c} (\widehat{\bm{W}}^{*(\ell)})_{q,0}$ and $\sqrt{\mathcal{I}_{\ell}^c} (\widehat{\bm{W}}^{*(\ell)})_{d+1,q}$ are asymptotically independent, and so are their corresponding $p$-values. 

Next, a closer look at the proof of Theorem \ref{thm1} shows that the type-I error rates can be uniformly controlled across different mediators. That is, 
\vspace{-0.1in} 
\begin{eqnarray}\label{uniformpvalues}
\max_{\ell\in \{1,2\}}\max_{q\in \{1,\ldots,d\}} \prob\Big\{ \widehat{p}_{\max}^{(\ell)}(q) \le \alpha \; | \; H_0(q) \; \textrm{holds} \Big\} \le \alpha+o(1),
\end{eqnarray}
for any given significance level $0 < \alpha < 1$. Following the proof of Theorem 1 in \cite{Djor2019}, and by \eqref{uniformpvalues}, we can show that, 
\vspace{-0.1in}
\begin{eqnarray}\label{uniformpvalues1}
\max_{\ell\in \{1,2\}}\max_{q\in \{1,\ldots,d\}} \prob\Big\{ \widehat{p}_{\max}^{(\ell)}(q)\le \alpha \; | \; H_0(q) \; \textrm{holds}, \;\; \widehat{p}_{\min}^{(\ell)}(q)\le c^{(\ell)} \Big\} \le \alpha+o(1),
\end{eqnarray}
for any significance level $\alpha$ and the critical value $c^{(\ell)}>0$. 

Then, similar to Theorem 1.3 of \cite{Ben2001}, and by \eqref{uniformpvalues1}, we have that FDR($\mathcal{H}^{(\ell)}$) is guaranteed at level $\alpha/2$. Recall $\mathcal{H}=\mathcal{H}^{(1)}\cup \mathcal{H}^{(2)}$ is the final set of our selected mediators, and $\mathcal{N}$ is the set of unimportant mediators. It follows that,
\begin{eqnarray*}
	\hbox{FDR}(\mathcal{H}) & = & \Mean \left(\frac{|\mathcal{N} \cup \mathcal{H}|}{\max(1,\mathcal{H})}\right)\le \sum_{\ell \in \{1,2\}}\Mean \left(\frac{|\mathcal{N} \cup \mathcal{H}^{(\ell)}|}{\max(1,\mathcal{H})}\right)\le \sum_{\ell \in \{1,2\}}\Mean \left(\frac{|\mathcal{N} \cup \mathcal{H}^{(\ell)}|}{\max(1,\mathcal{H}^{(\ell)})}\right) \\ 
	& = & \sum_{\ell \in \{1,2\}} \hbox{FDR}(\mathcal{H}^{(\ell)})=\alpha+o(1). 
\end{eqnarray*}
This completes the proof of Theorem \ref{thm3}.
\eop

\subsection{Proof of Proposition \ref{prop:oracle}}
\label{sec:proof-prop}

\textbf{Outline of the proof}: 
We choose $\lambda = \kappa \sqrt{\log n/n}$ for some sufficiently large constant $\kappa>0$. We divide the proof into three steps. In Step 1, we establish the uniform convergence rate of $\|\widetilde{\bm{W}}_j^{(\ell)}(\pi)-\bm{W}_{j}(\pi)\|_2$. Due to the  acyclic constraint,  $\widetilde{\bm{W}}^{(\ell)}=\widetilde{\bm{W}}^{(\ell)}(\pi)$ for some ordering $\pi$. In Step 2, we show that 
\vspace{-0.1in}
\begin{align}\label{proofofsecondstep}
\begin{split}
\min_{\pi\notin \Pi_0}\left(\sum_{i\in \mathcal{I}_{\ell}} \|\bm{X}_i-\widetilde{\bm{W}}^{(\ell)}(\pi) \bm{X}_i\|^2+\sum_{j=0}^{d+1}|\mathcal{I}_{\ell}| \lambda \|\widetilde{\bm{W}}_j^{(\ell)}(\pi)\|_1\right) \\
> \max_{\pi\in \Pi_0} \left(\sum_{i\in \mathcal{I}_{\ell}} \|\bm{X}_i-\bm{W}(\pi) \bm{X}_i\|^2+\sum_{j=0}^{d+1}|\mathcal{I}_{\ell}| \lambda \|\bm{W}_j(\pi)\|_1\right).
\end{split}
\end{align}
By definition, this implies that $\widetilde{\bm{W}}^{(\ell)}=\widetilde{\bm{W}}^{(\ell)}(\pi^*)$ for some true ordering $\pi^* \in \Pi^*$. This proves the first assertion of Proposition \ref{prop:oracle}. In the third step, we derive the convergence rate of $\widetilde{\bm{W}}^{(\ell)}$, and completes the second assertion of Proposition \ref{prop:oracle}.

\medskip
\noindent
\textbf{Step 1}:
For any $\pi,i,j$, let $\varepsilon_{i,j}(\pi)=X_{i,j}-\bm{X}_i^\top \bm{W}_j(\pi)$ denote the residual. By definition, 
\vspace{-0.1in}
\begin{eqnarray}
& &\sum_{j=0}^{d+1}\sum_{i\in \mathcal{I}_{\ell}} | X_{i,j}-\bm{X}_i^\top \widetilde{\bm{W}}^{(\ell)}_j(\pi)|^2 + \lambda |\mathcal{I}_{\ell} | \sum_{j=0}^{d+1}\|\widetilde{\bm{W}}_j^{(\ell)}(\pi)\|_1 \label{step1firsteq} \\
& \le & \sum_{j=0}^{d+1}\sum_{i\in \mathcal{I}_{\ell}} |X_{i,j}-\bm{X}_i^\top \bm{W}_j(\pi) |^2 + \lambda |\mathcal{I}_{\ell}| \sum_{j=0}^{d+1}\|\bm{W}_j(\pi)\|_1 
=  \sum_{j=0}^{d+1}\sum_{i\in \mathcal{I}_{\ell}} |\varepsilon_{i,j}(\pi)|^2 + \lambda |\mathcal{I}_{\ell}| \sum_{j=0}^{d+1}\|\bm{W}_j(\pi)\|_1. \nonumber 
\end{eqnarray}
Note that
\vspace{-0.1in}
\begin{eqnarray*}
	& & \sum_{i\in \mathcal{I}_{\ell}} |X_{i,j}-\bm{X}_i^\top \widetilde{\bm{W}}^{(\ell)}_j(\pi)|^2=\sum_{i\in \mathcal{I}_{\ell}} |\varepsilon_{i,j}(\pi)+\bm{X}_i^\top \bm{W}_j(\pi)-\bm{X}_i^\top \widetilde{\bm{W}}_j^{(\ell)}(\pi)|^2\\
	& = & \sum_{i\in \mathcal{I}_{\ell}} |\varepsilon_{i,j}(\pi)|^2 + 2\sum_{i\in \mathcal{I}_{\ell}} \varepsilon_{i,j}(\pi) \left\{ \bm{X}_i^\top \bm{W}_j(\pi)-\bm{X}_i^\top \widetilde{\bm{W}}^{(\ell)}_j(\pi) \right\} + \sum_{i\in \mathcal{I}_{\ell}} \left[ \bm{X}_i^\top \left\{\bm{W}_j(\pi)-\widetilde{\bm{W}}^{(\ell)}_j(\pi) \right\} \right]^2.
\end{eqnarray*}
Let $s_j(\pi)$ and $\tilde{s}_j^{(\ell)}(\pi)$ denote the number of nonzero elements in $\bm{W}_j(\pi)$ and $\widetilde{\bm{W}}^{(\ell)}_j(\pi)$, respectively. By Theorem 7.1 of \cite{van2013}, we have, 
\begin{align}\label{importantevent}
\begin{split}
\sum_{j=0}^{d+1} \sum_{i\in \mathcal{I}_{\ell}}2\left|\varepsilon_{i,j}(\pi) \bm{X}_i^\top \left\{ \bm{W}_j(\pi)- \widetilde{\bm{W}}^{(\ell)}_j(\pi) \right\}\right| -\delta \sum_{j=0}^{d+1} \sum_{i\in \mathcal{I}_{\ell}} \left| \bm{X}_i^\top \left\{ \bm{W}_j(\pi)- \widetilde{\bm{W}}^{(\ell)}_j(\pi) \right\} \right|^2 \\ 
\le \sum_{j=0}^{d+1}\frac{\kappa \left\{ s_j(\pi)+\tilde{s}_j^{(\ell)}(\pi) \right\} \log n }{\delta},
\end{split}
\end{align}
for some constant $\kappa>0$, any $\delta>0$, and any order $\pi$. Under the event defined in \eqref{importantevent}, 
\begin{align}\label{importantevent0}
\begin{split}
\sum_{i\in \mathcal{I}_{\ell}} & \Big| X_{i,j} - \bm{X}_i^\top \widetilde{\bm{W}}^{(\ell)}_j(\pi) \Big|^2 \ge \sum_{j=0}^{d+1} \sum_{i\in \mathcal{I}_{\ell}} |\varepsilon_{i,j}(\pi)|^2 \\
& +(1-\delta) \sum_{i\in \mathcal{I}_{\ell}} \sum_{j=0}^{d+1} \left[ \bm{X}_i^\top \left\{\bm{W}_j(\pi)-\widetilde{\bm{W}}^{(\ell)}_j(\pi) \right\}\right]^2
-\sum_{j=0}^{d+1}\frac{\kappa \left\{s_j(\pi)+\tilde{s}_j^{(\ell)}(\pi) \right\} \log n}{\delta}.
\end{split}
\end{align}
Setting $\delta=1/2$, together with \eqref{step1firsteq}, we have, 
\vspace{-0.1in}
\begin{eqnarray*}
	\frac{1}{2}\sum_{i\in \mathcal{I}_{\ell}} \sum_{j=0}^{d+1} \left[ \bm{X}_i^\top \left\{ \bm{W}_j(\pi)-\widetilde{\bm{W}}^{(\ell)}_j(\pi) \right\}\right]^2 + \lambda |\mathcal{I}_{\ell}| \sum_{j=0}^{d+1} \big\| \widetilde{\bm{W}}_j^{(\ell)}(\pi) \big\|_1\\
	\le \sum_{j=0}^{d+1} 2 \kappa \left\{ s_j(\pi)+\tilde{s}_j^{(\ell)}(\pi) \right\} \log n + \lambda |\mathcal{I}_{\ell}| \sum_{j=0}^{d+1} \big\| \bm{W}_j(\pi) \big\|_1.
\end{eqnarray*}
Using similar arguments in bounding $\max_{j_1,j_2,\ell} I_{N,1}(j_1,j_2,\ell)$ in Step 1 of the proof of Theorem \ref{thm1}, we can show there exists constants $\kappa^* \le \lambda_{\min}(\bm{\Sigma}_0)\le \lambda_{\max}(\bm{\Sigma}_0) \le 1/\kappa^*$ such that
\begin{align}\label{importantevent1.5}
\begin{split}
\kappa^* |\mathcal{I}_{\ell}| \sum_{j=0}^{d+1} \big\| \bm{W}_j(\pi)-\widetilde{\bm{W}}^{(\ell)}_j(\pi) \big\|_2^2 & \le \frac{1}{2}\sum_{i\in \mathcal{I}_{\ell}} \sum_{j=0}^{d+1} \left[ \bm{X}_i^\top \left\{ \bm{W}_j(\pi)-\widetilde{\bm{W}}^{(\ell)}_j(\pi) \right\} \right]^2 \\
& \le \frac{1}{\kappa^*} |\mathcal{I}_{\ell}| \sum_{j=0}^{d+1} \big\| \bm{W}_j(\pi)-\widetilde{\bm{W}}^{(\ell)}_j(\pi) \big\|_2^2,
\end{split}
\end{align}
for any order $\pi$. It then follows that, for any order $\pi$, 
\vspace{-0.1in}
\begin{align}\label{importantevent2}
\begin{split}
\kappa^* \sum_{j=0}^{d+1} \big\| \bm{W}_j(\pi) - \widetilde{\bm{W}}^{(\ell)}_j(\pi) \big\|_2^2 + \lambda \sum_{j=0}^{d+1} \big\| \widetilde{\bm{W}}_j^{(\ell)}(\pi) \big\|_1 
\le & \sum_{j=0}^{d+1} \frac{4c}{n} \left\{ s_j(\pi)+\tilde{s}_j^{(\ell)}(\pi) \right\} \log n \\
& + \lambda \sum_{j=0}^{d+1}\|\bm{W}_j(\pi)\|_1.
\end{split}
\end{align}

Next, define the $(d+2)$-dimensional vector, 
\vspace{-0.1in}
\begin{eqnarray*}
	\bm{\mu}(\pi) = \left\{ \widetilde{\bm{W}}_0^{(\ell)\top}(\pi)-\bm{W}_0^{(\ell)\top}(\pi), \widetilde{\bm{W}}_1^{(\ell)\top}(\pi)-\bm{W}_1^{(\ell)\top}(\pi),\ldots,\widetilde{\bm{W}}_{d+1}^{(\ell)\top}(\pi)-\bm{W}_{d+1}^{(\ell)\top}(\pi) \right\}^\top. 
\end{eqnarray*}
Let $\mathcal{M}(\pi)$ denote the support of $\bm{\mu}(\pi)$. Our goal is to bound $\|\bm{\mu}(\pi)\|_2 = \sum_{j=0}^{d+1} \|\widetilde{\bm{W}}^{(\ell)}_j(\pi) - \bm{W}_j(\pi)\|_2^2$. We next consider two cases. 

First, when $\|\bm{\mu}(\pi)\|_2 \le \sum_{j=0}^{d+1} 4c \left\{ s_j(\pi)+\tilde{s}_j^{(\ell)}(\pi) \right\}\log n/(n\kappa^*)$, it is already bounded.

Second, when $\|\bm{\mu}(\pi)\|_2 > \sum_{j=0}^{d+1} 4c \left\{ s_j(\pi)+\tilde{s}_j^{(\ell)}(\pi) \right\}\log n/(n\kappa^*)$, we aim to obtain a larger upper bound. Specifically, under the event defined in \eqref{importantevent2}, we have, for all $\pi$, 
\vspace{-0.1in}
\begin{eqnarray*}
	\sum_{j=0}^{d+1} \big\| \widetilde{\bm{W}}_j^{(\ell)}(\pi) \big\|_1 \le \sum_{j=0}^{d+1} \big\| \bm{W}_j(\pi) \big\|_1. 
\end{eqnarray*}
It follows that, for all $\pi$, 
\vspace{-0.2in}
\begin{eqnarray}\label{case2event}
\|\bm{\mu}(\pi)_{\mathcal{M}(\pi)}\|_1\le \|\bm{\mu}(\pi)_{\mathcal{M}^c(\pi)}\|_1,
\end{eqnarray}
where $\bm{\mu}(\pi)_{\mathcal{M}(\pi)}$ and $\bm{\mu}(\pi)_{\mathcal{M}^c(\pi)}$ denote the sub-vector formed by the elements in $\mathcal{M}(\pi)$ and $\mathcal{M}^c(\pi)$, respectively. Then, under the event defined in \eqref{case2event}, and by the definition of $s_0$,
\vspace{-0.1in}
\begin{eqnarray*}
	\sum_{j=0}^{d+1} \big\| \bm{W}_j(\pi) \big\|_1-\sum_{j=0}^{d+1} \big\| \widetilde{\bm{W}}_j^{(\ell)}(\pi) \big\|_1 & \le & \|\bm{\mu}(\pi)_{\mathcal{M}(\pi)}\|_1+\|\bm{\mu}(\pi)_{\mathcal{M}^c(\pi)}\|_1\le 2\|\bm{\mu}(\pi)_{\mathcal{M}(\pi)}\|_1\\
	& \le & 2\sqrt{|\mathcal{M}(\pi)|} \|\bm{\mu}(\pi)_{\mathcal{M}(\pi)}\|_2\le 2\sqrt{s_0 (d+2)} \|\bm{\mu}(\pi)_{\mathcal{M}(\pi)}\|_2.
\end{eqnarray*}
By \eqref{importantevent2}, we have that, for all $\pi$, 
\vspace{-0.1in}
\begin{eqnarray*}
	\kappa^* \|\bm{\mu}(\pi)\|_2^2 & \le & \sum_{j=0}^{d+1} \frac{4c}{n } \left\{ s_j(\pi)+\tilde{s}_j^{(\ell)}(\pi) \right\} \log n + 2\lambda \sqrt{s_0 (d+2)} \|\bm{\mu}(\pi)_{\mathcal{M}(\pi)}\|_2, \\
	\kappa^* \|\bm{\mu}(\pi)_{\mathcal{M}(\pi)}\|_2^2 & \le & \sum_{j=0}^{d+1} \frac{4c}{n } \left\{ s_j(\pi)+\tilde{s}_j^{(\ell)}(\pi) \right\} \log n +  2\lambda \sqrt{s_0 (d+2)} \|\bm{\mu}(\pi)_{\mathcal{M}(\pi)}\|_2.
\end{eqnarray*}
It then follows that, for all $\pi$ and some positive constant $O(1)$, 
\vspace{-0.1in}
\begin{eqnarray*}
	\|\bm{\mu}(\pi)_{\mathcal{M}(\pi)}\|_2\le O(1)\left(\lambda \sqrt{s_0 (d+2)}+\sqrt{\sum_{j=0}^{d+1} \frac{4c}{n} \left\{ s_j(\pi)+\tilde{s}_j^{(\ell)}(\pi) \right\} \log n} \; \right). 
\end{eqnarray*}
Under the event defined in \eqref{case2event}, and by Equations (A5), (A6) in \cite{zhou2009}, we have, 
\begin{eqnarray}\label{importantevent3}
\|\bm{\mu}(\pi)\|_2\le O(1)\left(\lambda \sqrt{s_0 (d+2)}+\sqrt{\sum_{j=0}^{d+1} \frac{4c}{n} \{s_j(\pi)+\tilde{s}_j^{(\ell)}(\pi)\}\log n}\right),
\end{eqnarray}
for all $\pi$.  

Therefore, in both cases, we show that $\|\bm{\mu}(\pi)\|_2$ is uniformly bounded for all $\pi$. This completes Step 1.

\bigskip
\noindent
\textbf{Step 2}:
Under the events defined in \eqref{importantevent1.5} and \eqref{importantevent3}, we have that, for all $\pi$, 
\vspace{-0.1in}
\begin{eqnarray*}
	\sum_{i\in \mathcal{I}_{\ell}} \sum_{j=0}^{d+1} \left[ \bm{X}_i^\top \left\{ \bm{W}_j(\pi)-\widetilde{\bm{W}}^{(\ell)}_j(\pi) \right\}\right]^2 \le O(1)d^2\log n.
\end{eqnarray*}
Combining this together with \eqref{importantevent0} yields that, for all $\pi$, 
\begin{eqnarray}\label{importantevent4}
\sum_{i\in \mathcal{I}_{\ell}} \left| X_{i,j}-\bm{X}_i^\top \widetilde{\bm{W}}^{(\ell)}_j(\pi) \right|^2 \ge \sum_{j=0}^{d+1} \sum_{i\in \mathcal{I}_{\ell}} |\varepsilon_{i,j}(\pi)|^2+O(1)d^2 \log n.
\end{eqnarray}
Under the omega-min condition in (A6), using similar arguments in the proof of Lemma 7.8 of \cite{van2013}, we have, 
\vspace{-0.1in}
\begin{eqnarray*}
	\min_{\pi\notin \Pi_0}\sum_{j=0}^{d+1} \sum_{i\in \mathcal{I}_{\ell}} |\varepsilon_{i,j}(\pi)|^2\ge \max_{\pi\in \Pi_0}\sum_{j=0}^{d+1} \sum_{i\in \mathcal{I}_{\ell}} |\varepsilon_{i,j}(\pi)|^2 + \kappa (d+2) n,
\end{eqnarray*}
for some constant $\kappa>0$. Under the event defined in \eqref{importantevent4}, we have, 
\vspace{-0.1in}
\begin{eqnarray*}
	\min_{\pi\notin \Pi_0}\left\{ \sum_{i\in \mathcal{I}_{\ell}} \|\bm{X}_i-\widetilde{\bm{W}}^{(\ell)}(\pi) \bm{X}_i\|^2+\sum_{j=0}^{d+1}|\mathcal{I}_{\ell}| \lambda \big\| \widetilde{\bm{W}}_j^{(\ell)}(\pi) \big\|_1\right\} \ge \min_{\pi\notin \Pi_0}\sum_{i\in \mathcal{I}_{\ell}} \big\| \bm{X}_i -\widetilde{\bm{W}}^{(\ell)}(\pi) \bm{X}_i \big\|^2\\
	\ge \max_{\pi\in \Pi_0}\sum_{j=0}^{d+1} \sum_{i\in \mathcal{I}_{\ell}} |\varepsilon_{i,j}(\pi)|^2 + \kappa (d+2) n - O(1) d^2 \log n.
\end{eqnarray*}
Under the Condition $\|\bm{W}_0\|_2=O(1)$ for any $\pi\in \Pi^*$, we have $\|\bm{W}_j(\pi)\|_2=\|\bm{W}_{0,j}\|_2=O(1)$. As a result, 
\vspace{-0.1in}
\begin{eqnarray*}
	\max_{\pi\in \Pi_0}\sum_{j=0}^{d+1} \lambda |\mathcal{I}_{\ell}| \|\bm{W}_j(\pi)\|_1 = \sum_{j=0}^{d+1} \lambda |\mathcal{I}_{\ell}| \|\bm{W}_{0,j}\|_1 \le \sqrt{d+2} \sum_{j=0}^{d+1}|\mathcal{I}_{\ell}| \lambda \|\bm{W}_{0,j}\|_2 = O\left( d^{3/2}n^{1/2}\sqrt{\log n} \right).
\end{eqnarray*}
It follows that,
\vspace{-0.1in}
\begin{eqnarray*}
	& & \min_{\pi\notin \Pi_0}\left(\sum_{i\in \mathcal{I}_{\ell}} \|\bm{X}_i-\widetilde{\bm{W}}^{(\ell)}(\pi) \bm{X}_i\|^2+\sum_{j=0}^{d+1}|\mathcal{I}_{\ell}| \lambda \|\widetilde{\bm{W}}_j^{(\ell)}(\pi)\|_1\right) \\
	& & \ge \; \max_{\pi\in \Pi_0}\sum_{j=0}^{d+1} \left(\sum_{i\in \mathcal{I}_{\ell}} |\varepsilon_{i,j}(\pi)|^2+\sum_{j=0}^{d+1}|\mathcal{I}_{\ell}| \lambda \|\bm{W}_j(\pi)\|_1\right) \\
	& & + \; \kappa (d+2)n-O(1) d^2 \log n - O\left( d^{3/2}n^{1/2}\sqrt{\log n} \right),
\end{eqnarray*}
where the last line is strictly positive. As a result, we obtain \eqref{proofofsecondstep}. This completes Step 2.

\bigskip
\noindent
\textbf{Step 3}:
Built on the results obtained in Step 2, we have $\widehat{\pi}\in \Pi_0$, where $\widehat{\pi}$ is the estimated ordering. To complete the proof, it suffices to show that $\max_{\pi\in \Pi_0} \max_{j\in \{0,\ldots,d+1\}} \|\widetilde{\bm{W}}_{j}^{(\ell)}-\bm{W}_{0,j}\|_2 = O\left( n^{-1/2}\sqrt{s_0\log n} \right)$. Since the number of elements in $\Pi^*$ is bounded by $O(n^{\kappa^*})$ for some $\kappa^*>0$, by Bernstein's inequality and Bonferroni's inequality, we have that,
\begin{eqnarray*}
	\max_{\pi\in\Pi_0} \max_{k\in\{0,1,\ldots,j-1\}}|\sum_{i\in \mathcal{I}_{\ell}}\varepsilon_{i,j}(\pi) X_{i,\pi_k}| = O\left( n^{1/2} \sqrt{\log n} \right).
\end{eqnarray*}
The explicit convergence rate of $\max_{\pi\in \Pi_0} \max_{j\in \{0,\ldots,d+1\}} \|\widetilde{\bm{W}}_{j}^{(\ell)}-\bm{W}_{0,j}\|_2$ can then be similarly derived following Theorem 3.1 of \cite{zhou2009}. This completes Step 3.

\medskip
\noindent
\change{\textbf{Remark}: Finally, we discuss the possibility of relaxing the condition $d=o(n)$, by imposing some sparsity conditions on $\widetilde{\bm{W}}^{(\ell)}(\pi)$, and the population limit of $\widetilde{\bm{W}}^{(\ell)}(\pi)$, i.e., $\bm{W}_0(\pi)$. In this case, the condition $d\ll n$ can be relaxed to $d=O(n^{\kappa_1})$, for some constant $\kappa_1>0$. Since we do not require $\kappa_1<1$, the dimension $d$ can grow at a much faster rate than $n$.
	
	Specifically, suppose the number of nonzero elements in each column of $\widetilde{\bm{W}}^{(\ell)}(\pi)$ and $\bm{W}_{0}(\pi)$ satisfies that $\max_{j,\pi} [\max\{s_j(\pi),\tilde{s}_j^{(\ell)}(\pi)\}]=O(n^{\kappa_*})$, for some constant $0<\kappa_*<1$. It follows from \eqref{importantevent0} and \eqref{importantevent3} that
	\begin{align*}
	\begin{split}
	\sum_{i\in \mathcal{I}_{\ell}} \Big| X_{i,j} - \bm{X}_i^\top \widetilde{\bm{W}}^{(\ell)}_j(\pi) \Big|^2 
	\ge \; & \sum_{j=0}^{d+1} \sum_{i\in \mathcal{I}_{\ell}} |\varepsilon_{i,j}(\pi)|^2 + (1-\delta) \sum_{i\in \mathcal{I}_{\ell}} \sum_{j=0}^{d+1} \left[ \bm{X}_i^\top \left\{\bm{W}_j(\pi)-\widetilde{\bm{W}}^{(\ell)}_j(\pi) \right\}\right]^2 \\
	& - \sum_{j=0}^{d+1}\frac{O(1) n^{\kappa_*} \log n}{\delta}, \\
	\|\bm{\mu}(\pi)\|_2 \le \; & O(1)\left(\lambda \sqrt{d}n^{\kappa_*/2}+\sqrt{d}n^{\kappa_*/2-1/2}\sqrt{\log n}\right),
	\end{split}
	\end{align*}
	for all $\pi$, where $O(1)$ denotes some positive constant. Therefore, we have
	\begin{eqnarray*}
		\sum_{i\in \mathcal{I}_{\ell}} \left| X_{i,j}-\bm{X}_i^\top \widetilde{\bm{W}}^{(\ell)}_j(\pi) \right|^2 \ge \sum_{j=0}^{d+1} \sum_{i\in \mathcal{I}_{\ell}} |\varepsilon_{i,j}(\pi)|^2+O(1)d n^{\kappa_*} \log n,
	\end{eqnarray*}
	for all $\pi$. The rest of the proof follows using similar arguments as in Steps 2 and 3.}
\eop

\subsection{Additional numerical results}
\label{sec:add-numerical}

\change{First, we provide more details about the simulation setup. Table \ref{tab:strength} reports the corresponding mediators with nonzero mediation effects, and their associated $\delta(q)$, where $\delta(q) = (\bm{W}_0^*)_{d+1,q}(\bm{W}_0^*)_{q,0}$, $\bm{W}_0^*$ is constructed based on $\bm{W}_0$, $q = 1, \ldots, d$. Meanwhile, Figure \ref{fig-1} shows an example of the adjacent matrix when $d=100$.

	\begin{table}[t!]
		\footnotesize
		\centering
		\begin{tabular}{c} \toprule
			Scenario A: $(d,p_1,p_2)=(50,0.05,0.15)$, $n=100,200$ \\ \midrule
			\begin{tabular}{@{}c|c|c|c|c|c|c@{}}
				$q$ & 10 & 12 & 20 & 28 & 30 & 41\\ \hline
				$\delta(q)$ & 1.06 & 1.03 & 0.63 & 1.08 & 0.64 & 1.31 \\ 
			\end{tabular} \\ \midrule
			Scenario B: $(d,p_1,p_2)=(100,0.025,0.075)$, $n=250,500$ \\ \midrule
			\begin{tabular}{@{}c|c|c|c|c|c|c|c|c|c@{}}
				$q$ & 5 & 14 & 17 & 20 & 30 & 71 & 80 & 82 & 89\\ \hline
				$\delta(q)$ & 0.99 & 0.33 & 0.92 & 0.85 & 0.85 & 1.37 & 0.72 & 0.72 & 0.72\\ 
			\end{tabular}\\\midrule
			Scenario C: $(d,p_1,p_2)=(150,0.02,0.05)$, $n=250,500$ \\ \midrule
			\begin{tabular}{@{}c|c|c|c|c|c|c|c|c|c|c@{}}
				$q$ & 19 &  23 &  51 & 52 & 54 &56 & 60 & 81 & 92 & 93 \\ \hline
				$\delta(q)$ & 1.24 & 0.93 & 1.71 & 1.28 & 1.90 & 1.90 & 0.94 & 0.88 & 0.88 & 1.02\\ 
			\end{tabular}\\ \bottomrule
		\end{tabular}
		\vspace{-0.1in}
		\caption{Mediators with nonzero mediation effects, $q=1,\ldots,d$.}
		\label{tab:strength}
	\end{table}

	\begin{figure}[b!]
		\centering
		\includegraphics[width=6.5cm]{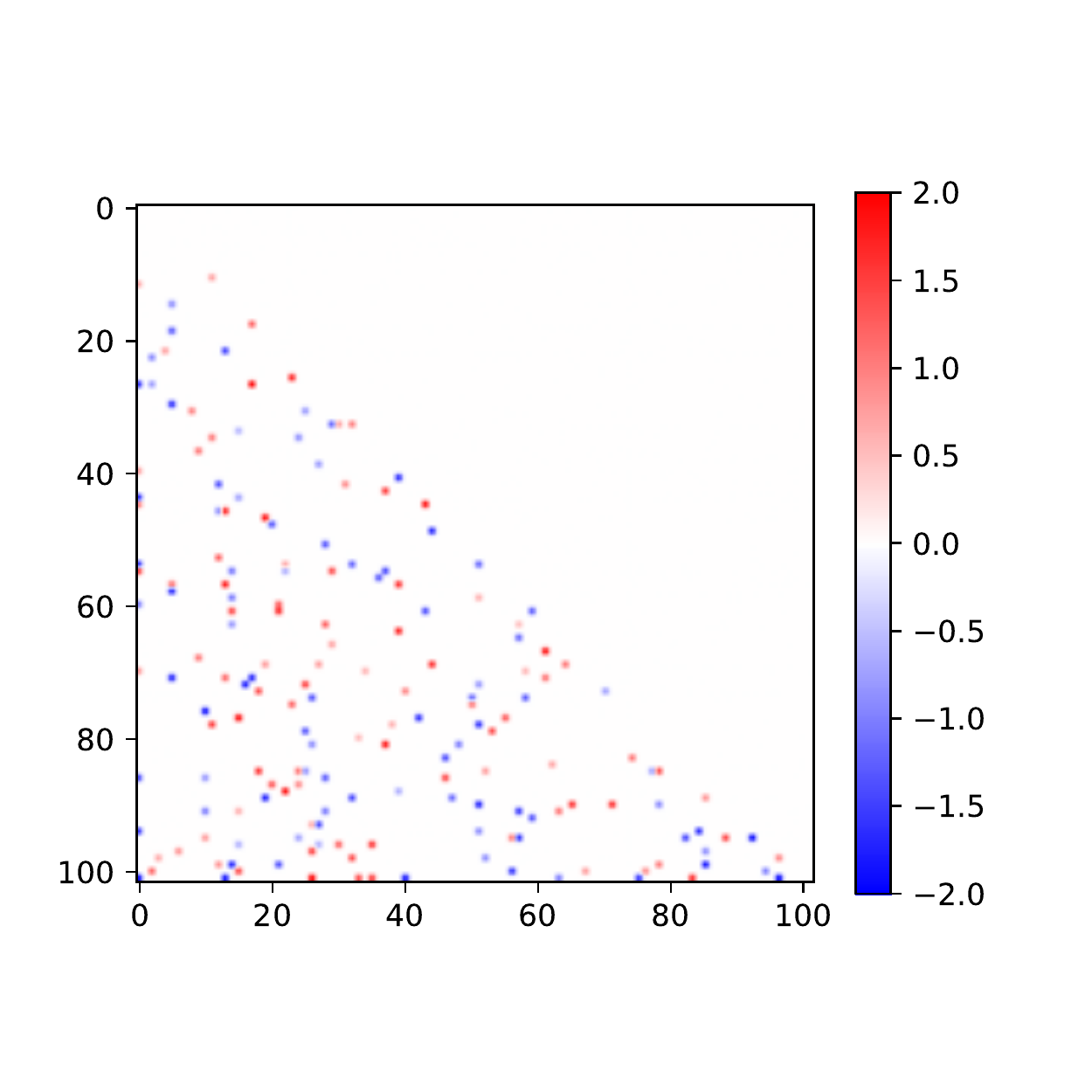} 
		\caption{The example adjacent matrix with $d=100$.}\label{fig-1}
	\end{figure}

	Next, we report the simulation results for testing of a single mediator.  Figures \ref{figAB} and \ref{figEF} show the results for $d=100$ and $d=150$, respectively, which complement the results for $d=50$ reported in Section \ref{sec:simulations} of the paper. We observe a similar qualitative pattern that our proposed test achieves a valid size under the null hypothesis, and achieves a larger empirical power under the alternative hypothesis.

	\begin{figure}[t!]
		\centering
		\includegraphics[width=15cm]{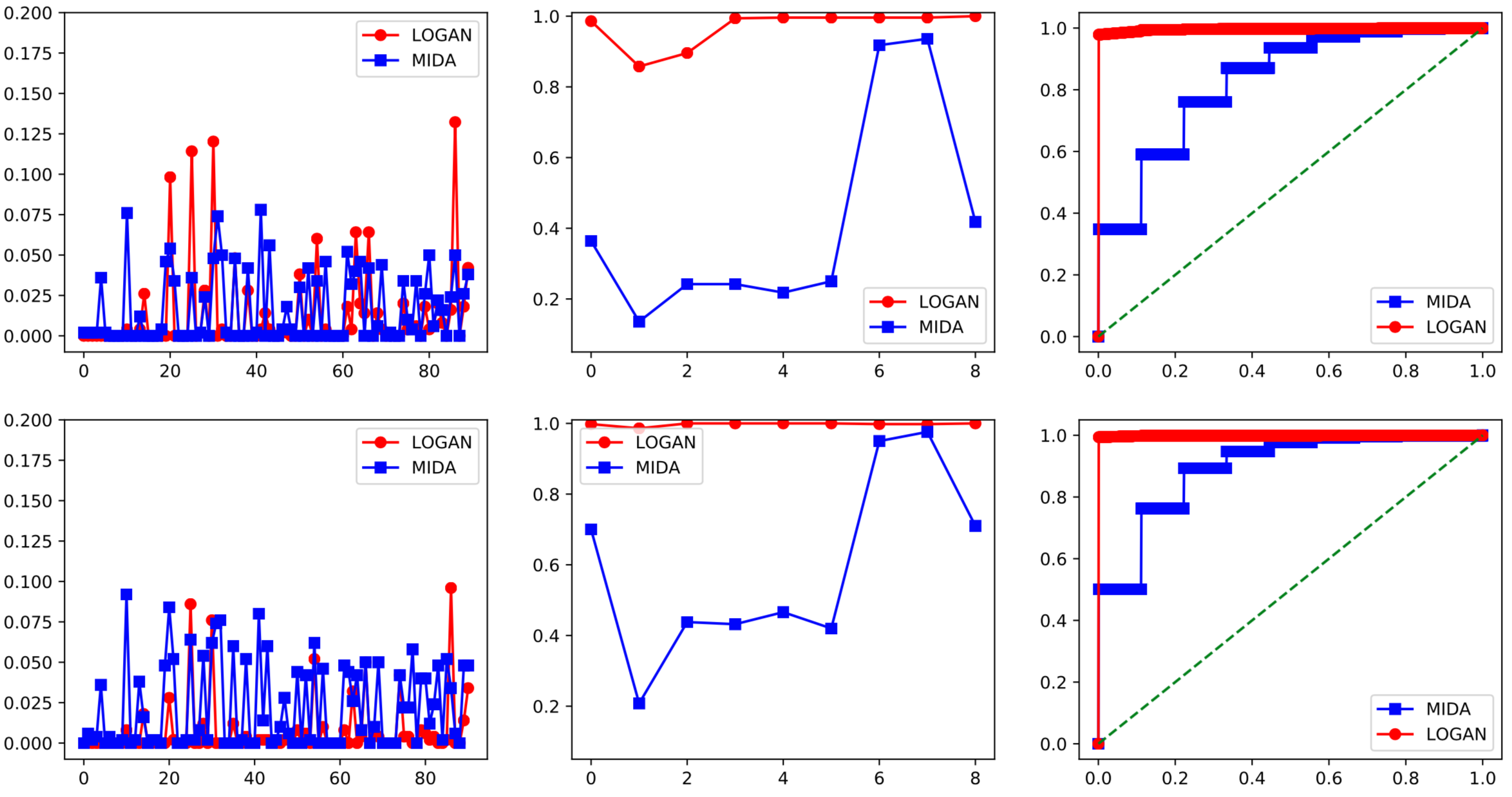}
		\caption{Empirical rejection rate and ROC curve of the proposed test, LOGAN, and the test of \cite{LHZ2018}, MIDA, when $d=100$. The upper panels: $n=250$, and the bottom panels: $n=500$. The left panels: under $H_0$, the middles panels: under $H_1$, where the horizontal axis is the mediator index, and the right panels: the average ROC curve.}
		\label{figCD}
	\end{figure}

	\begin{figure}[b!]
		\centering
		\includegraphics[width=15cm]{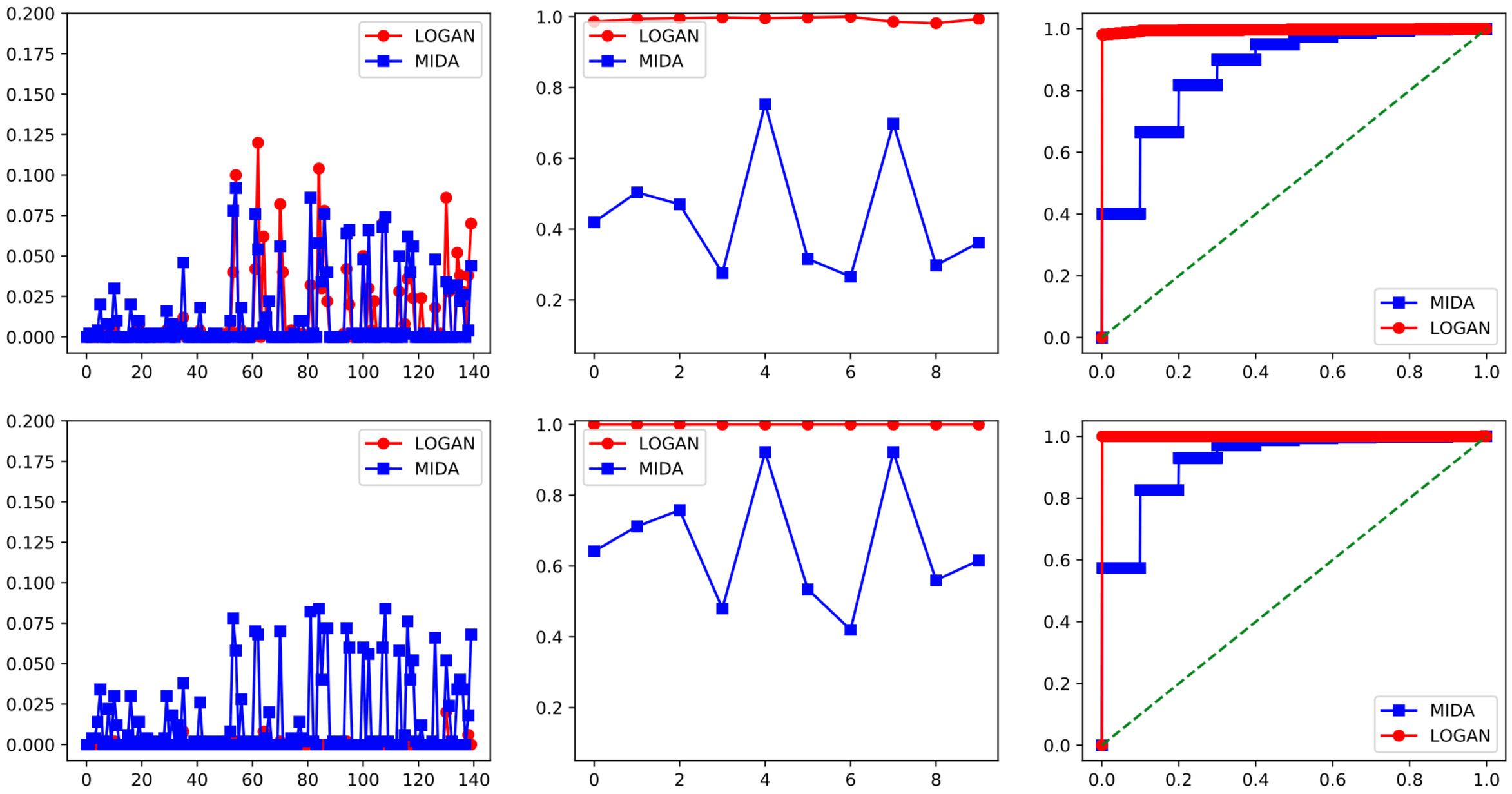}
		\caption{Empirical rejection rate and ROC curve of the proposed test, LOGAN, and the test of \cite{LHZ2018}, MIDA, when $d=150$. The upper panels: $n=250$, and the bottom panels: $n=500$. The left panels: under $H_0$, the middles panels: under $H_1$, where the horizontal axis is the mediator index, and the right panels: the average ROC curve.}
		\label{figEF}
	\end{figure}

	Next, we report the simulation results for multiple testing. Figure \ref{fig:fdr-supp} show the results for $d=100$ and $d=150$, respectively, which complement the results for $d=50$ reported in Section \ref{sec:simulations} of the paper. We again observe a similar qualitative pattern that both our test and the standard Benjamini-Yekutieli (BY) procedure achieve a valid false discovery control, whereas our method is more powerful than BY.

	\begin{figure}[t!]
		\centering
		\includegraphics[width=16cm]{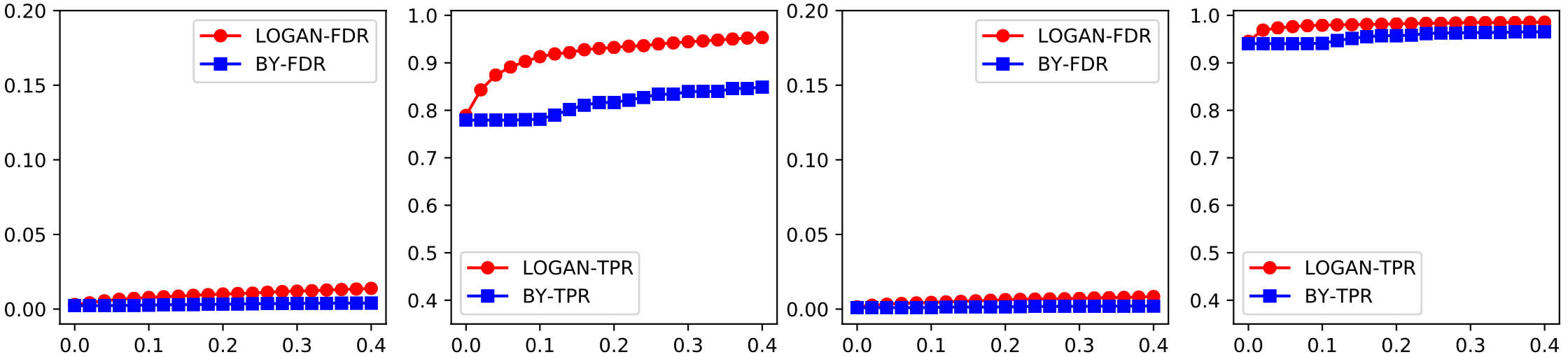}
		\includegraphics[width=16cm]{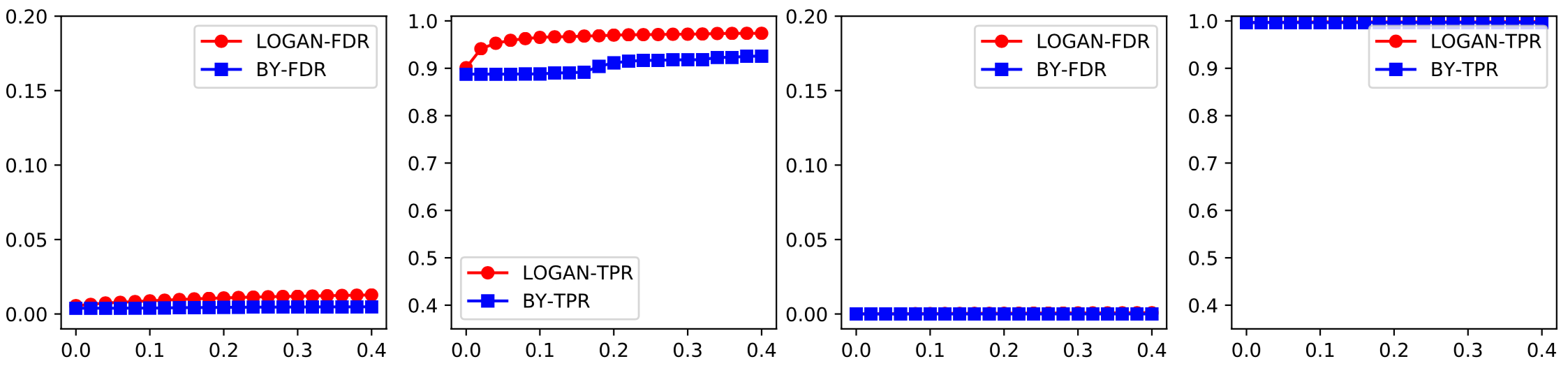}
		\caption{False discover rate and true positive rate of the proposed method and the  Benjamini-Yekutieli procedure. The horizontal axis corresponds to the significance level $\alpha$. The upper panels: $d=100$, and the bottom panels: $d=150$. The left two panels: $n=250$, and the right two panels: $n=500$.}
		\label{fig:fdr-supp}
	\end{figure}

	Finally, we evaluate the constant variance condition for the real data example. Following a similar idea of \cite{LiShen2019}, we compute and plot the residuals $X_{i,j}-\widetilde{\bm{W}}^{(\ell)}\bm{X}_i$, for $i\in \mathcal{I}_{\ell}^c$, $j\in\{0,\ldots,d+1\}$, $\ell=1,2$. Figure \ref{fig0-resid} shows the boxplots of such residuals for the amyloid negative and amyloid positive groups, respectively. It is seen that the lengths of boxes and whiskers are similar across different variables, suggesting no strong evidence against the constant variance condition.

	\begin{figure}[t!]
		\centering
		\includegraphics[width=16cm]{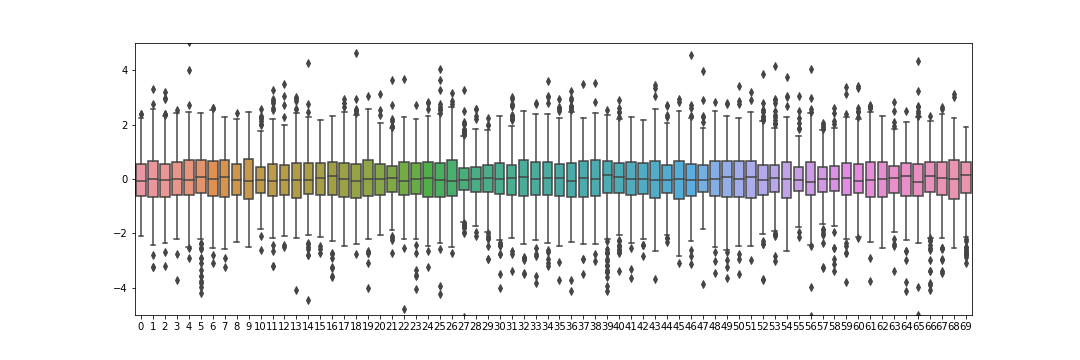}
		\includegraphics[width=16cm]{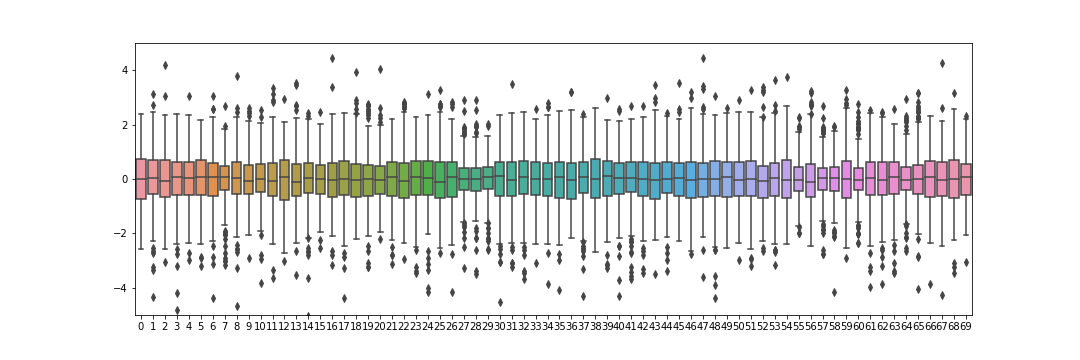}
		\caption{Residual plots for model diagnosis. The top panel: the amyloid negative group, and the bottom panel: the amyloid positive group.}
		\label{fig0-resid}
	\end{figure}  
}

\clearpage

\baselineskip=14pt
\bibliographystyle{apa}
\bibliography{ref-medtest}

\end{document}